\documentclass[longauth]{aa} 

\usepackage{graphicx}
\usepackage{xcolor}  
\graphicspath{{./plots/}} 
\usepackage{txfonts}
\newcommand{\msun}{M$_{\odot}$}

\newcommand{\kms}{km~s$^{-1}$}
\newcommand{\ergs}{erg s$^{-1}$}
\newcommand{\Ha}{H$\alpha$}

\newcommand{\HeI}{He~{\sc i}}

\newcommand{\OI}{O~{\sc i}}
\newcommand{\Oneb}{[O~{\sc i}]}
\newcommand{\CII}{C~{\sc ii}}
\newcommand{\NaI}{Na~{\sc i}}
\newcommand{\MgII}{Mg~{\sc ii}}
\newcommand{\MgI}{Mg~{\sc i}}

\newcommand{\SiII}{Si~{\sc ii}}

\newcommand{\CaII}{Ca~{\sc ii}}
\newcommand{\CaiiF}{[Ca~{\sc ii}]}
\newcommand{\TiII}{Ti~{\sc ii}}
\newcommand{\CrII}{Cr~{\sc ii}}

\newcommand{\FeII}{Fe~{\sc ii}}
\newcommand{\FeIII}{Fe~{\sc iii}}

\newcommand{\ArII}{Ar~{\sc ii}}
\newcommand{\CI}{C~{\sc i}}
\newcommand{\Fefs}{$^{56}$Fe}
\newcommand{\Cofs}{$^{56}$Co}
\newcommand{\Nifs}{$^{56}$Ni}
\newcommand{\mej}{$M_\mathrm{ej}$}
\newcommand{\ek}{$E_\mathrm{k}$}
\newcommand{\vph}{$v_\mathrm{ph}$}
\newcommand{\vsc}{$v_\mathrm{sc}$}
\newcommand{\lp}{$L_\mathrm{p}$}

\newcommand{\tdecay}{$t_{+1/2}$}
\newcommand{\eom}{$E_\mathrm{k}/M_\mathrm{ej}$}
\newcommand{\lam}{$\lambda$}

\newcommand{\Eh}{$E\left(B-V\right)_\mathrm{host}$}
\newcommand{\Emw}{$E\left(B-V\right)_\mathrm{MW}$}

\newcommand{\mni}{$M_\mathrm{Ni}$}

\newcommand{\tp}{$t_\mathrm{p}$}

\newcommand{\taum}{$\tau_\mathrm{m}$}

\begin{document}

   \title{The rise and fall of an extraordinary Ca-rich transient\thanks{Partially based on observations collected at the European Organisation for Astronomical Research in the Southern Hemisphere under ESO programmes 199.D-0143 and 0102.D-0137(A).}}

   \subtitle{The discovery of ATLAS19dqr/SN~2019bkc}
   \authorrunning{S. J. Prentice et al.}
   \author{S.~J.~Prentice
          \inst{1,2},
          K.~Maguire\inst{1},
          A.~Fl\"ors\inst{3,4,5},
          S.~Taubenberger\inst{3},
          C.~Inserra\inst{6},
          C.~Frohmaier\inst{7},
          T.~W.~Chen\inst{8},
          J.~P.~Anderson\inst{9},
          C.~Ashall\inst{10},
          P.~Clark\inst{2},
          M.~Fraser\inst{11},
          L.~Galbany\inst{12},
          A.~Gal-Yam\inst{13},
          M.~Gromadzki\inst{14},
          C.~P.~Guti\'errez\inst{15},
          P.~A.~James\inst{16},
          P.~G.~Jonker\inst{17},
          E.~Kankare\inst{19},
          G.~Leloudas\inst{20},
          M.~R.~Magee\inst{1},
          P.~A.~Mazzali\inst{16,3},
          M.~Nicholl\inst{21,22},
          M.~Pursiainen\inst{15},
          K.~Skillen\inst{1},
          S.~J.~Smartt\inst{2},
          K.~W.~Smith\inst{2},
          C.~Vogl\inst{3},
          \and D.~R.~Young\inst{2}.
          }

   \institute{$^1$School of Physics, Trinity College Dublin, The University of Dublin, Dublin 2, Ireland\\
              \email{sipren.astro@gmail.com}\\
   $^2$Astrophysics Research Centre, School of Mathematics and Physics, Main Physics Building, Queen's University Belfast, Belfast, County Antrim, BT7 1NN, United Kingdom\\
   $^3$Max-Planck-Institut f\"ur Astrophysik, Karl-Schwarzschild-Stra\ss e 1, D-85748 Garching bei M\"unchen, Germany\\
   $^4$European Southern Observatory, Karl-Schwarzschild-Stra\ss e 2, D-85748 Garching bei M\"unchen, Germany\\
   $^5$Physik-Department, Technische Universit\"at M\"unchen, James-Franck-Stra\ss e 1, D-85748 Garching bei M\"unchen, Germany\\
   $^{6}$School of Physics \& Astronomy, Cardiff University, Queens Buildings, The Parade, Cardiff, CF24 3AA, UK\\
   $^{7}$Institute of Cosmology and Gravitation, University of Portsmouth, Portsmouth PO1 3FX, UK\\
   $^{8}$Max-Planck-Institut f{\"u}r Extraterrestrische Physik, Giessenbachstra\ss e 1, 85748, Garching, Germany\\
   $^{9}$European Southern Observatory, Alonso de C\'ordova 3107, Casilla 19, Santiago, Chile\\
   $^{10}$Department of Physics, Florida State University, Tallahassee, FL 32306, USA\\
   $^{11}$School of Physics, O’Brien Centre for Science North, University College Dublin, Belfield, Dublin 4, Ireland\\
   $^{12}$Departamento de F\'isica Te\'orica y del Cosmos, Universidad de Granada, E-18071 Granada, Spain\\
   $^{13}$Department of Particle Physics and Astrophysics, Weizmann Institute of Science, Rehovot 76100, Israel\\
   $^{14}$Astronomical Observatory, University of Warsaw, Al. Ujazdowskie 4, 00-478 Warszawa, Poland\\
   $^{15}$Department of Physics and Astronomy, University of Southampton, Southampton, SO17 1BJ, UK\\
   $^{16}$Astrophysics Research Institute, Liverpool John Moores University, IC2, Liverpool Science Park, 146 Brownlow Hill, Liverpool L3 5RF, UK\\
   $^{17}$SRON, Netherlands Institute for Space Research, Sorbonnelaan 2, 3584 CA, Utrecht, The Netherlands\\
   $^{18}$Department of Astrophysics/IMAPP, Radboud University, P.O. Box 9010, 6500 GL, Nijmegen, The Netherlands\\
   $^{19}$Tuorla Observatory, Department of Physics and Astronomy, University of Turku, FI-20014 Turku, Finland\\
   $^{20}$DTU Space, National Space Institute, Technical University of Denmark, Elektrovej 327, 2800 Kgs. Lyngby, Denmark\\
   $^{21}$Institute for Astronomy, University of Edinburgh, Royal Observatory, Blackford Hill, EH9 3HJ, UK \\
   $^{22}$Birmingham Institute for Gravitational Wave Astronomy and School of Physics and Astronomy, University of Birmingham, Birmingham B15 2TT, UK \\
              }


   \date{Received xxx; accepted xxx}

 
  \abstract
{This work presents the observations and analysis of ATLAS19dqr/SN 2019bkc, an extraordinary rapidly evolving transient event located in an isolated environment, tens of kiloparsecs from any likely host. Its light curves rise to maximum light in $5-6$ d and then display a decline of $\Delta m_{15} \sim5$ mag. With such a pronounced decay, it has one of the most rapidly evolving light curves known for a stellar explosion. 
The early spectra show similarities to normal and `ultra-stripped' type Ic SNe, but the early nebular phase spectra, which were reached just over two weeks after explosion, display prominent calcium lines, marking SN 2019bkc as a Ca-rich transient. The Ca emission lines at this phase show an unprecedented and unexplained blueshift of 10\,000 -- 12\,000 \kms. 
Modelling of the light curve and the early spectra suggests that the transient had a low ejecta mass of $0.2 - 0.4$ \msun\ and a low kinetic energy of $ (2-4)\times 10^{50}$ erg, giving a specific kinetic energy \eom\ $\sim1$ [$10^{51}$ erg]/\msun.
The origin of this event cannot be unambiguously defined. While the abundance distribution used to model the spectra marginally favours a progenitor of white dwarf origin through the tentative identification of \ArII, the specific kinetic energy, which is defined by the explosion mechanism, is found to be more similar to an ultra-stripped core-collapse events. SN 2019bkc adds to the diverse range of physical properties shown by Ca-rich events. }


   \maketitle
%


\section{Introduction}
The advent of large-scale high-cadence transient surveys in the past decade has led to the discovery of new classes of transients that lie significantly outside the parameter space of traditional classes of supernovae (SNe). In particular, the many repeat visits to the same patch of sky has led to a dramatic increase in the number of transients that are much faster evolving than typically seen in more traditional events, such as SNe Ia, Ibc, II. Some of these are highly luminous and rapidly evolving \citep{Prentice2018b,Pursiainen2018,Drout2014,Poznanski2010}, while others are `faint' and fast evolving events \citep[e.g.][]{Perets2010,Kasliwal2010,Drout2013,Valenti2014,Inserra2015}. These faint events were missed in shallower, lower cadence surveys and recent results suggest there may be large populations of these events that previously went undetected \cite[e.g.][]{Frohmaier2018}.

The untargeted nature of these recent transient searches compared to previous galaxy-targeted surveys has also resulted in a more complete census of transients occurring in locations that are significantly offset ($>$10 kpc) from their host galaxy centres. The class of so-called `Ca-rich' transients are of particular interest, they are named after the unusually strong forbidden \CaiiF\ emission (relative to typical [\OI] lines) that develop in their spectra a few months after maximum \citep[for a definition, see][]{Kasliwal2012}. At early times, the spectra of most Ca-rich events show \HeI\ features with SN-like velocities of $\sim$10000 \kms\ and they are most commonly classified as peculiar Type Ib SNe from maximum-light spectroscopy \citep{Perets2010,Gal-Yam2017}. However, their fast evolution to an optically thin regime (`nebular phase') and large \CaiiF\ to \Oneb\ ratio differentiates them from typical Ib SNe. Although the sample size of Ca-rich events is small because of their faintness and fast evolution, the intrinsic rate of Ca-rich transients is measured to be very high at 33--94\%\ of the local volumetric SN Ia rate \citep{Frohmaier2018}. 

The majority of Ca-rich transients occur in an unusual parameter space of galaxy environments; they prefer galaxy group environments with large offsets from early-type galaxies, which suggests an association with an old stellar population \citep{Perets2010,Lyman2013,Lyman2016b,Lunnan2017}. Strict limits on underlying dwarf galaxies or globular clusters have also been placed \citep{Lyman2013,Lyman2016b}. Their preference for offsets from their potential host galaxies can be explained by a very old metal-poor stellar population  \citep{Yuan2013}.  The observed (and unexplained) Ca enrichment of the intra-cluster medium can be explained with a significant rate of Ca-rich transients \citep{Mulchaey2014,Frohmaier2018}.

The origin of Ca-rich events is an area under much debate and a number of explosion channels have been suggested. The most obvious explanation given their preference for remote locations in galaxy groups and clusters is a system involving a white dwarf. Proposed models involving a white dwarf include a He-shell detonation on the surface of a white dwarf \citep{Perets2010,Shen2010,Woosley2011,Waldman2011,Sim2012}, the tidal disruption (and detonation) of a white dwarf by an intermediate-mass black hole \citep{Rosswog2008,Macleod2014,Sell2015}, or the merger of a white dwarf with a neutron star \citep{Metzger2012,Margalit2016}. 
The white tidal disruption model is expected to produce an X-ray signature \cite[e.g.][]{Sell2015}, which has not been observed in a Ca-rich event as of yet. 
To explain the large offset and the lack of underlying stellar population, it has been suggested that the objects were not formed at their explosion positions but travelled to these locations perhaps due to a SN kick \citep{Lyman2014b} or interaction with a supermassive black hole \citep{Foley2015}. However, one argument in favour of this `kicked' scenario based on the line shift of the \CaiiF\ feature \citep{Foley2015} has been questioned by \citet{Mili2017} because of the lack of a coeval observed shift in the \OI\ lines. 
Nethertheless, there is still a relative lack of Ca-rich SNe within galaxies which would need to be explained.
The need for formation at a different location to explosion is somewhat mitigated by the discovery of the association of these events with group or cluster environments since the intracluster space may contain more stars due to galaxy-galaxy interaction and galaxy mergers than typically found at large distances from galaxies in non-cluster environments. 
Additionally, SNe Ia can be found in these environments \citep{GalYam2013}.

A minority of events that show Ca-rich spectra at late times have also been detected in star-forming galaxies, such as the unusually slowly evolving and brighter than normal iPTF15eqv that also displayed \Ha\ emission in its late-time spectra \citep{Mili2017} or with double-peaked light curves \citep[SN 2014ft/iPTF14gqr and iPTF16hgs;][]{De2018,De2018b}. 
It is likely that the events associated with higher star-formation environments are not members of the strict class of `Ca-rich' events as defined \cite[see][]{Kasliwal2012, Lunnan2017} and their progenitor may be towards the lower end of the core-collapse progenitor mass range \citep[$<12$ \msun, see][]{Suh2011,Mili2017,De2018b}, perhaps ones that are highly stripped in binary systems \citep{Tauris2015,Moriya2017}.

In this work we present the observations and analysis of the rapidly evolving Ca-rich transient, ATLAS19dqr/SN 2019bkc, which shows, for the first time in a Ca-rich event, a very large shift in its \CaiiF\ lines of nearly 11\,000 \kms\ at $\sim$3 weeks after explosion. An analysis of SN 2019bkc was presented in \citet{Chen2020} in which they noted its rapid light curve and classification as a ``peculiar SN I'' but lacked the late-time spectra that confirm it to be of the Ca-rich class. Stricter pre-detection photometry limits presented here also place tighter constraints on the evolution of the early-light curve (Section~\ref{sec:lcs}).
In Section~\ref{sec:obs} we discuss the discovery and host-environment, while the photometric and spectroscopic data are presented in Sections~\ref{sec:lcs} and ~\ref{sec:spectra} respectively. In Section~\ref{sec:models} the photospheric spectra are modelled in order to determine physical characteristics of the explosions. Section~\ref{sec:discuss} covers the discussion of our results and the possible scenarios that may have led to this explosion. finally, in Section~\ref{sec:conc} we bring together our conclusions and summarise the paper.

\begin{figure*}
    \centering
    \includegraphics[scale=0.33, ]{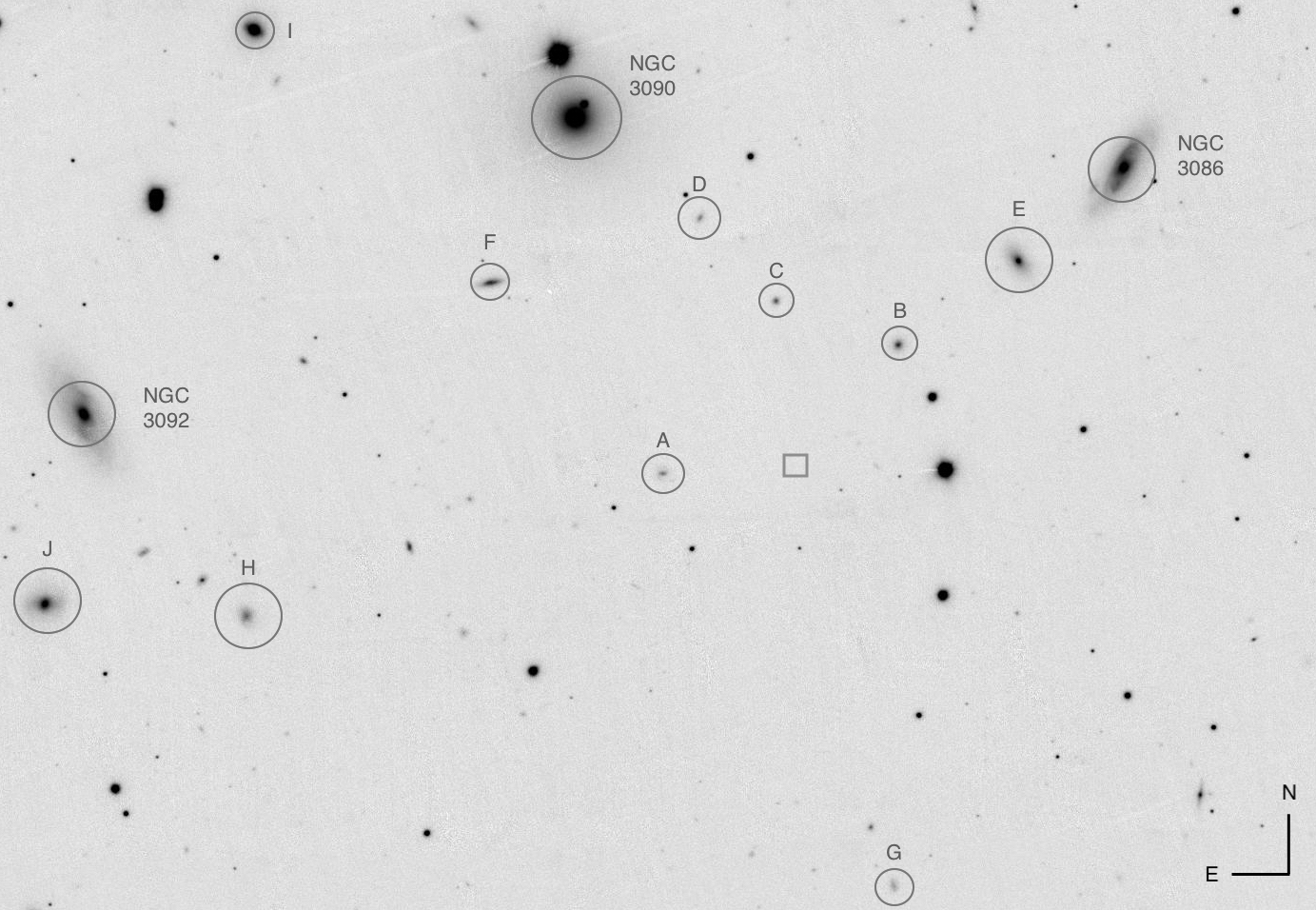}
    \caption{Stacked archive PanSTARRS $g$-band image showing galaxies in the region around SN 2019bkc (square). The labelled galaxies are at the redshift of the NGC 3090 galaxy group ($z=0.020$), apart from two significantly more distant galaxies at $z=0.188$ (labelled A and C) that are within 100 arcsec of the position of SN 2019bkc. }
    \label{fig:gals}
\end{figure*}

\begin{table*}[]
    \caption{Galaxies with measured redshifts within 400 arcsec of the position of SN 2019bkc. The labels (A--J) correspond to those shown on Fig.~\ref{fig:gals}. }
    \label{tab:gals}
    \centering
    \begin{tabular}{lccc}
    \hline
    Galaxy  &   $z$     &   Angular separation  &   Projected distance    \\
            &                      &   [arcsec]           &   [kpc]     \\
    \hline
    A - 2dFGRS TGN216Z080  &    0.188 & 69.29 & 306.7 \\
    B - 2dFGRS TGN216Z084      &          0.019      &  84.58     &  34.1  \\
    C - 2dFGRS TGN216Z082  &  0.188 & 88.19 & 393.7 \\
    D - 2dFGRS TGN216Z079     &        0.018     &    140.28   &  52.0       \\
    E - 2dFGRS TGN216Z089   &     0.022    &  160.51     &  74.5       \\
    F - 2dFGRS TGN217Z150     &       0.023      &  187.02     &   91.0      \\
    NGC 3090     &       0.020       & 219.02      &  94.6       \\
    *G - 2dFGRS TGN216Z085     &      0.043     &  225.02     &   209.8      \\
    NGC 3086     &        0.023     &  234.56     &   113.0      \\
    
    $\dag$6dFGS gJ100027.0-025650  &        0.022   &  270.67     &   128.9      \\
    H - 2dFGRS TGN216Z069      &        0.023     &  297.15     &     145.9    \\
    I - 2dFGRS TGN216Z072  & 0.020 & 365.04  & 159.0 \\
    $\dag$2dFGRS TGN216Z103 &  0.022& 365.86  & 171.2 \\
    
    NGC 3092    &       0.020     &  373.83     &  158.0       \\
     J - 2dFGRS TGN217Z145    &       0.019     &  399.60     & 147.7         \\
    \hline
    \multicolumn{4}{p{\textwidth}}{$\dag$Outside the field of view of Fig.~\ref{fig:gals}}\\
    \multicolumn{4}{p{\textwidth}}{*This is an updated redshift obtained with the LT:SPRAT. The previously reported redshift from \citet{Colless2001} was $z=0.0003$.}\\

    \end{tabular}
\end{table*}

\section{Observations}\label{sec:obs}
ATLAS19dqr was discovered by the ATLAS survey \citep{Tonry2018} on 2019-03-02 10:24:57 UT (MJD 58544.43) at a magnitude of 17.59 in the ATLAS \textit{o} band \citep{Tonry19bkcTNS}. It was subsequently reported to the Transient Name Server\footnote{https://wis-tns.weizmann.ac.il/} (TNS) and was given the designation of SN 2019bkc, which we use throughout.
There were robust non-detections in the ATLAS o-band on the four individual 30 second
frames taken around the mean time of MJD 58540.45. Forced photometry on each of these provided a flux limit and the weighted mean and standard deviation of these flux measurements was $5.2\pm14.4$ $\mu$Jksy. This corresponds to a 3$\sigma$ upper limit of $o > 19.8$ AB mag.
The transient had been observed prior to the ATLAS discovery by the Zwicky Transient Facility \citep[ZTF;][]{Bellm2019} on 2019-02-27 06:07:40 UT (MJD 58541.26) at $g = 18.9\pm{0.1}$ mag, obtained from the LASAIR broker \citep{Smith2019}.

Multi-colour $ugriz$ photometry  with the IO:O imager on the Liverpool Telescope \citep[LT;][]{Steele2004}  commenced on 2019-03-03 22:34:54 UT (MJD 58546.02) and was reduced using the standard IO:O pipeline. Aperture photometry was performed on the field using a series of custom {\sc python} scripts utilising {\sc pyraf} and calibrated to the Sloan Digital Sky Survey photometric system \citep{Ahn2012}.

SN 2019bkc was also observed as part of the GREAT survey \citep{Chen2018}, using the Gamma-Ray Burst Optical/Near-Infrared Detector \citep[GROND;][]{Greiner2008}, a 7-channel imager that collects multi-colour photometry simultaneously in the $g'r'i'z'JHKs$ bands, which is mounted at the 2.2 m MPG telescope at European Southern Observatory (ESO) La Silla Observatory in Chile. The images were reduced by the GROND pipeline \citep{Kruhler2008}, which applies de-bias and flat-field corrections, stacks images, and provides astrometric calibration.

Optical spectroscopy of SN 2019bkc using SPRAT \citep{Piascik2014} on the LT was also obtained starting 1.5 d after discovery and were reduced using the standard SPRAT pipeline \citep{Piascik2014} with a further flux calibration to a standard star from the \citet{Oke1990} catalogue. Two epochs of optical spectroscopy and one epoch of photometry were obtained with the EFOSC2 imaging spectrograph \citep{Buzzoni1984} on the ESO New Technology Telescope as part of the extended Public ESO Spectroscopic Survey of Transient Objects \citep[ePESSTO;][]{Smartt2015}. At each epoch both grism \#11 (13.8 \AA\ resolution) and grism \#16 (13.4 \AA\ resolution) spectra were obtained and were reduced using the custom pipeline described in \citet{Smartt2015}.

X-Shooter \citep{vernet2011} on ESO's Very Large Telescope (VLT) was used to obtain two epochs of optical and near-infrared (NIR) spectroscopy. The spectra were initially reduced using the Reflex software package \citep{
Modigliani2010,Freudling2013} to produce two-dimensional spectra, which were then extracted using a custom-built pipeline\footnote{https://github.com/jselsing/xsh-postproc}. Spectroscopic observations will be made publicly available on The Weizmann Interactive Supernova Data Repository (WISeREP) \footnote{https://wiserep.weizmann.ac.il/}.

\subsection{Swift X-ray observation}
We observed the field of SN 2019bkc on MJD 58583.09 with the Neil Gehrels Swift Observatory (Swift) X-ray telescope (XRT)
The XRT data were reduced using the HEASARC FTOOLS pipeline {\sc xrtpipeline}. In a circle with radius of 47\arcsec\, (this is the 90\% encircled energy radius at 1.5 keV) centred on the optical source position, we find two X-ray photons in an effective exposure time of 2767 s. In a source-free circular region on the sky close to the source with a radius of 200\arcsec, we found 52 background counts, making the expectation value for the background at the location of SN 2019bkc 2.9 counts. Following \citet{Kraft1991}, we derived a 95\% confidence level upper limit of 4.5 source counts or $1.6\times10^{-3}$ counts s$^{-1}$. Using W3PIMMS\footnote{https://heasarc.gsfc.nasa.gov/cgi-bin/Tools/w3pimms/w3pimms.pl} with N$_H$ $=5\times 20$ cm$^{-2}$ and a power law model with an index of 2.0 for the assumed source spectral energy distribution (SED), we converted this limit to an unabsorbed 0.2--10 keV flux of $6.6\times 10^{-14}$ erg cm$^{-2}$ s$^{-1}$, corresponding to a luminosity of $<6.3\times 10^{40}$ \ergs.

\subsection{Host environment and redshift}\label{sec:host}
SN 2019bkc was observed in a region densely populated by galaxies (Fig~\ref{fig:gals}, properties tabulated in Table~\ref{tab:gals}), which contains a galaxy group with members in the redshift range of 0.018 -- 0.022. 
There is no host visible at the position of SN 2019bkc and pre-explosion imaging from the Dark Energy Camera Legacy Survey\footnote{https://datalab.noao.edu/} \citep[DECaLS;][]{Dey2019} puts a limit of $g\sim 24.7$ mag, $r\sim 23.9$ mag, $z\sim 23.0$ mag on any underlying non-detected host. The brightest galaxy in the MKW1 group, NGC 3090, is a giant elliptical galaxy with a redshift of 0.0203 \citep{Morgan1975,Faber1989}. 

Similar to \citet{Chen2020}, we assumed that SN 2019bkc is associated with the galaxy group, MKW1 \citep{Morgan1975}, and have adopted a redshift of 0.020 for the transient for two reasons: firstly, this is the redshift of the most massive galaxy in the cluster and therefore the most likely host, and secondly, this is the mean of the redshifts of the galaxies in the cluster.  There is a more distant group of galaxies at $z\sim0.2$ immediately surrounding the transient but exclude any of these as the potential host on the basis that the photospheric-phase spectra are a better match for SNe at $z\sim0.02$ (see Section~\ref{sec:spectra}). 

Using Nine-year Wilkinson Microwave Anisotropy Probe cosmological parameters $H_0=69.32$ km s$^{-1}$ Mpc $^{-1}$, $\Omega_m=0.286$, $\Omega_\Lambda=0.714$ \citep{Hinshaw2013} and a redshift of $z=0.020$ gives a luminosity distance, $D_L=89.1$ Mpc and a distance modulus, $\mu = 34.75$ mag. The Milky Way extinction in the direction of SN 2019bkc is \Emw$=0.05$ mag \citep{Schlafly2011} with an assumed $R_V$ of 3.1.  We take \Eh\ $=0$ mag due to the lack of obvious underlying host and lack of \NaI\ D absorption lines. 
At $z=0.02$, the absolute magnitude of any underlying host galaxy or globular cluster would be fainter than $M_g\sim -10.1$ mag, $M_r\sim -10.9$ mag, and $M_z\sim -11.5$ mag, which is insufficiently deep to rule out most globular clusters \citep[e.g.][]{Lyman2016b}.

\section{Light curves}\label{sec:lcs}

\begin{figure}
    \centering
    \includegraphics[scale=0.6]{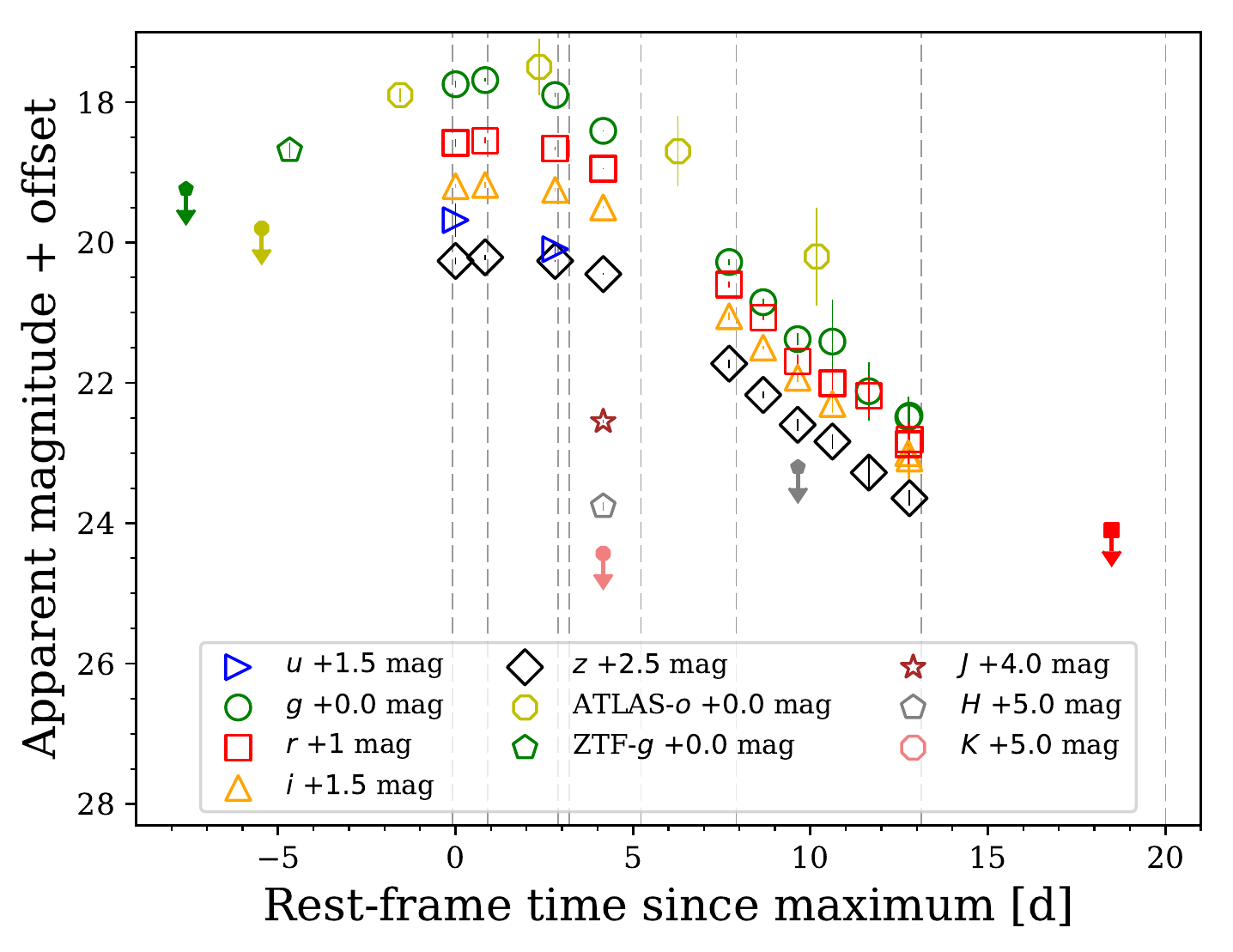}
    \includegraphics[scale=0.63]{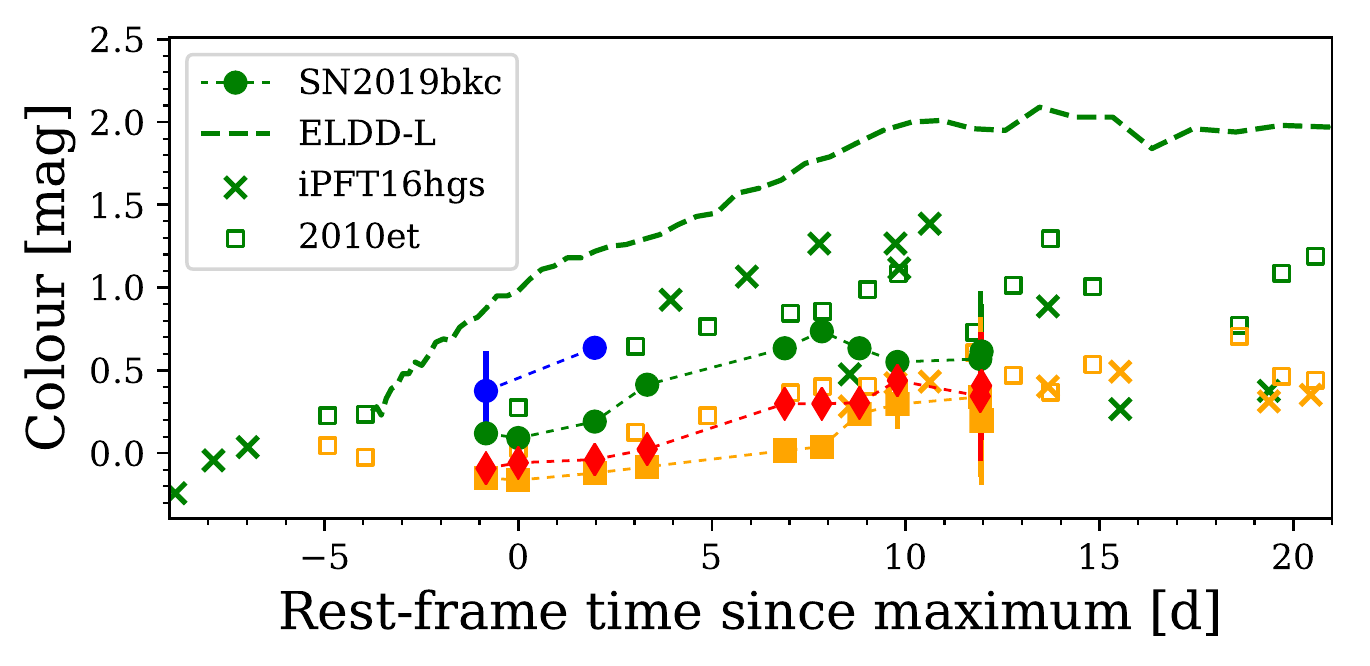}
    \caption{(Top) The multi-colour light curves of SN 2019bkc from ZTF, ATLAS, LT, GROND, and NTT. NIR observations are from GROND and the LT. The grey dashed lines denote spectroscopic observations. (Bottom) Colour evolution in $u-g$ (blue), $g-r$ (green), $r-i$ (orange), and $i-z$ (red). The dashed line is the $g-r$ colour curve of the \citet{Sim2012} low-mass edge-lit double detonation (ELDD-L) model and SN 2019bkc is seen to be significantly bluer than this model. Also shown are the  $g-r$ and $r-i$ colour curves of iPTF16hgs \citep{De2018b} and SN 2010et \citep{Kasliwal2012}, with the latter requiring interpolation to match the observed dates.  }
    \label{fig:lcs}
\end{figure}

\begin{table}[]
    \centering
    \caption{Table of optical and NIR photometry}
    \begin{tabular}{lcccc}
    \hline
    MJD &  Band &  Apparent magnitude  & Facility \\
    \hline
     58546.02 & u & 18.1 $\pm{0.2}$ & LT \\
 58548.88 & u & 18.59 $\pm{0.07}$ & LT \\
 58546.02 & g & 17.74 $\pm{0.05}$ & LT \\
 58546.87 & g & 17.68 $\pm{0.02}$ & LT \\
 58548.88 & g & 17.89 $\pm{0.03}$ & LT \\
 58553.88 & g & 20.27 $\pm{0.04}$ & LT \\
 58554.86 & g & 20.85 $\pm{0.05}$ & LT \\
 58555.85 & g & 21.37 $\pm{0.08}$ & LT \\
 58556.85 & g & 21.4 $\pm{0.6}$ & LT \\
 58557.89 & g & 22.1 $\pm{0.4}$ & LT \\
 58559.03 & g & 22.4 $\pm{0.3}$ & LT \\
 58546.02 & r & 17.58 $\pm{0.05}$ & LT \\
 58546.87 & r & 17.54 $\pm{0.04}$ & LT \\
 58548.88 & r & 17.66 $\pm{0.03}$ & LT \\
 58553.88 & r & 19.59 $\pm{0.04}$ & LT \\
 58554.86 & r & 20.06 $\pm{0.03}$ & LT \\
 58555.85 & r & 20.69 $\pm{0.01}$ & LT \\
 58556.85 & r & 21.0 $\pm{0.2}$ & LT \\
 58557.89 & r & 21.1 $\pm{0.3}$ & LT \\
 58559.03 & r & 21.8 $\pm{0.3}$ & LT \\
 58564.87 & r & >23.1  & LT \\
 58546.02 & i & 17.69 $\pm{0.03}$ & LT \\
 58546.87 & i & 17.68 $\pm{0.04}$ & LT \\
 58548.88 & i & 17.74 $\pm{0.02}$ & LT \\
 58553.88 & i & 19.55 $\pm{0.06}$ & LT \\
 58554.86 & i & 19.99 $\pm{0.03}$ & LT \\
 58555.85 & i & 20.42 $\pm{0.06}$ & LT \\
 58556.85 & i & 20.8 $\pm{0.1}$ & LT \\
 58559.03 & i & 21.5 $\pm{0.4}$ & LT \\
 58546.02 & z & 17.76 $\pm{0.04}$ & LT \\
 58546.87 & z & 17.71 $\pm{0.04}$ & LT \\
 58548.88 & z & 17.76 $\pm{0.02}$ & LT \\
 58553.88 & z & 19.22 $\pm{0.06}$ & LT \\
 58554.86 & z & 19.67 $\pm{0.05}$ & LT \\
 58555.85 & z & 20.09 $\pm{0.08}$ & LT \\
 58556.85 & z & 20.3 $\pm{0.1}$ & LT \\
 58557.89 & z & 20.7 $\pm{0.3}$ & LT \\
 58555.86 & H & >18.2  & LT \\
 58559.06 & g & 22.4 $\pm{0.1}$ & NTT \\
 58559.06 & r & 21.8 $\pm{0.2}$ & NTT \\
 58559.06 & i & 21.5 $\pm{0.3}$ & NTT \\
 58559.06 & z & 21.1 $\pm{0.1}$ & NTT \\
 58550.26 & g & 18.41 $\pm{0.02}$ & GROND \\
 58550.26 & r & 17.95 $\pm{0.02}$ & GROND \\
 58550.26 & i & 18.00 $\pm{0.02}$ & GROND \\
 58550.26 & z & 17.95 $\pm{0.02}$ & GROND \\
 58550.26 & J & 18.55 $\pm{0.03}$ & GROND \\
 58550.26 & H & 18.76 $\pm{0.06}$ & GROND \\
 58550.26 & K & >19.4  & GROND \\
 58540.45 & o & >19.8  & ATLAS \\
 58544.43 & o & 17.9 $\pm{0.1}$ & ATLAS \\
 58548.42 & o & 17.5 $\pm{0.4}$ & ATLAS \\
 58552.41 & o & 18.7 $\pm{0.5}$ & ATLAS \\
 58556.40 & o & 20.2 $\pm{0.7}$ & ATLAS \\

    \hline
    \end{tabular}
    \label{tab:phottab}
\end{table}

\begin{table}[]
    \centering
    \caption{Multi-colour light curve properties including the absolute magnitude at peak for each of the bands and the time for each light curve to decay to half its peak luminosity. }
    \begin{tabular}{ccc}
       \hline
       Band  & $M_\mathrm{peak}$ & \tdecay  \\
          & [mag] & [d]  \\
       \hline
       $u$ &    $-17.3\pm{0.1}$              &    -       \\
       $g$ &     $-17.19\pm{0.02}$              &    3.8$\pm{0.1}$       \\
       $r$ &    $-17.33\pm{0.03}$               &    4.3$\pm{0.1}$       \\
       $i$ &    $-17.15\pm{0.05}$               &    4.4$\pm{0.3}$       \\
       $z$ &    $-17.09\pm{0.03}$               &    4.5$\pm{0.4}$       \\

        \hline    
    \end{tabular}
    
    \label{tab:photstats}
\end{table}

The multi-colour photometry and colour curves of SN 2019bkc are shown in Fig.~\ref{fig:lcs} and the calibrated photometry is presented in Table~\ref{tab:phottab}. The ATLAS non-detection and ZTF detection are less than 20 hours apart (0.80 d) but are separated by $\sim$1 magnitude in brightness (albeit in different bands, ZTF-\textit{g} and ATLAS-\textit{o}). This suggests that the transient exploded some time close to the non-detection.

The LT $g$-band light curve rises to a peak on MJD $58546.9\pm{0.5}$, which is $5-6$ d after the ZTF detection. The $riz$ bands also peak around this time. The $u$-band is not observed on the rise, but from the earliest $ugriz$ SED we estimate a temperature of $\sim8700$ K, which suggests that  maximum light in the $u$-band occurred close to the first observation. 
We set our fiducial reference point throughout as the time of pseudo-bolometric maximum light, MJD $58546.02\pm{0.25}$ (see Section \ref{bolo}). 

GROND observations at $+$4.24 d after pseudo-bolometric peak brightness provide the only detection in the NIR and show that the object was not NIR-bright -- when the SED is fitted with a black body, no significant NIR excess is seen. NIR imaging from the LT with IO:I provided a non-detection in \textit{H} with a limiting magnitude of $>18.2$ mag at $+$9.84 d.
Further observations showed a decay rate of $\sim 0.5$ mag d$^{-1}$ and a total decline of $3.4 - 4.8$ mag in $\sim12$ d across $griz$ with the redder bands decaying slower.
Extrapolation of the $g$-band light curve gives $\Delta m_{15} (g) \sim 5$ mag. 

The $I$-band light curve point of SN 2019bkc at $+18$ d after maximum from \citet{Chen2020} is consistent with a \Cofs\ tail. 
The absolute peak magnitudes and times for the luminosities in each band to decay by half their peak luminosity are given in Table \ref{tab:photstats}.

\subsection{Comparison of light curves}\label{sec:lightcurves}
\begin{figure}
    \centering
    \includegraphics[scale=0.65]{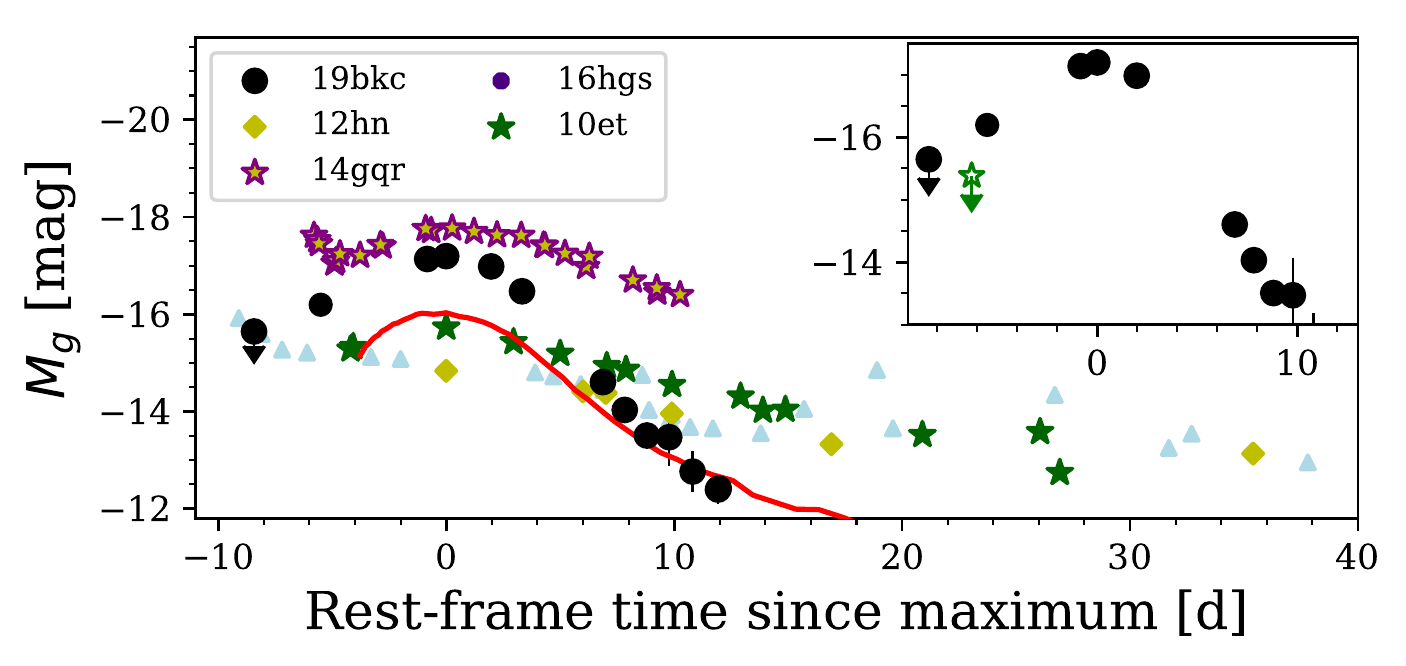}
    \includegraphics[scale=0.65]{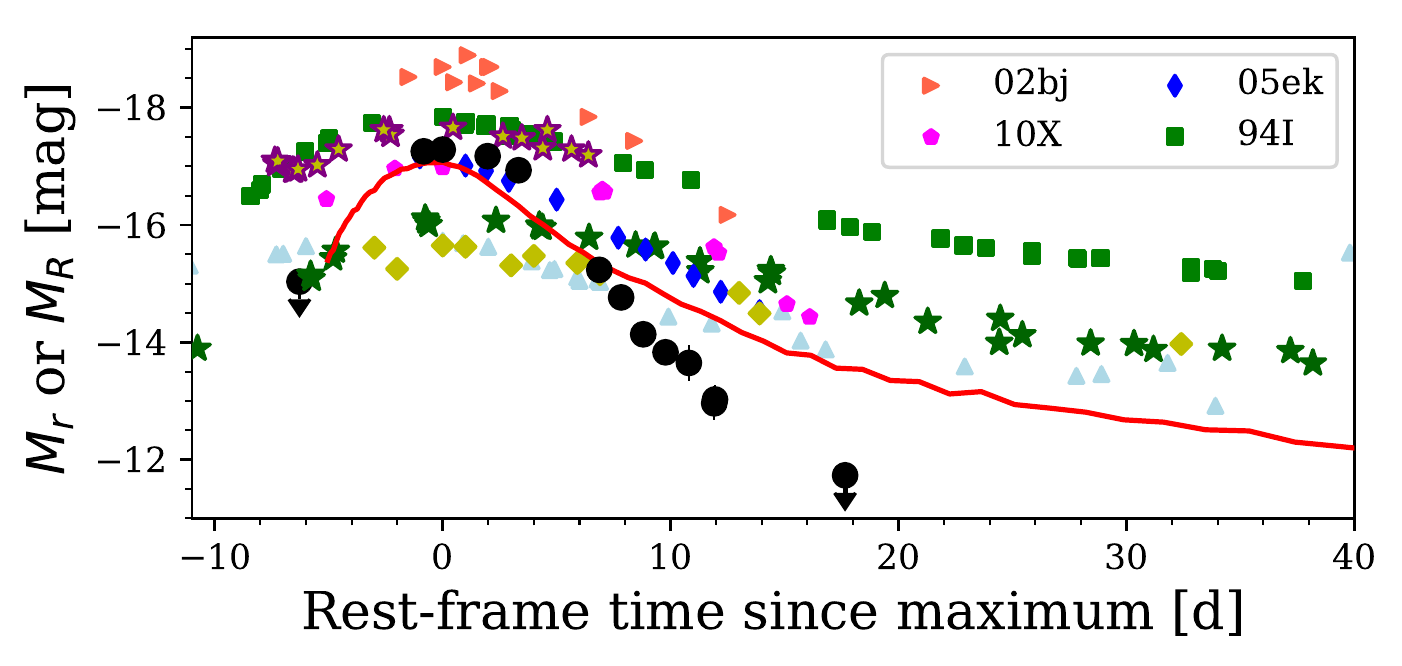}
    \includegraphics[scale=0.65]{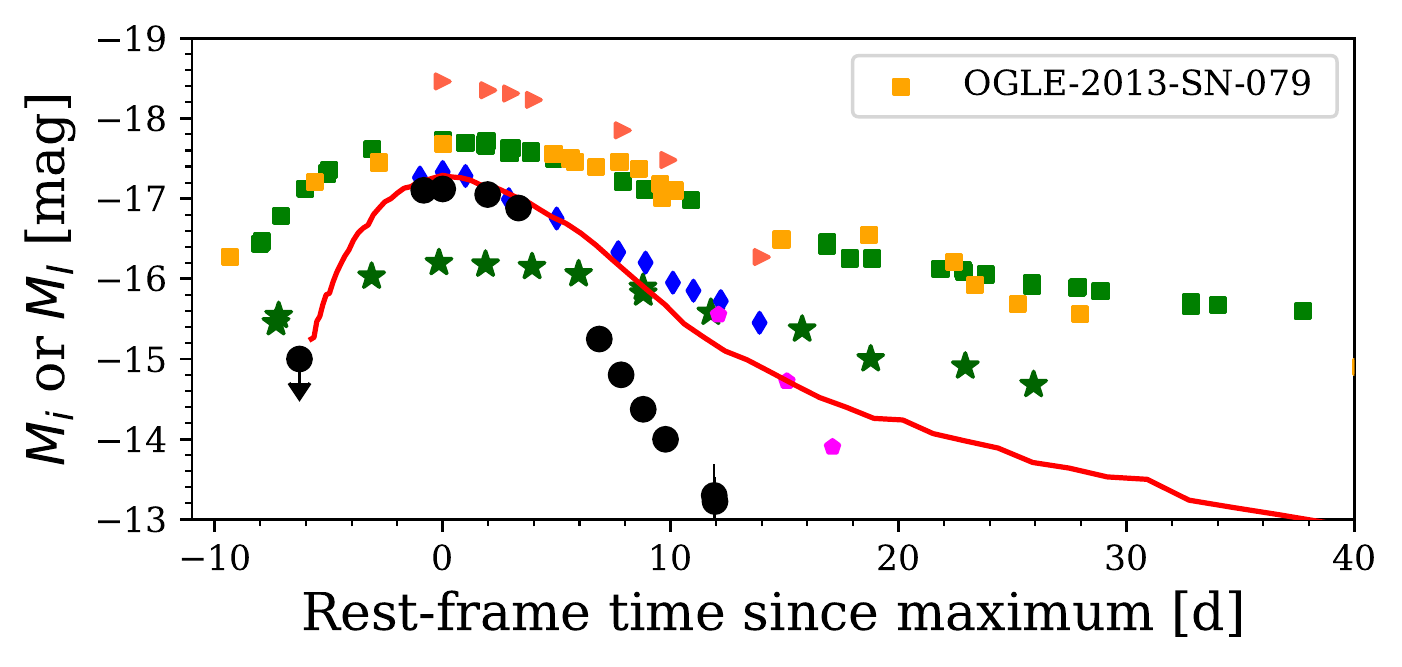}
    \caption{(Top) $g$-band light curves of various fast transients; SN 2019bkc and the Ca-rich SNe 2010et , 2012hn, iPTF14gqr, and iPTF16hgs. The red line is the ELDD-L model from \citet{Sim2012}. The inset shows the \textit{g} band light curve with an estimated upper limit of 19.5 mag (green star) derived from the  ATLAS-$o$ band magnitude of 19.8 assuming the SED is a black body at $\sim20000$ K. (Middle) $r$ or $R$ band light curves including SNe Ic 2005ek and 1994I,  and the unusual transients SNe 2010X and 2002bj. (Bottom) $i$ or $I$ band comparison with the addition of the Ti-dominated  transient OGLE-2013-SN-079. The initial ATLAS-$o$ non-detection is treated as if it was in $r$ and $i$ respectively.}
    \label{fig:bands}
\end{figure}

\begin{figure}
    \centering
    \includegraphics[scale=0.63]{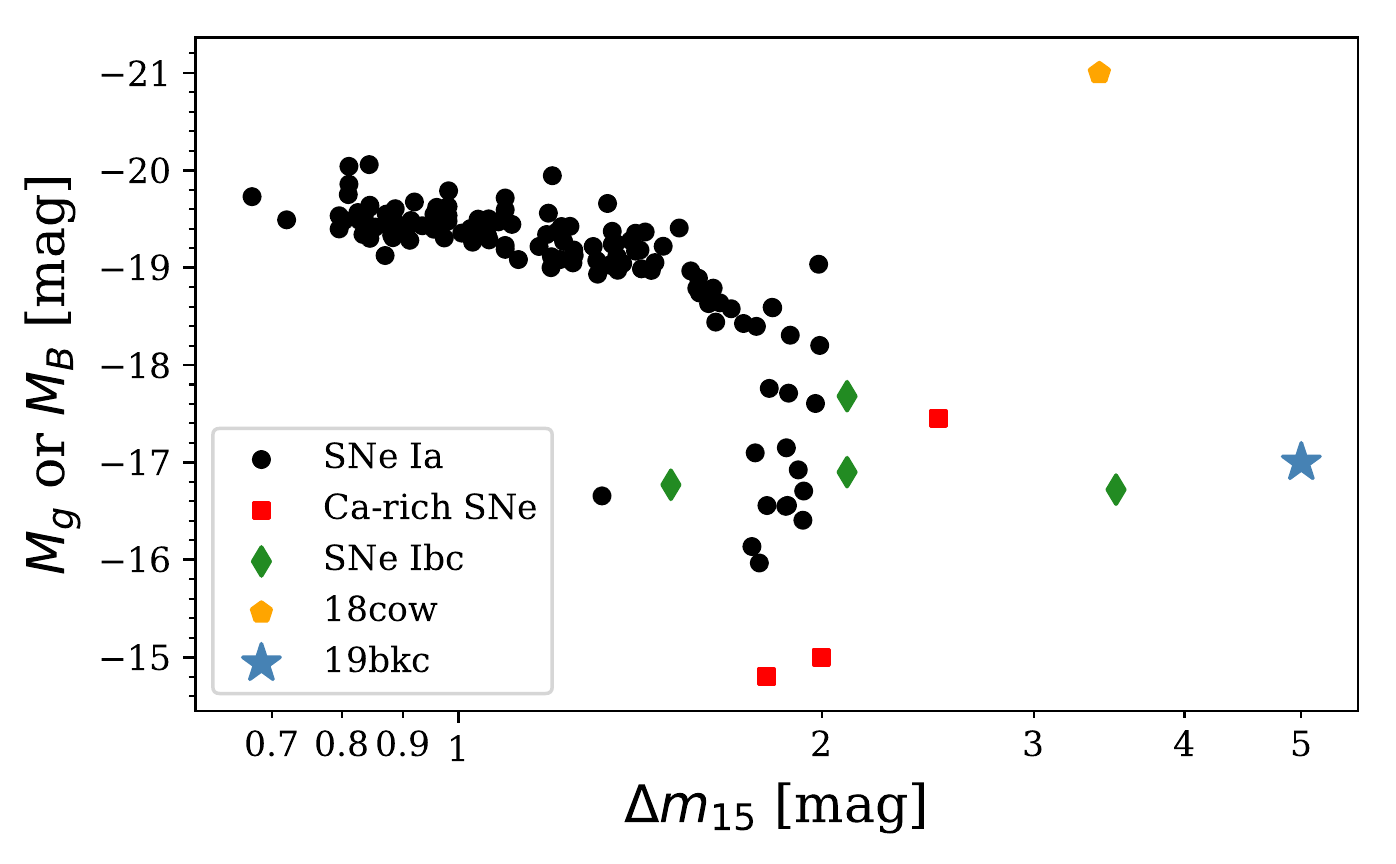}
    \caption{Peak absolute magnitude in $B$ or $g$ as a function of decline rate for a selection of transients and SN 2019bkc. Differences in magnitude system and filter bandpass are negligible for the purpose of this plot which shows that the rapid decline of SN 2019bkc is unprecedented compared to the comparison objects. The SNe~Ia data are from \citet{Galbany2019}.}
    \label{fig:dm15}
\end{figure}

In Fig.~\ref{fig:bands}, we compare the \textit{gri}-band light curves of SN 2019bkc to those of various fast transients: the rapidly evolving SNe Ic 1994I \citep{Richmond1994} and 2005ek \citep{Drout2013}, the Ca-rich events iPTF14gqr \citep{De2018}, iPTF16hgs \citep{De2018b}, SN 2012hn \citep{Valenti2014}, and SN 2010et \citep{Kasliwal2012}, the rapidly evolving unexplained transients, SNe 2010X \citep{Kasliwal2010} and 2002bj \citep{Poznanski2010}, and the unusual Ti-dominated  OGLE-2013-SN-079 \citep{Inserra2015}.  Fig.~\ref{fig:dm15} shows how extraordinarily fast the $\Delta m_{15}$ decline of SN 2019bkc is in relation to other transients. SN 2019bkc has an incredibly fast decline, with the previously most rapidly declining objects SN 2005ek and SN 2010X seen to be significantly slower. 

The light curves of the low-mass edge-lit double detonation (ELDD-L) model of a He-shell on the surface of a white dwarf from \citet{Sim2012} is also shown in Fig.~\ref{fig:bands}. This model has a short rise time which match the early $r$ and $i$ light curves of SN 2019bkc but fails to replicate the $g$ band. The model is also broader in $i$, cannot match the rapid decline seen in the three bands, and reaches the \Cofs\ decay tail earlier than SN 2019bkc.

Additionally, the shape of the light curve between the ATLAS-\textit{o} non-detection and the ZTF-\textit{g} band detection is of interest. We estimate the $g$-band upper-limit from the observed ATLAS-\textit{o} band limit using the assumption that the underlying spectrum is a black body, which appears reasonable since the spectrum at peak is still blue and relatively featureless. This results in a synthetic $g$ magnitude limit of $19.8>g>19.1$ mag for temperatures in the range, $10^4<T<10^5$ K. At $T=25000$ K the $g$ limit is $\sim 19.5$ mag.
The inset in the top panel of Fig.~\ref{fig:bands} shows that SN 2019bkc had a very sharp rise from non-detection to detection before levelling off, this is difficult to reconcile with a simple blast-wave approximation of $L \propto t^2$, and we note that the light curve of Ca-rich iPTF14gqr showed a double-peaked profile \citep{De2018}.

\subsection{Pseudo-bolometric light curve}
\label{bolo}
 Fig.~\ref{fig:bols} shows the $griz$ ($4000-10000$ \AA) and $ugriz$ ($3000-10000$ \AA) pseudo-bolometric light curves of SN 2019bkc, constructed following the method in \citet{Prentice2016}. 
A simple polynomial fit to the light curve shows it reached a $griz$ peak luminosity \lp\ $=(1.41\pm{0.06})\times10^{42}$ \ergs\ in $6.1\pm{0.4}$ d after the estimated explosion date, and $ugriz$ peak of $(1.9 \pm{0.1})\times 10^{42}$ \ergs\ in $5.0\pm{0.4}$ d. 
The former is chosen to allow direct comparison with other objects, as this wavelength range encompasses the most commonly observed bands. 
The latter provides a better estimate of the bolometric luminosity by utilising all of the available photometry.
A true bolometric light curve would encompass the UV to NIR, fortunately our $u$-band observations indicate that the peak of the SED of the transient is between $u$ and $g$ at our first epoch of multi-colour photometry. The derivatives of the two light curves at these times also suggest that we are not missing any significant contribution from the UV because in this case the light curves would show a steep decline. 
The absence of NIR photometry is harder to rectify, but in \citet{Prentice2016} it was found that the NIR flux over the range $10000-24000$ \AA\ is typically around 10\% of the flux in $3000-10000$ \AA\ at maximum light, for SE-SNe. GROND NIR observations a few days after the first epoch of multi-colour photometry show that this transient is not NIR bright.
Thus, assuming that $u$ captures the majority of the flux blueward of $g$, and taking a 10\% NIR contribution, the peak bolometric luminosity could be expected to be in the region $\sim 2.2 \times 10^{42}$ \ergs.

The time for the $griz$ light curve to decay by half its peak luminosity is $4.01\pm{0.04}$ d.
SN 2019bkc reached a similar peak luminosity to \Nifs-driven core-collapse events but approximately ten times lower than that of thermonuclear SNe~Ia.
A rise time from explosion to peak \tp\ of $\sim 5-6$ d is short but not excessively so, and is similar to iPTF14gqr. However,
its subsequent rapid decline is only rivalled by that of kilonova AT 2017gfo \citep[e.g.][]{Drout2017,Smartt2017}, but the two are not spectroscopically similar. 

\subsubsection{Estimating \mni}
On the assumption that the light curve peak is powered by \Nifs\ decay (and excluding in this scenario the presence of short-lived radioisotopes), we use the analytical form of ``Arnett's rule'' \citep{Arnett1982} as given in \citet{Stritzinger2005} to estimate the \Nifs\ mass \mni\ required to give these peak luminosities 

\begin{equation}
\begin{split}
\frac{M_{\mathrm{Ni}}}{{\rm M}_\odot}= & L_{\mathrm{p}}\times\left(10^{43} \textrm{erg s}^{-1} \right)^{-1} \\ & \times \left(6.45\times e^{-t{_\mathrm{p}}/8.8}+1.45\times e^{-t{_\mathrm{p}}/111.3}\right)^{-1}, 
\end{split}
\label{eq:Mni}
\end{equation}

Where \lp\ and \tp\ are the peak luminosity and rise-time respectively.
For the $griz$ bolometric light curve we take \tp\ to be $6.1\pm{0.4}$ and \lp\ $=(1.41\pm{0.06})\times10^{42}$ \ergs, which gives \mni\ $=0.031\pm{0.001}$ \msun.
For the $ugriz$ bolometric light curve we use \lp $=(1.9 \pm{0.1})\times 10^{42}$ \ergs\ and \tp\ $=5.0\pm{0.4}$ d, this returns \mni\ $=0.038\pm{0.001}$ \msun.
for the $griz$ and $ugriz$ luminosities, respectively. 
This method also assumes that the \Nifs\ is located centrally, which may not be valid for this event \citep[see][]{Khatami2019}, thus these values should be considered upper limits. 

\begin{figure}
    \centering
    \includegraphics[scale=0.6]{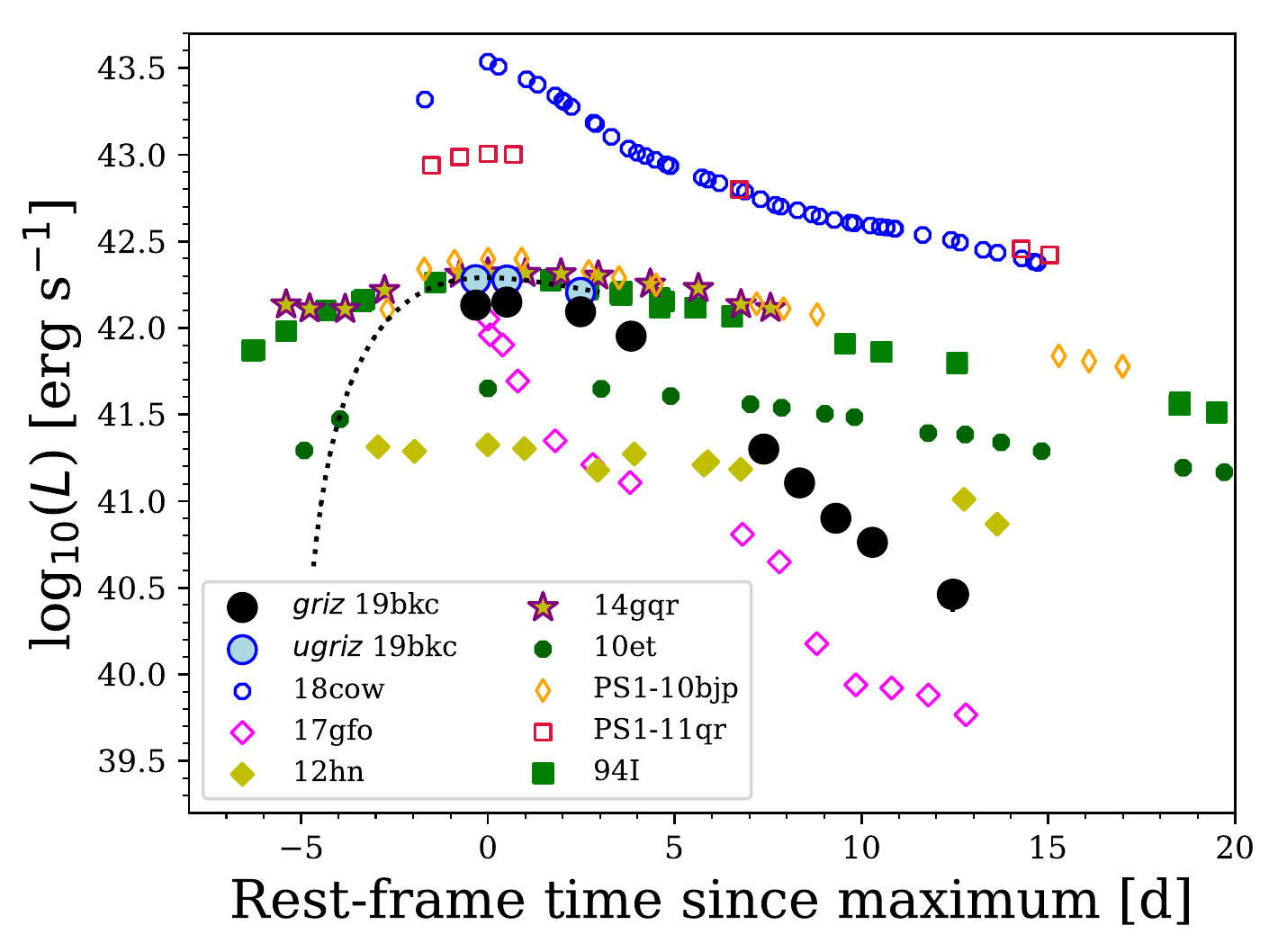}
    \caption{$griz$ (black solid circles) and $ugriz$ (open black pentagons) pseudo-bolometric light curves of SN 2019bkc (black) shown in comparison to the $griz$ pseudo-bolometric light curves of Type Ic SN 1994I (green), kilonova AT 2017gfo \citep{Smartt2017}, AT2018cow \citep{Prentice2018b}, and fast blue transients PS1-10bjp and PS1-11qr \citep{Drout2014}, Ca-rich SNe 2012hn \citep{Valenti2014} and 2010et \citep{Kasliwal2012}, and ``ultra-stripped'' Ca-rich iPTF14gqr \citep{De2018}.
    The dotted black line shows the best fit analytic \Nifs\ powered light curve model of \citet{Valenti2008}, fit to the $ugriz$ bolometric light curve until 9 days after explosion.
    }
    \label{fig:bols}
\end{figure}

 A simple model fit to the $ugriz$ light curve, using the photospheric model of \citet{Valenti2008}, is also shown in Fig.~\ref{fig:bols}. 
The fit omits the ZTF detection, as per the discussion of the early light curve shape in Section~\ref{sec:lightcurves}.
It extends to 9 days after explosion, after this time ejecta transparency leads to deviation between the observations and the model \citep[see][]{Valenti2008}.  The resulting physical parameters derived from the fits, \tp\ $=4.8\pm^{1.2}_{0.03}$ days and \mni\ $= 0.039\pm^{0.003}_{0.001}$ \msun\ are consistent with that found above. 

\subsubsection{Estimating the ejecta mass}
We also estimated the ejecta mass \mej\ using the formulation given in \citet{Arnett1982} and expressed as
\begin{equation}
M_\mathrm{ej} = \frac{1}{2}\left( \frac{\beta c}{\kappa} \right)\tau_\mathrm{m}^{2} v_\mathrm{sc}
\label{eqn:arnett}
\end{equation}
where $\beta \approx 13.7\pm^{0.1}_{3.4}$ is dimensionless factor which varies for different density profiles \citep{Arnett1980, Arnett1982}, \taum\ is a timescale of the light curve model which can be approximated by the rise time, \vsc\ is a scale velocity which we take to be a measured line velocity around maximum light, and $\kappa$ is a constant optical opacity.

The choice of $\kappa$ has some bearing on the value of \mej\ as calculated in Equation~\ref{eqn:arnett}. 
Although taken as a constant in various model light curves, the actual optical opacity varies with time \citep[e.g.][]{Chugai2000,Nagy2018}, but tests have shown that that a constant opacity can be a reasonable assumption at early times \citep[see][]{Mazzali2000,Taddia2018}.
SN 2019bkc is a H/He-deficient SN, so to select an appropriate value for $\kappa$, we search the literature literature and consider only SNe of this type.
For SNe Ia, where the large abundance of Fe-group elements is a source of significant opacity, \citet{Cappellaro1997} used $\kappa=0.15$ cm$^2$ g$^{-1}$, while \citet{Arnett1982} used $\kappa=0.08$ cm$^2$ g$^{-1}$.
The light curve of the broad lined SNe Ic 1997ef was modelled in \citet{Mazzali2000} using the same code as \citet{Cappellaro1997}, this time a lower constant opacity of 0.08 cm$^2$ g$^{-1}$ was used owing to the lower Fe abundance in this SN type.
The average opacity of model CC-SNe light curves was investigated by \citet{Nagy2018}, where it was found that $\kappa=0.1$ cm$^2$ g$^{-1}$ was suitable for their H/He-deficient models.
\citet{Taddia2018} demonstrated that fitting the light curves of H-deficient CC-SNe with a simple analytical model and a constant opacity $\kappa =0.07$ cm$^2$ g$^{-1}$ returned physical parameters consistent with more complex hydrodynamic light curve models.
\citet{Prentice2019} also found that using $\kappa =0.07$ cm$^2$ g$^{-1}$ gave comparable ejecta masses to both photospheric phase and nebular phase spectroscopic modelling of H-deficient CC-SNe, with the exception of some gamma-ray burst SNe.
From Section~\ref{sec:models} we find that the outer ejecta is not rich in Fe-group elements, and there is no indication of helium. 
From the early-nebular spectra in Section~\ref{neb_spec} we see no indication of strong \FeII\ or \FeIII\ line emission, demonstrating that the ejecta is poor in Fe-group elements relative to normal SNe Ia and is more like H/He-deficient CC-SNe. 
This justifies the use of $\kappa = 0.07$ cm$^2$ g$^{-1}$, but to show the range of possible \mej, we consider the range 0.07--0.1 cm$^2$ g$^{-1}$.

Thus, using the values and uncertainties associated with $\beta$ and $\kappa$, along with the $ugriz$ pseudo-bolometric rise time of $5.0\pm{0.4}$ d and a scale velocity $v_\mathrm{sc} = 14\,500\pm{500}$ \kms\ from the \SiII\ \lam 6355 line (see Section~\ref{sec:spectra}), we find \mej\ $=0.4\pm^{0.1}_{0.2}$ \msun.

\section{Spectroscopy}\label{sec:spectra}
\begin{figure*}
    \centering
    \includegraphics[scale=0.52]{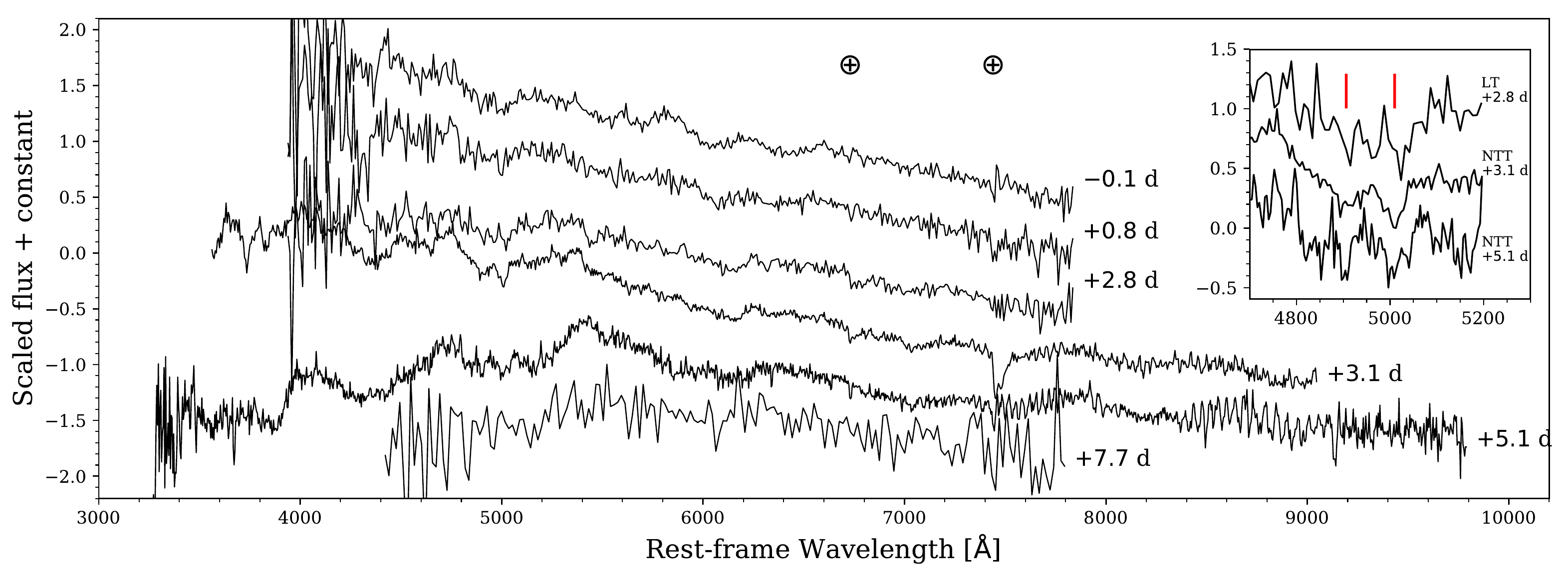}
    \includegraphics[scale=0.52]{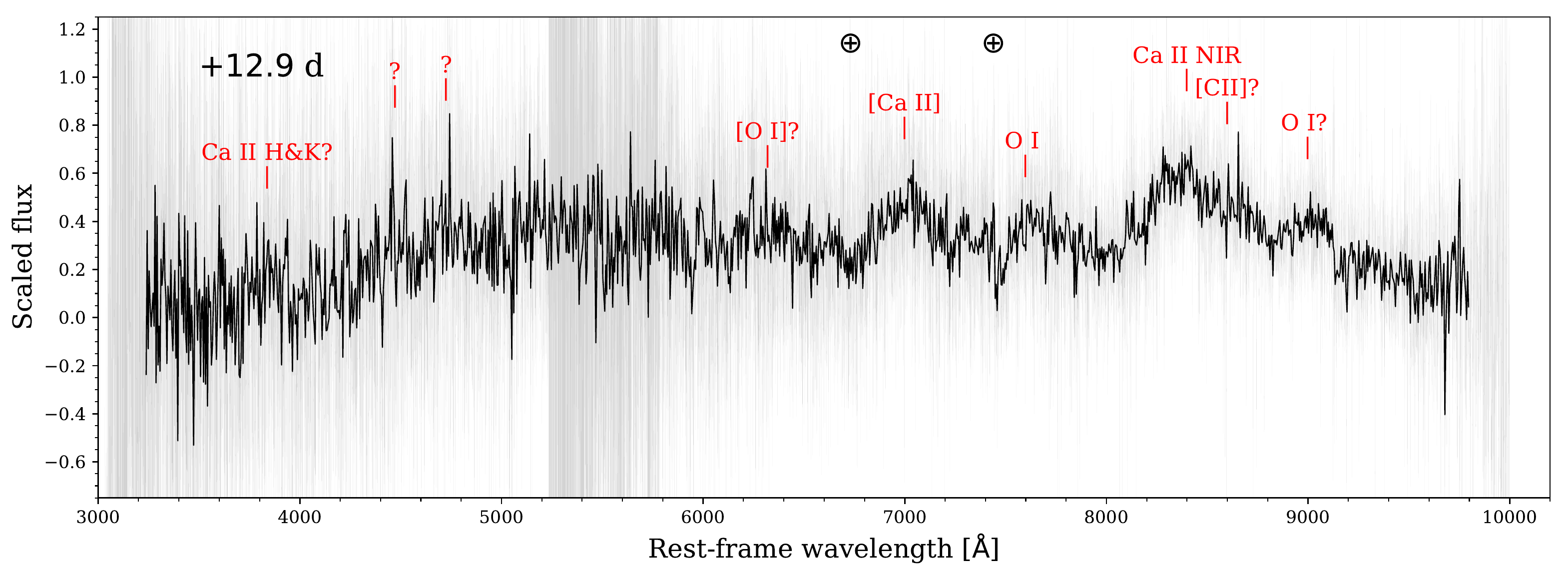}
    \includegraphics[scale=0.52]{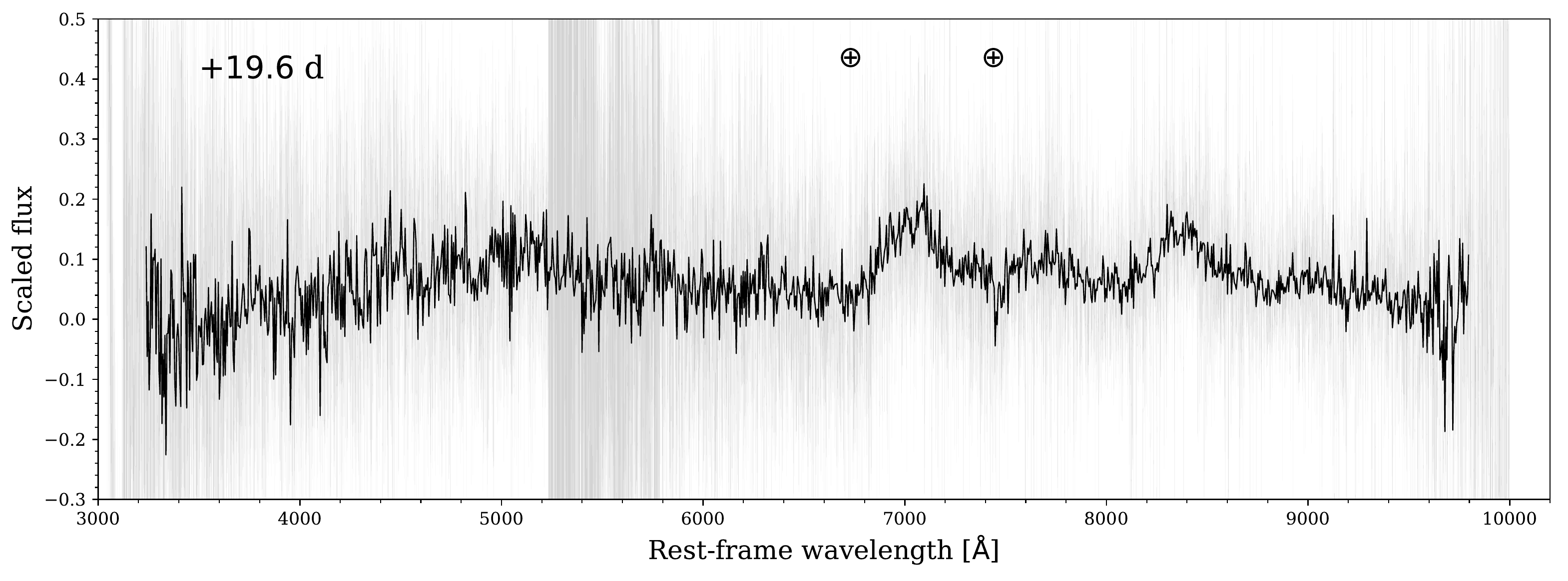}
    \caption{(Top) Optical spectra from LT:SPRAT and NTT:EFOSC2. The phases are with respect to maximum light. The inset highlights some narrow features in the spectra that remain unidentified. (Middle and bottom) The raw VLT:X-Shooter UV and optical spectra (light grey) and binned to 5 \AA\ (black). The outer edges of the UVB and VIS arms have been omitted from the binning process due to excess noise. Note that the region between 5200 and 5800 \AA\ is the inner boundary between the UVB and VIS arms. Due to the low signal to noise ratio (S/N) any features here are unlikely to be real. }
    \label{fig:spectra}
\end{figure*}

Spectra of SN 2019bkc were taken over a 20 day period covering from shortly before maximum light until well into the decay phase (Fig.~\ref{fig:spectra}). The epochs that spectra were taken at are denoted with respect to the light curve in Fig.~\ref{fig:lcs}. The initial spectra appeared photospheric but there was a rapid evolution to an optically thin `nebular regime', beginning at just 13 d post maximum. Analysis of the unusual late-time velocity shifts is presented in Section \ref{neb_spec}, while modelling of the photospheric phase spectra is performed in Section \ref{sec:models}.

\subsection{Photospheric phase spectra}

The first spectra around maximum light are blue with broad, shallow absorption features in the range of 6000 -- 7000 \AA, with these features becoming narrower between our $-0.1$ d and $+0.8$ d spectra (top panel of Fig.~\ref{fig:spectra}). 
We identify the blueward feature as \SiII\ \lam 6355 (for the redward feature, we tentatively suggest \ArII, see Section~\ref{sec:models}) at a velocity of $\sim -15\,000$ \kms. 
One day later, $+0.8$ d after maximum, the \SiII\ velocity has decreased to $\sim -13\,000$ \kms\ and the redward feature of the two has weakened.
Two days later, the spectra are still blue but less so than previously seen and the \SiII\ absorption line profile with $v \sim -10\,000$ \kms appears to have narrowed, while the feature to the red of it is very weak. 
There is evidence of P-Cygni-like emission from the \CaII\ NIR triplet at $3.1$ d. 
Absorption by the \OI\ \lam777 lines may be obscured by atmospheric {\sc O}$_2$ absorption, which occurs in the 7441 -- 7509 \AA\ region when corrected for $z=0.02$. This corresponds with oxygen blueshifts of 10000--13000 \kms.

\begin{figure}
    \centering
    \includegraphics[scale=0.6]{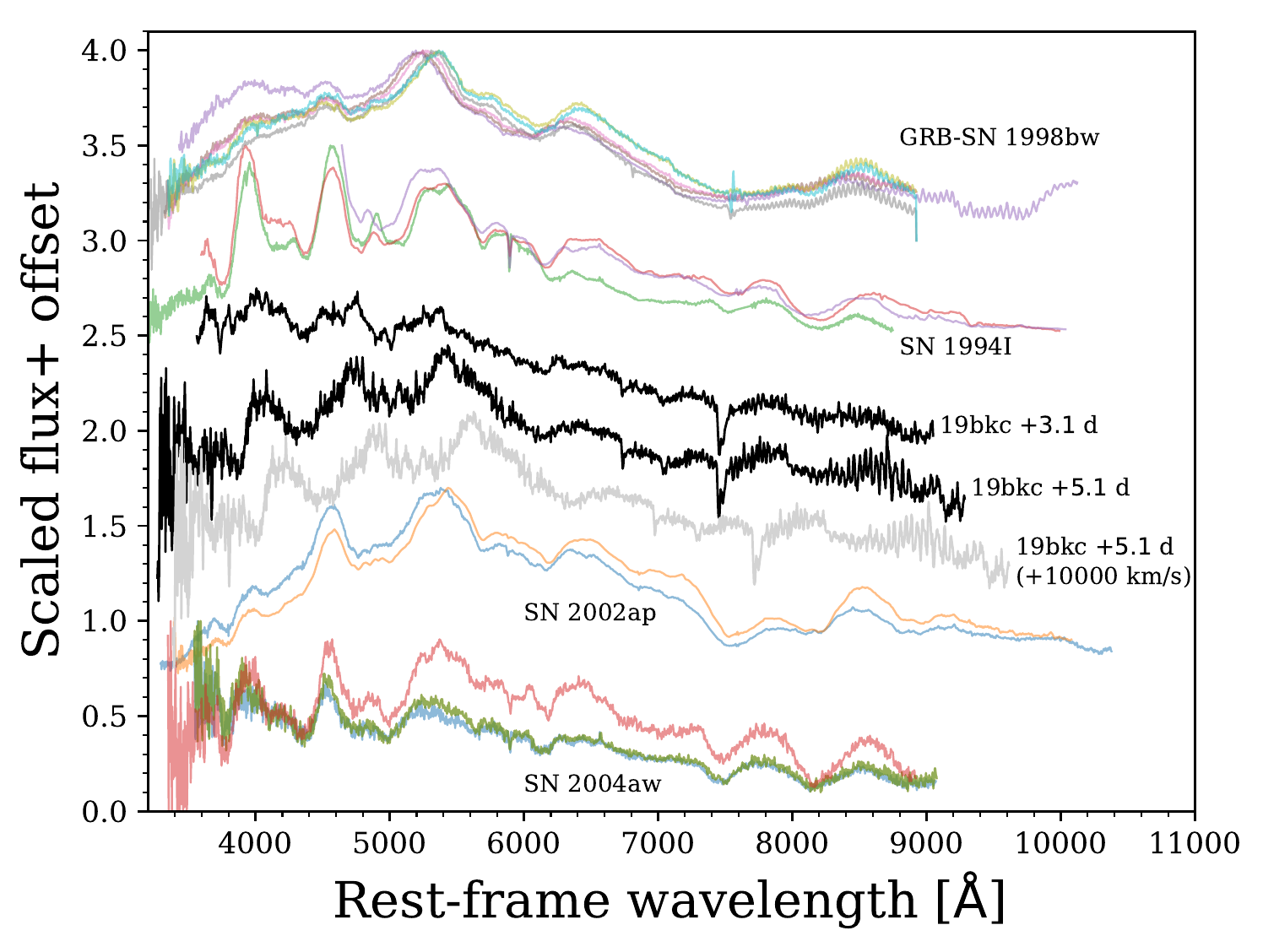}
    \caption{SN 2019bkc $+3.1$ d and $+5.1$ d spectra compared with a sample of SNe Ic \citep[data from][]{Patat2001,Filippenko1995,Foley2003,Modjaz2014}. The comparison objects are shown at phases from $0$ to $+10$ d after maximum light. There are similarities between SN 2019bkc and these objects, especially in the $+5.1$ d spectrum. Without our nebular spectra this transient could have been considered a peculiar type~Ic. Also plotted is the rest-frame $+5.1$ d spectrum red-shifted 10\,000 \kms\ (grey). This now poorly matches the comparison SNe, especially in the regions redward of 7000 \AA. It demonstrates that the earlier spectra are likely not affected by whatever process causes the velocity shifts in the nebular spectra and that a redshift significantly lower than $z=0.02$ is incompatible. }
    \label{fig:Ic}
\end{figure}

By $+5.1$ d, the spectrum of SN 2019bkc is substantially different. The 6000 -- 7000 \AA\ features have given way to a single feature centred on the \SiII\ line, which is broader with an absorption minimum at $v\sim -11\,000$ \kms, slightly greater than that measured a few days before. 
This broadening of absorption lines and ``plateau'' of the measured line velocities was also see in SN 2012hn \citep{Valenti2014}.
This spectrum at this time shows remarkable similarity to SNe Ic spectra 0 -- 10 d after maximum light, as is shown in Fig.~\ref{fig:Ic}. 
Unlike SNe Ic, the spectra of SN 2019bkc show considerable complexity in the region blueward of $\sim 5000$ \AA\ with a series of broad features within which there are narrower features. These features are not just noise, and the inset of Fig~\ref{fig:spectra} shows how this region evolves from $+2.8$ d to $+5.1$ d, which covers the period the object was observed with the NTT.
Two absorption features at 4900 \AA\ and 5010 \AA\ are seen and show no significant velocity evolution during the three observations ($\sim 2.3$ d).
These two features have a similar wavelength difference as \FeII\ \lam\lam 4924, 5018.
If we hypothetically attribute the absorption to these lines then the measured expansion velocity would be $\sim -1000$ \kms, which is significantly lower than the $\sim -11\,000$ \kms\ measured for the \SiII\ line. 
There is no indication of a feature associated with the much stronger \FeII\ \lam 5169 line however, which suggests that the wavelength difference is coincidental. Additionally, if this line is associated with the two absorption features then there is nothing which coincides with the rest of the \FeII\ lines, making this association unlikely.

We have taken this investigation a step further however.
Comparison of the $+5.1$ d spectrum with those of SN 1994I between maximum light and $+10$ d in Fig.~\ref{fig:Ic} shows a curious offset between the region dominated by Fe-group elements ($<5500$ \AA) and the intermediate mass elements (IME). In SN 1994I, the line profiles in the region 4000 -- 5400 \AA\ at these epochs are attributable to \FeII, \MgI, \TiII\ \citep{Sauer2006}. 
If the features were to be formed by the same line transitions in SN 2019bkc then they are some 8000 \kms\ slower than in SN 1994I, effectively at rest.
This contrasts with the lines formed by IMEs -- \SiII\ \lam 6355, \OI\ \lam 7772, and the \CaII\ NIR triplet -- which are seen at a higher velocity in SN 2019bkc.
This would require that the line forming region of the Fe group elements is below that of the IMEs, contrary to what is seen in SNe Ic \citep{Prentice2017}.

Finally, a low S/N spectrum was obtained with the LT less than three days later at $+7.7$ d after maximum light. Very little can be obtained from this spectrum other than it shows no obvious signs of [\CaII] and \CaII\ emission lines, which appear strong within two to three weeks of maximum in SNe Ic.

\subsection{The extraordinary Ca-rich early nebular phase}
\label{neb_spec}
\begin{figure}
    \centering
    \includegraphics[scale=0.54]{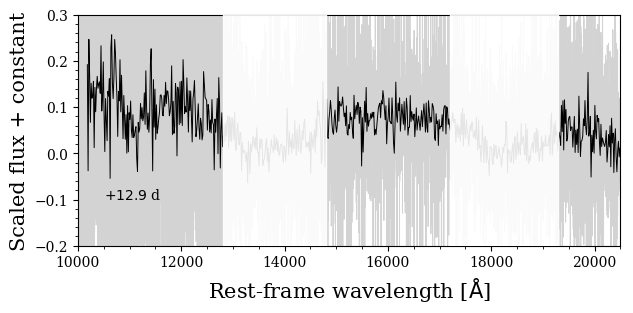}
    \includegraphics[scale=0.54]{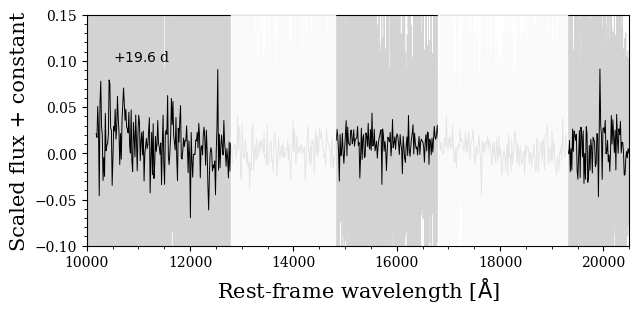}
    \caption{The X-Shooter NIR spectra at phases $+12.9$ d (top, grey) and $+19.6$ d (lower, grey) and the binned spectra (black). Regions with telluric absorption have been masked. }
    \label{fig:NIRspectra}
\end{figure}

We obtained two X-Shooter spectra at $+12.9$ and $+19.6$ d after maximum covering the near-UV through NIR wavelengths, the UV and optical spectra are shown in the bottom panel of Fig.~\ref{sec:spectra} and the NIR spectra in Fig.~\ref{fig:NIRspectra}. Both optical spectra show a flux excess above the continuum level in the range of 4500 -- 6000 \AA, which becomes more prominent in the later spectrum. They also show three evolving emission lines at $\sim7000$, 7700, and 8400 \AA, and possibly three at 3838 \AA, 4472 \AA, and 4726 \AA.
The NIR spectra on these days are noisy and somewhat featureless when telluric features are accounted for, with the only line being an absorption around $\sim 11\,200$ \AA. This feature remains unexplained as there are no obvious lines in this aside from [\FeII] in this region, but these do not result in absorption,  see Fig.~\ref{fig:NIRspectra}.

The positioning of the optical emission features between 6500 and 8800 \AA\ is difficult to reconcile with the rest wavelength of any strong emission line that is commonly seen in SNe. 
 The most probable candidates for the 7000 \AA\ feature are blueshifted [\FeII] \lam 7155 and [\CaII] \lam\lam 7921, 7324. These lines are seen in the nebular phase spectra of most types of SNe, with the former being prominent in the spectra of SNe Ia, but less so in CC-SNe \citep[See][]{Jerkstrand2014,Mazzali2015}. 
In Fig.~\ref{fig:spectra} it can be seen that the 7000 \AA\ feature is prominent, and is at least as strong as any emission blueward of it. If it were to be dominated by the \FeII\ \lam 7155 transition then it could be expected that the region around 4000 -- 5000 \AA\ would be dominated by strong [\FeII] and [\FeIII] lines, as in SNe Ia \citep{Mazzali2015,Mazzali2018}.
In instances where the [\FeII] \lam7155 is comparable to the flux in the blueward region, emission by forbidden calcium dominates the spectrum in this region \citep{Jerkstrand2014,Jerkstrand2015}.
This argument leads to the conclusion that the $\sim$7000 \AA\ feature is dominated by [\CaII] \lam\lam 7921, 7324.

In the region around 8400 \AA\ the strongest nebular emission lines in SNe are associated with \CaII\ \lam\lam 8498, 8542, 8662, [\FeII] \lam 8617, [\FeII] \lam 8892, and [\CI] \lam 8727.
The [\FeII] emissions typically dominates in SNe Ia \citep{Mazzali2011}, but is either absent, or present as an equal component in CC-SNe \citep{Mazzali2010,Jerkstrand2015}.
As argued previously, the lack of strong iron lines elsewhere in the ejecta suggests that this feature is principally composed of the \CaII\ NIR triplet.

\begin{figure}
    \centering
    \includegraphics[scale=0.6]{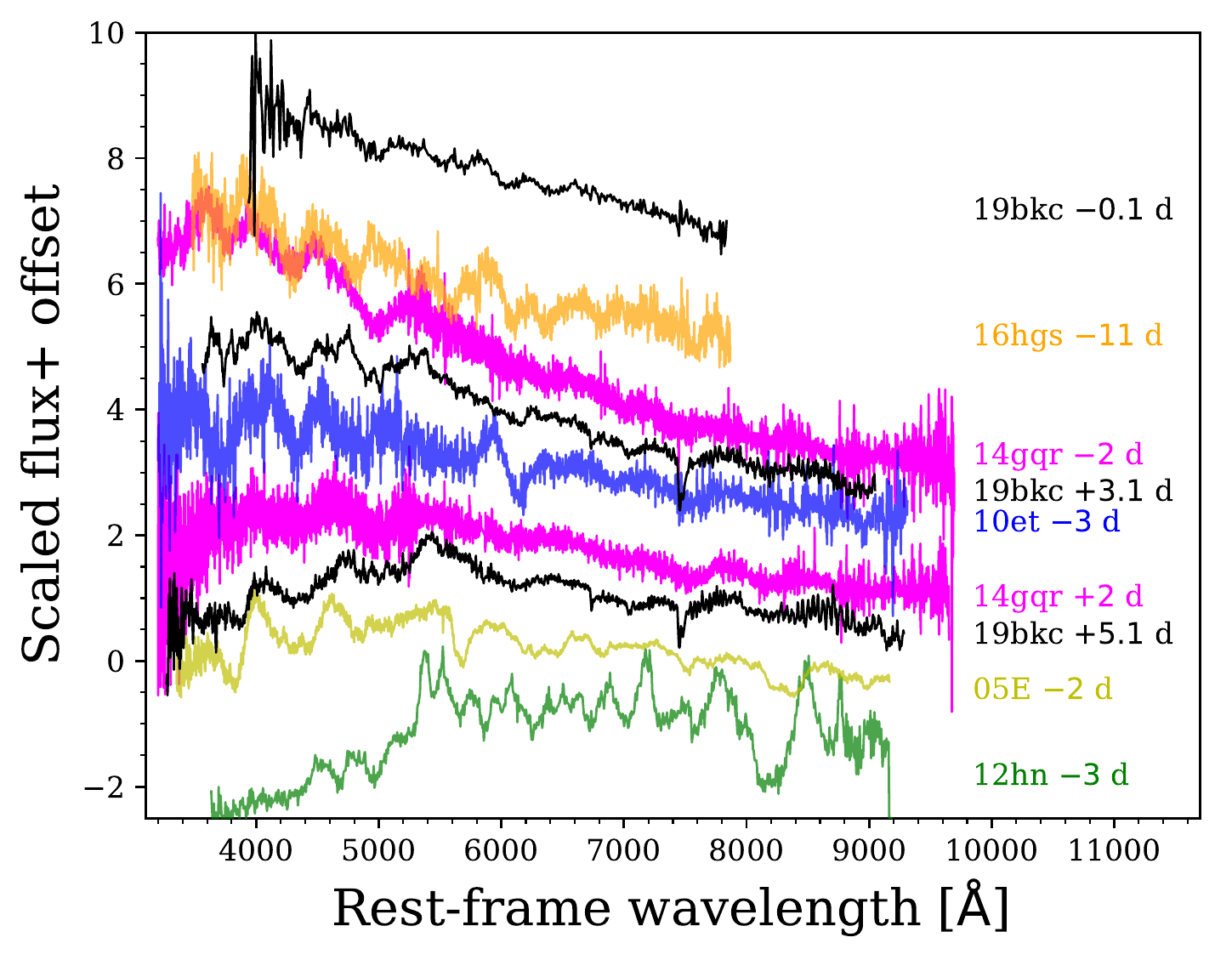}
    \includegraphics[scale=0.6]{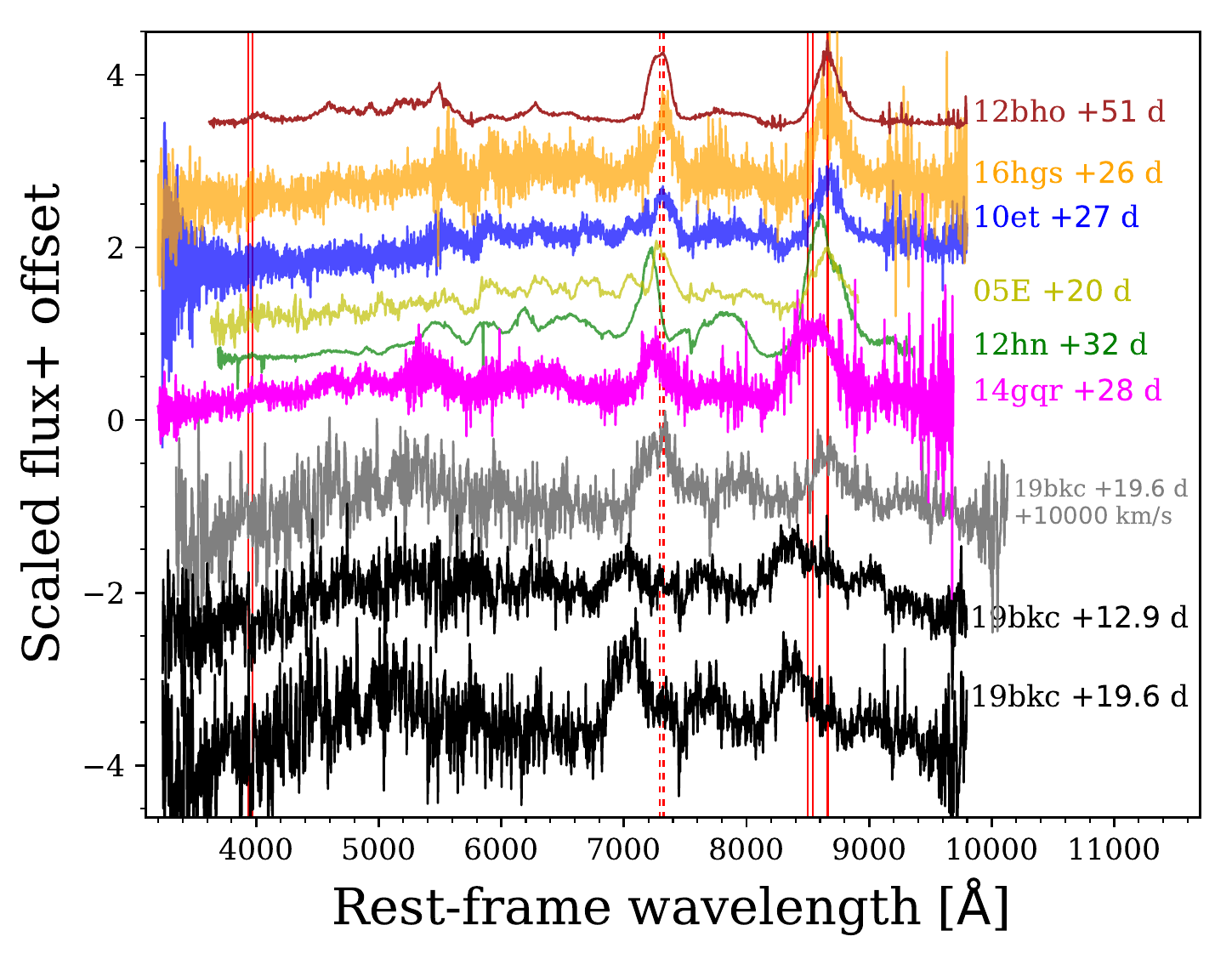}
    \caption{(Top) Photospheric spectra of SN 2019bkc compared with the 
    Ca-rich transients SN 2012hn, SN 2005E, iPTF16hgs, SN 2010et, and ``ultra-stripped SN'' iPTF14gqr. The spectra at this phase are heterogeneous and SN 2019bkc is most similar to iPTF14gqr. (Lower) Nebular phase spectra of SN 2019bkc compared with the same transients plus PTF12bho.  \CaII\ and \CaiiF\ lines are denoted in red solid and dashed lines respectively. To match the $+19.6$ d spectrum of SN 2019bkc to that of SN 2005E requires a velocity shift of $\sim10\,000$ \kms\ (grey). }
    \label{fig:comparespectra}
\end{figure}

These apparent calcium-dominated features are considerably blueward of the respective emission rest-wavelengths.
However, if a redshift of $\sim 10\,000$ \kms\ from the rest-frame is applied to the spectra then the main emission lines align with those seen in Ca-rich transients (bottom panel of Fig.~\ref{fig:comparespectra}), which correspond with [\CaII] \lam\lam 7291, 7324, \OI\ \lam 7772, the \CaII\ NIR. 
There are also two weaker features that may be [\CII] \lam 8727 \citep{Perets2010} and OI \lam 9263 \citep{Jerkstrand2015}. If these identifications are correct, the former is blueshifted by $\sim 4000$ \kms\ and the latter by $\sim 9000$ \kms. 
Additionally, if \OI\ \lam 7774 and \OI\ \lam 9263 are present then this may also suggest the presence of \OI\ \lam 8446 within the 8400 \AA\ feature, although this is not typically identified in models of nebular spectra.

In the blue, the apparent emission line at $\sim 3838$ \AA\ could be \CaII\ H\&K blueshifted by 10\,000 \kms. The two remaining features around 4472 \AA\ and 4726 \AA\ remain unidentified as we find no line that is positioned at these wavelengths in the rest frame or with a velocity of $\sim -10\,000$ \kms.
A final point on the line identification is that, for all the lines transitions and possible blends considered, the observed features require a significant blueshift from any of their rest-wavelengths.
The velocity shifts are discussed further in Section~\ref{sec:vshift}.

Given that the nebular spectra imply that SN 2019bkc belongs to the diverse family of Ca-rich transients, in the top panel of Fig.~\ref{fig:comparespectra} we compare its photospheric spectra with four Ca-rich events: SN 2012hn \citep{Valenti2014}, SN 2005E \citep{Perets2010}, SN 2010et \citep{Kasliwal2012}, iPTF16hgs \citep{De2018b}, as well as the ``ultra-stripped SN'' iPTF14gqr \citep{De2018} . The best spectroscopic match is to iPTF14gqr (at $+7.2$ and $+10.96$ d). We can find no good match with the other objects, even when we adjust the spectra of SN 2019bkc over the redshifts within the galaxy group ($z = 0.018-0.022$). We do note there is significant heterogeneity in the Ca-rich spectra, with SN 2012hn being a clear outlier in terms of appearance, where there is no indication of the apparent \HeI\ features seen in iPTF16hgs or SN 2010et. 

The lower panel of Fig.~\ref{fig:comparespectra} shows the $+12.9$ and $+19.6$ d spectra of SN 2019bkc compared to the same sample of Ca-rich nebular phase spectra, with the addition of PTF12bho \citep{Lunnan2017}. The emission lines at $\sim$7000 and $\sim$8400 \AA\ seen in the spectra of SN 2019bkc look similar to those seen in the Ca-rich events (and associated with \CaiiF\ and \CaII, respectively). However, the \CaiiF\ and \CaII\ lines in SN 2019bkc display a significant blueshift of $\sim 10\,000$ \kms\ compared to the rest wavelength of these features, which is discussed in Section \ref{sec:vshift}. The $+19.6$ d spectrum, corrected for this offset velocity, is shown in Fig.~\ref{fig:comparespectra} in grey and the peaks of the Ca features are seen to agree much more closely with the other objects after this correction.

\begin{figure*}
    \centering
    \includegraphics[scale=0.7]{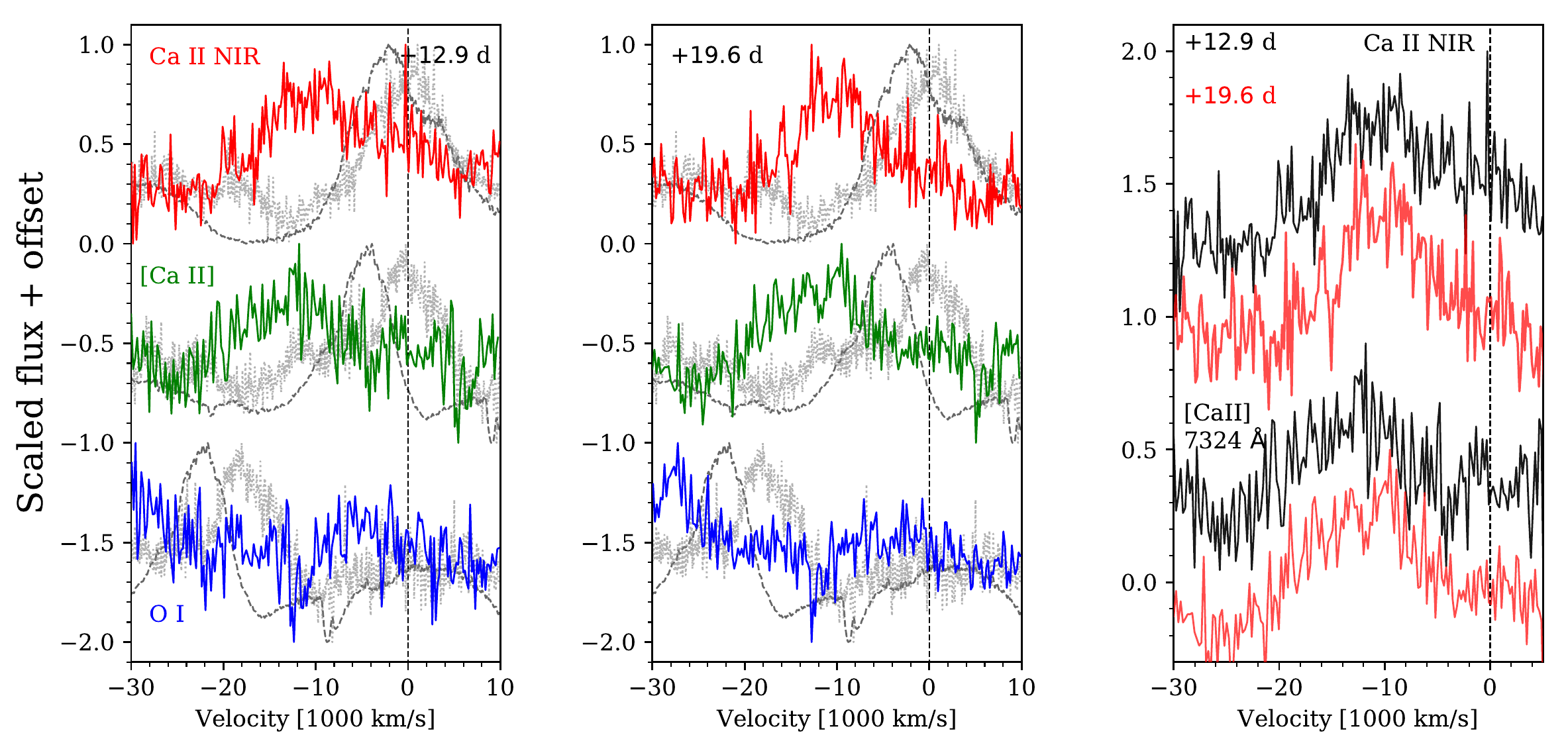}
    \caption{(Left and centre) The spectra of SNe 2019bkc (line), 2012hn (black dashed), and 2010et (grey dotted) centred in velocity space on the rest position of \CaII\ NIR (red), [\CaII] \lam 7324 (green), and \OI\ \lam 7772 (blue) respectively for $+12.9$ d (left) and $+19.6$ d (centre). A telluric feature is present in the \OI\ feature at $\sim 11\,000$ \kms\ and the peaks at $-23000$ and $-18000$ \kms\ for SNe 2012hn and 2010et, respectively, are the \CaiiF\ features located to the blue of this \OI\ feature.  (Right) The \CaII\ NIR (top) and [\CaII] (bottom) lines plotted in sequence for the two epochs. The \CaII\ NIR triplet is consistent with no velocity evolution, while the \CaiiF\ line appears to decrease in velocity. Given the S/N of the spectrum this may not be real.   } 
    \label{fig:vshift}
\end{figure*}

\subsection{Velocity shifts}\label{sec:vshift}
Fig.~\ref{fig:vshift} shows the nebular-phase spectra centred on the positions of \CaII\ \lam 8662\ NIR, [\CaII] \lam 7324 and \OI\ \lam 7772 in rest-frame velocity space for SN 2019bkc compared to SN 2012hn and SN 2010et. SN 2019bkc shows a blueshift in both the \CaiiF\ and \CaII\ features, with an offset of $\sim 10\,000$ \kms\ from their rest-frame positions. A blueshift in the \CaiiF\ line in the range of 3500--5000 \kms\ was previously identified for SN 2012hn \citep{Valenti2014}. As seen in Fig.~\ref{fig:vshift}, the spectra of SN 2010et are consistent with no discernible blueshift, while the lines of SN 2012hn appear blueshifted $\sim 4000$ \kms. Foley (2015) investigated velocity shifts in the \CaiiF\ lines of 13 Ca-rich events and found velocity shifts of up to 1700 \kms\ but nothing close to the $\sim -10\,000$ \kms\ identified for SN 2019bkc here.

The right hand panel of Fig.~\ref{fig:vshift} shows that there is no evidence of velocity evolution in the \CaII\ NIR line between the two observations. The relative position of this emission line in other Ca-rich transients and the rest-frame wavelength of the NIR triplet here gives a velocity offset of $\sim-10\,500$ \kms, which we adopt as the typical velocity shift. 
Alternatively, aligning [\CaII] \lam 7324 with the red peak of the observed line gives a velocity of $\sim -12\,000$ \kms\ at $+12.9$ d and $\sim -10\,000$ \kms\ at $+19.6$ d, which are consistent with the \CaII\ NIR velocity offset. 

In both SN 2019bkc and SN 2012hn there is a feature in the form of a broad emission ``bump" between [\CaII] and \CaII\ NIR. \citet{Valenti2014} do not attribute this emission to anything in particular but its position is consistent with \OI\ \lam7772. Emission from this line is seen in the early nebular phases of H-deficient core-collapse SNe, so we continue on the assumption that the feature is due primarily to \OI\ \lam 7772 and measure a velocity shift in the region of $-8\,000$ to $-10\,000$ \kms\ for this line. At redder wavelengths, we have identified a feature at $\sim 9000$ \AA\ that may be \OI\ \lam 9263 \citep{Jerkstrand2015}. If the identification of \OI\ \lam 9263 is correct, then the velocity shift would be $\sim -9000$ \kms, which would be in rough agreement with the velocity of the assumed \OI\ \lam 7772 line. This would suggest the \OI -emitting region is also located at a similar velocity to the Ca-emitting regions. No double-peaked profile is identified for the \OI\ feature but the region is strongly contaminated by a strong telluric feature  $\sim -11\,000$ \kms\ from 7772 \AA.

Finally, we have excluded dust as being responsible for the velocity offset. In circumstances involving absorption by dust, the red wing of a symmetric emission profile is suppressed, leading to a line that is peaked blueward of the rest wavelength. Here, the extreme velocity shift observed would require the original emission lines to have a full width half maximum of $\sim 20\,000$ \kms, suggesting it is very unlikely that dust is causing this velocity offset to the blue.

\subsection{Estimating [\CaII]/\Oneb}
Calculating the ratio of [\CaII] to \Oneb\ line strength is a common analytical tool for Ca-rich transients to show how strong the [\CaII] line is compared with that in core-collapse (CC) SNe.
In previous Ca-rich SNe, the \Oneb\ \lam\lam 6300, 6363 line appeared several tens of days after maximum light \citep{Valenti2014,Mili2017}. 
In Fig.~\ref{fig:spectra} it appears as if there may be emission features around $\sim 6300$ \AA\ in the $+12.9$ and $+19.6$ d spectra. However, on investigation these are found not to be significantly above the noise level of the spectra, thus we conclude there is no clear emission from \Oneb. 
This may be because the spectra are too early, even accounting for the rapid evolution of the light curve, or that the S/N is just too poor. 

We estimate an upper limit to the \Oneb\ flux by assuming that the line is present and accounts for the entirety of the flux in a region approximated by a Gaussian with the same width as that of the \CaiiF\ \lam 7291 line. Measurements were taken at two velocity offsets; 0 \kms\ and the velocity measured for the [\CaII] of $-10\,000$ \kms. For the $+12.9$ d spectrum these upper limits gave [\CaII]/\Oneb\ $>5.4$ and $>5.4$, respectively, and for $+19.6$ d they were $>14.1$ and $>10.8$. 
The similarity in the values supports the argument that the observable ``features'' in the spectra at $\sim 6300$ \AA\ are not real.

\begin{figure}
    \centering
    \includegraphics[scale=0.6]{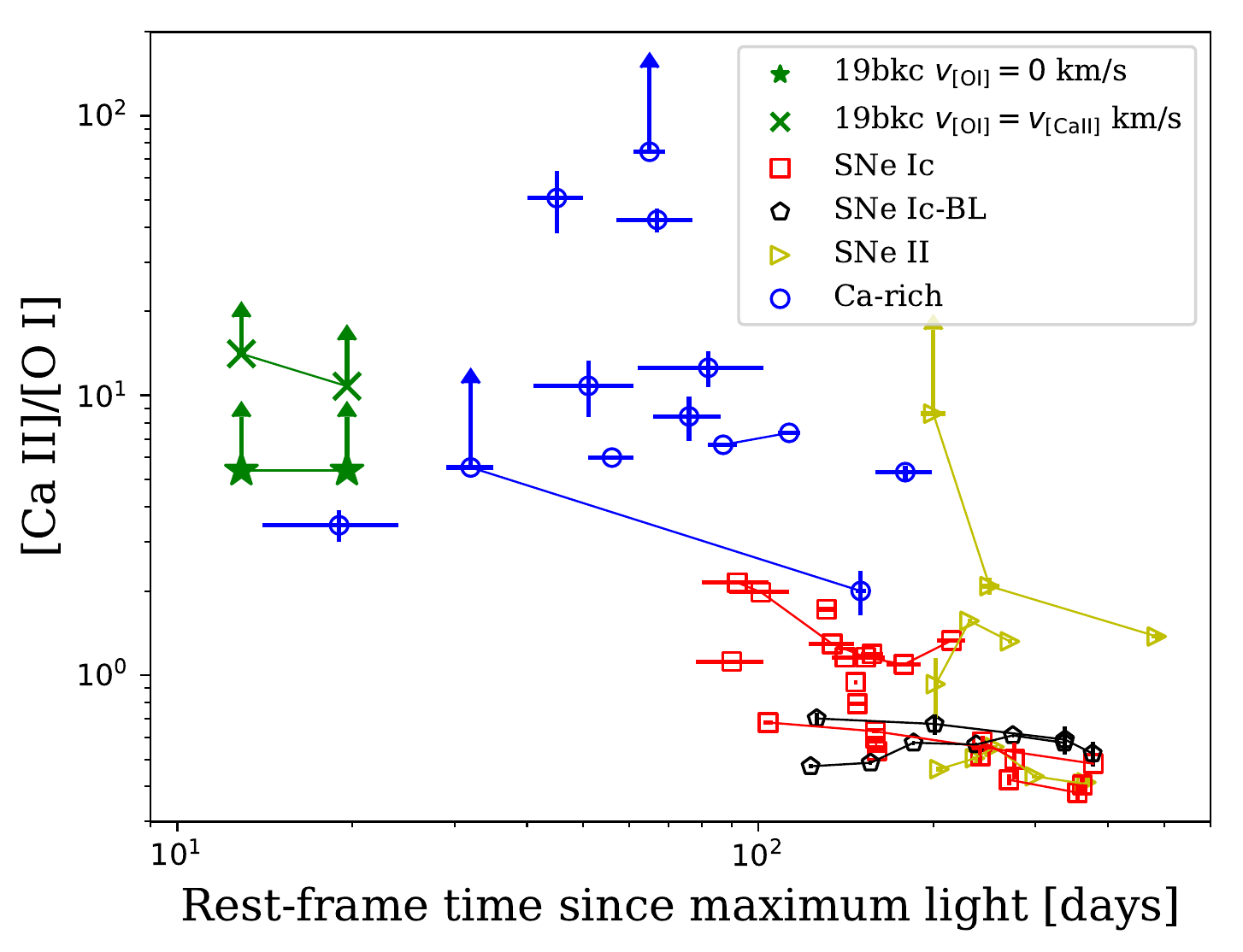}
    \caption{The ratio of [\CaII] \lam\lam 7291, 7324 to \Oneb\ \lam\lam 6300, 6363 for a sample of core-collapse SNe and Ca-rich events; the ratio is significantly larger in Ca-rich objects. SN 2019bkc is shown in green calculated for two velocity positions, 0 \kms and if it has the same offset as the \CaiiF\ and \CaII\ lines. Ancillary data from \citet{Valenti2014}.}
    \label{fig:ratios}
\end{figure}

Fig~\ref{fig:ratios} demonstrates that with these upper-limits, SN 2019bkc sits within the Ca-rich region of the [\CaII]/\Oneb\ parameter space and well outside that of CC events where the ratio is typically $<1-2$ \citep{Valenti2014,Mili2017,Nicholl2019}. \Oneb\ is not seen in normal SNe Ia but has been detected in one subluminous event \citep{Taubenberger2013}.

SN 2019bkc displays an absence of \Oneb\ and a 1:1 ratio of [\CaII]/\CaII. In SNe Iax, this has been attributed to a high-density environment \citep{Jha2006,McCully2014}. However, SNe Ibc typically have a [\CaII]/\CaII\ ratio of $\sim$ 1:1 but with stronger \Oneb\ emission in the early nebular phase.
The He-shell detonation models of \citet{Dessart2015} are also strong in \CaII\ and weak in \Oneb\ due to a very low oxygen mass. 
The nebular spectra of SN 2005E showed \Oneb\ emission at $\sim 2$ months after explosion. Nebular modelling of these phases reveals that oxygen is $\sim 13\%$ of the total ejecta mass \citep{Perets2010}. In the case of SN 2019bkc, it may be too early for the \Oneb\ line to have developed due to the high density of the ejecta at this phase or it may be that the ejecta is intrinsically oxygen deficient.

\section{Spectral modelling}\label{sec:models}

We model the photospheric phase spectra at $-0.1$, $+3.1$ d, and $5.1$ d with a Monte-Carlo 1D radiative transfer code \citep{Mazzali1993,Lucy1999,Mazzali2000}. 
This code has previously been used to model SNe Ia \citep[e.g.][]{Mazzali2014, Ashall2014}, stripped-envelope SNe \citep[e.g.][]{Mazzali2017,Prentice2018a}, and the Ca-rich event SN 2005E \citep{Waldman2011}. 
The model spectra on \citet{Waldman2011} used the results of He detonation models, specifically an explosion model of 0.45 \msun\ CO core with a 0.2 \msun\ He envelope.
The resultant model spectra were seen to be rich in narrow \TiII\ lines and weak in \SiII\ lines.

The approach taken here is somewhat different as we have assumed no specific origin for this object (e.g.~white dwarf, massive star). Rather than produce model spectra for a pre-determined explosion profile, we start with a density profile and attempt to match the shape of spectra of SN 2019bkc at three epochs as best as possible by varying the luminosity $L$. The time since explosion ($t-t_\mathrm{exp}$) is also a key parameter in the model and fortunately, the explosion time of SN 2019bkc is very well constrained by our non-detection ($t_\mathrm{exp} =$ MJD $58540.9\pm{0.4}$ ). Following the approach of previous works \citep[e.g.][]{Sauer2006,Prentice2018a}, we then aimed to match the line velocities and the features present by varying the photospheric velocity, \vph, and the abundance of elements above the photosphere. 

The observed spectra were absolute flux calibrated using the available photometry and corrected for \Emw\ and the assumed redshift of $z=0.02$. The distance to the  object is not well constrained due to the affects of peculiar motions within the cluster on the actual redshift of SN 2019bkc. We used a distance modulus of $34.71$ mag as stated in Section~\ref{sec:host}, but note that as $T \propto L^{1/4} \propto D^{1/2}_L$, then the distance uncertainty adds an additional uncertainty to the model parameters.  

Given the similarity of some of the early spectra to SNe Ic, the density profile used for the models is CO21, which was used to model type Ic SN 1994I \citep{Nomoto1994,Iwamoto1994,Sauer2006}  
This density profile has \mej\ $\sim1$ \msun\ and \ek\ $\sim1\times 10^{51}$ erg. 
SN 2019bkc has a rise time approximately half that of SN 1994I, so in order to match the spectral evolution of SN 2019bkc, the CO21 density profile was scaled in mass to give total masses of 0.8, 0.6, 0.4, and 0.2 \msun. Only for the latter two mass models did the density drop sufficiently fast with time to form the spectra with an acceptable photospheric velocity, temperature, and luminosity at the appropriate epochs. Of these two models, the lowest mass density profile \mej\ $\sim 0.2$ \msun\ provided marginally better models. 
The mass scaling preserves the specific kinetic energy \ek/\mej, giving this model \ek\ $\sim 2\times 10^{50}$ erg.
Thus, from the models we find that \mej\ $=0.2-0.4$ \msun, consistent with the mass derived from the pseudo-bolometric light curve using  Equation~\ref{eqn:arnett}, and with \ek\ $=2-4\times 10^{50}$ erg and \eom\ $\sim 1$ (in units of [$10^{51}$ erg]/[\msun]).

We assumed that the ejecta is completely defined by this model setup and did not include a He envelope as used by \citet{Waldman2011}. The lack of He lines in the observed spectra suggests that either He is not present or the conditions are not right to non-thermally excite the He present \citep{Lucy1991}. In the latter case we cannot rule out some dynamical effect on the ejecta from transparent He.

\subsection{$-0.1$ d spectrum}
The $-0.1$ d ($t-t_\mathrm{exp} = 5.0$ d) spectrum is approximately at maximum light.
The SPRAT spectrum does not extend blueward of $\sim 4000$ \AA, so to constrain the blueward part of the model we calculated a synthetic $u$-band using the LT SDSS-$u$ filter profile and match the spectrum to the dereddened photometry.
This spectrum (Fig.~\ref{fig:models}) is modelled with a photospheric velocity of 14\,500 \kms\ and a black body temperature, $T_\mathrm{bb} \sim 11\,000$ K.
We find that the spectra are much better represented with the heavier elements found in the \citet{Waldman2011} abundance distributions than those for SNe~Ic, where the vast majority of the ejecta is C and O with only trace Fe-group elements.
The main features in the bluest part of the model spectra are produced by \FeIII, \CrII. Although \TiII\ lines are present they do not contribute significantly to line formation in any region of the observed spectrum. 
These elements make up $\sim4\%$ of the abundance distribution for this model. The broad double feature between 6000 and 6600 \AA\ is replicated on the blue side by \SiII\ \lam 6355  and a series of \ArII\ lines, mainly \ArII\ \lam\lam 6638, 6639, 6643, on the red side. For the latter, \CII\ \lam 6580 was found to appear too far blueward and the absence of a significant contribution from \HeI\ \lam 5876 to the spectrum rules out \HeI\ \lam 6678. Importantly, Ar constitutes $1-10 \%$ of the ejecta in the \citet{Waldman2011} He detonation model, and $10\%$ in this model, but it is not present in abundance distributions of SNe~Ic \citep[see][]{Sauer2006,Mazzali2017}.

\subsection{$+3.1$ d spectrum}
The $+3.1$ d ($t-t_\mathrm{exp} = 8.2$ d) spectrum is the first observed with the NTT and is the first to extend blueward of 4000 \AA.
This spectrum has a higher S/N ratio and resolution compared with the earlier spectra and extends to shorter wavelengths. It also shows a degree of complexity not seen in ``normal'' Type I events due to the number of features at wavelengths shorter than 5400 \AA. 
The model spectrum has $T_\mathrm{bb} \sim 9300$ K and a photospheric velocity of $10\,500$ \kms.
The last $u$-band observation was 0.3 d prior to this at $\sim 18.6$ mag. The synthetic $u$ calculated from the model is $\sim 30$\% dimmer than the extrapolated value, which itself has an uncertainty of at least 10\%. This may suggest that the temperature needs to be higher to increase the model flux in this region, or that the quantity of Fe-group elements needs to be reduced to decrease the effect of line-blanketing. Such changes would affect the model spectrum in the observed region but without knowledge of the actual spectrum in the $u$-band region the kind of fine tuning required to investigate this further cannot be performed.

Despite the presence of Ar as 10\% of the ejecta above the photosphere at this phase, the \ArII\ lines seen in the earlier spectra are not replicated, which is consistent with the evolution of the $\sim6400$ \AA\ feature in the observed spectra. As before, the features in the blue side of the spectrum are predominately formed from \CaII\ H\&K, \MgII, \CrII, \TiII, and now \FeII rather than \FeIII\ oweing to the lower temperature. On the red side of the spectrum we see more lines formed from IMEs, \SiII\ \lam 6355, \OI\ \lam 7772, and \CaII\ NIR.
The strength and position of the \OI\ line is ambiguous however, owing to contamination by the O$_2$ telluric feature. 
The abundance of Fe-group elements in the model at this phase is $\sim 4\%$, the majority of which is \Nifs\ and Cr and $\sim 1\%$  Fe and Ti.

\subsection{$+5.1$ d spectrum}
The $+5.1$ d ($t-t_\mathrm{exp} = 10.2$ d) spectrum was modelled with a photospheric velocity of 7500 \kms\ and $T_\mathrm{bb} \sim 8300$ K. 
At this phase the model is producing narrow lines while those in the spectrum are broader, which suggests that the slope of the density profile above the photosphere boundary at this point should be flatter (i.e. the line forming region more extended) than we have used. This behaviour is especially unusual however, as broad lines are seen to give way to narrower lines in normal SNe. 
Also, as noted in Section~\ref{sec:spectra}, this spectrum is rather peculiar, as the \SiII\ line is broader and at a higher velocity than earlier. 
It was also noted that, when compared with SN 1994I, the velocity of features associated with Fe-group elements in the blue are apparently well below that of the IME velocities in the red. 
We see that this is not reflected in the model spectrum, the IME lines form at too low a velocity, while that of the Fe-group lines is too high. 
As a consequence, the model cannot recreate the redward extent of the blue wing of the pseudo-emission peak at $\sim 5500$ \AA.
The IME line velocities could be better replicated by increasing the total mass of the model used, which would allow the photosphere to form at a larger velocity co-ordinate (as a test, setting \vph\ $=10\,500$ \kms\ proved to position these lines well), but this adversely affects other aspects of the model. Firstly, it would further increase the velocity of the line-forming regions of the Fe-group elements, then to conserve the flux within the spectrum (and therefore the luminosity), the temperature would have to decrease which may lead to problems in matching the spectral shape. Additionally, these affects would be seen in the earlier two models.  
The difficulty in finding a model to match the spectrum at this phase suggests that the structure of the ejecta is complex, and it must be stressed that the spectra are effectively nebular $<7$ d after this observation so this spectra may not be fully photospheric, which is a requirement of the model setup.

\begin{figure}
    \centering
    \includegraphics[scale=0.5]{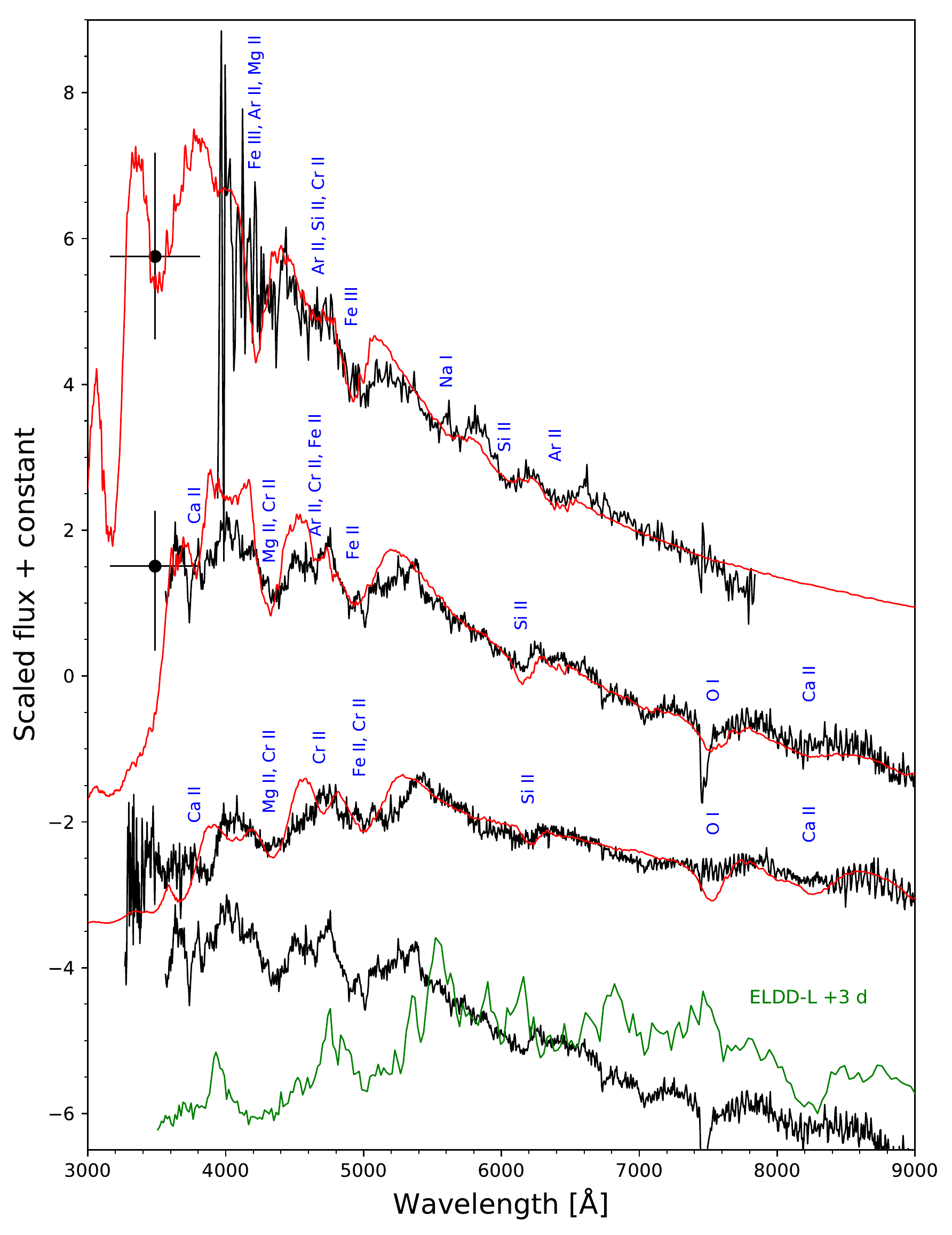}
    \caption{Spectroscopic modelling of SN 2019bkc (red) using a density profile with \mej\ $\sim 0.2$ \msun\ and \ek\ $\sim2\times 10^{50}$ erg compared with observed spectra at $-0.1$, $+3.1$, and $+5.1$ d (black). Prominent elements are noted along with the $u$-band fluxes used to constrain the models. Also included is the scaled ELDD-L model spectrum from \citet{Sim2012} (green) at $\sim$3 d after maximum compared to our spectra of SN 2019bkc at a similar phase.}
    \label{fig:models}
\end{figure}

\subsection{Comparison of model properties}
In Table~\ref{tab:modcomp} the properties derived from the spectroscopic models of SN 2019bkc are compared with the He-detonation CO.45He.2 explosion model of \citet{Waldman2011} and the low-mass edge-lit double-detonation explosions model EDDL-L of \citet{Sim2012}, both of which are modelled from CO white dwarf progenitors with a He envelope. 
We also compare with the ultra-stripped model of \cite{De2018}, which is of core-collapse origin.

The total ejecta mass for our model of SN 2019bkc is smaller by a factor of $\sim2$ compared with the two white dwarf explosion models, but is similar to that for the `ultra-stripped' model.
The specific kinetic energy is most similar to the `ultra-stripped' model at $\sim 1$ and is marginally larger compared with ELDD-L but is larger by a factor of $\sim0.3$ compared with CO.45He.2. These results marginally favour a massive star progenitor.
However, we tentatively detected \ArII\ in the spectra, which is in agreement with the abundance profile of CO.45He.2 and is perhaps the strongest evidence for a white-dwarf progenitor. This is further supported by the difficulty obtaining a suitable model spectrum using a stripped-envelope SN-like abundance, although we note that the abundance distribution used is weaker in Fe-group elements (especially Ti) compared with CO.45He.2/ELDD-L \citep{Waldman2011,Sim2012}.
Consequently, we cannot say for certain that our results favour one scenario over another.

\begin{table*}[]
    \centering
    \caption{Comparison of model properties.}
    \begin{tabular}{lccccc}
    \hline
    Model  & Progenitor  &\mej\ & \ek & \ek/\mej &   Reference \\
      &   &[\msun] & [erg] & [$10^{51}$ erg]/[\msun] &  \\
    \hline
    2019bkc & - &0.2--0.4  & 2 -- 4$\times10^{50}$ & $\sim 1$ & This work \\
    CO.45He.2$^{\dag}$ & White dwarf & 0.65  & $1.8\times10^{50}$ & $\sim0.3 $ & \citep{Waldman2011} \\
    ELDD-L & White dwarf & 0.66 & $\sim 6\times10^{50}$  & $\sim 0.9$  & \citep{Sim2012} \\
    iPTF14gqr & Massive star &0.2 & $\sim2\times10^{50}$ & $\sim1$ & \citep{De2018} \\
    \hline
    \multicolumn{6}{p{\textwidth}}{$^{\dag}$This model was used to create synthetic spectra which were then compared with SN~2005E in \citet{Waldman2011}. The nebular phase spectra of SN~2005E were modelled in \citet{Perets2010} and gave an ejecta mass of $\sim0.3$ \msun. }\\
    \end{tabular}
    \label{tab:modcomp}
\end{table*}

\section{Discussion}\label{sec:discuss}
SN 2019bkc has shown itself to be rather unusual in several ways, from the rapidly declining light curve and early spectra with similarities to normal type Ic SNe to the high velocity shift of $\sim 10\,000$ \kms\ seen in the \CaiiF, \CaII and \OI\ features in the nebular spectra. We now discuss possible explanations for its physical appearance and the environment that hosts it.

\subsection{Explaining the blueshifted calcium emission lines}

Shifts in the central positions of emission lines are seen in the late-time spectra of SNe Ia with values in the range of $\pm$2000 \kms\ \citep[e.g.][]{Maeda2010, Maguire2018}. Velocity offsets in forbidden lines of Ca and O have also been identified in Type Ib/c SNe but the magnitudes of the shifts are significantly lower with values of up to $\sim$1000 \kms\ \citep{Taubenberger2009}.   The Ca-rich transient sample studied to date showed velocity offsets in their \CaiiF\ features with respect to their rest-frame positions ranging from zero to 5000 \kms\ and have been seen as both redshifts and blueshifts \citep{Valenti2014,Foley2015}. However, the blueshifts seen in the \CaiiF\ and \CaII\ emission lines of SN 2019bkc of $\sim$10\,000 \kms\  are the highest seen to date in any Ca-rich event and we tentatively identify a similar blueshift in the \OI\ 7772 \AA\ feature. In this section, we discuss some scenarios to explain these extreme velocity shifts.

\subsubsection{SN 2019bkc as an asymmetric explosion}

A very asymmetric explosion is a potential way to explain the large velocity shifts seen in the nebular phase spectra of SN 2019bkc, with the Ca-rich (and O-rich) ejecta, or perhaps $^{56}$Ni ejecta with trace amounts of Ca and O, ejected towards us at high velocity. The velocity shift measured is the {\it minimum} total velocity, this is in the case of the asymmetry being directly towards us. Any other viewing angle increases the total velocity, which increases the kinetic energy without increasing the mass, so the specific kinetic energy increases. If the remnant is a compact object then momentum could be conserved without the need for mass ejected in the opposing direction, which we see no evidence of. Asymmetry also affects the observed light curve in both duration and peak magnitude \citep[e.g.][]{Sim2007,Barnes2018}. If the explosion was significantly asymmetric this could explain the rapidly declining light curve through a reduction in the photon diffusion time and the velocity shift seen in the nebular spectra.  The main argument against this model is that the photospheric-phase spectra are inconsistent with any significant velocity shift when compared with other objects, see Fig~\ref{fig:Ic} and the emitting region at early times appears symmetric.

Alternatively, an asymmetric explosion gives an asymmetric distribution of elements. 
3D explosion models of core-collapse SNe have shown that explosions happen through symmetry breaking in the gain region, low pressure regions at the gain layer allow shocked material pass through and unbind the star. The result of this process is plumes of material being ejected \citep{Wongwathanrat2013,Summa2018}.
If the radioactive material synthesised in the explosion was preferentially ejected as a ``blob'' moving $\sim 10\,000$ \kms\ then the nebular emission could be explained by the radioactive material exciting the material around it. It also explains the fast light curve on the basis of reduced photon diffusion time as the column density above the material is greatly reduced of the centrally-located case. 
This simplistic explanation is not without its own problems, however, as it requires no radioactive elements anywhere else in the ejecta as there is no emission at velocities significantly away from $\sim 10\,000$ \kms. The underlying explosion scenario resulting in this type of ejecta structure is also unexplained.

\subsubsection{SN 2019bkc as an extreme line-of-sight motion object}\label{sec:SMBH}
Ca-rich transients have a preference for remote locations far from the centres of their host galaxies. This has been confirmed by \citet{Frohmaier2018} not to be a selection affect due to the setup of transient surveys but something intrinsic to the explosions. It has been suggested that Ca-rich events result from systems that have previously been kicked from the centre of their host galaxies by interaction with a central supermassive black hole \citep[SMBH;][]{Foley2015}. A potential correlation was identified between the offset from the likely host galaxy and the velocity offset seen in the \CaiiF\ emission lines, with less offset objects displaying larger velocity offsets.  If we assume that the most likely host galaxy for the object is the brightest cluster galaxy, NGC 3090 at an offset distance of 95 kpc from SN 2019bkc, then it was predicted it should have a velocity offset roughly consistent with zero. However, it has the highest offset seen to date for a Ca-rich event of $\sim$10\,000 \kms\ and does not fit in this picture. 
Even if it was associated with 2dFGRS TGN216Z084 (galaxy `B' in Fig.~\ref{fig:gals}) at a redshift of 0.019, it would have a separation of 34 kpc. 

\citet{Mili2017} also discussed that if interaction with a SMBH was responsible for the velocity offset seen in the Ca features then it should also be seen in the O features and typically this is not the case, but our results suggest this may be the case for SN 2019bkc where a potential \OI\ feature blueshifted by $8\,000 - 10\,000$ \kms\ may be present. \citet{Coughlin2018} has explored the kick velocities that would be expected for systems interacting with SMBH and found that some systems could have velocities greater than 10\,000 \kms.  
This in an attractive solution for the velocity offset in the nebular phase but is inconsistent when applied to explain the photospheric phase spectra that do not require a velocity offset. Therefore, we conclude that this scenario is unlikely to explain the properties of SN 2019bkc.

\subsection{Explosion scenarios for SN 2019bkc}
The analysis presented so far has not resulted in a firm identification of a particular explosion scenario or progenitor for SN 2019bkc. In this section we discuss possible progenitor scenarios and the pros and cons for each.

\subsubsection{SN 2019bkc as a double-detonation event}
In Fig.~\ref{fig:lcs}, we compared SN 2019bkc to a low-luminosity edge-lit double-detonation model (ELDD-L) of a sub-Chandrasekhar mass white dwarf by \citet{Sim2012}. Even compared to this extreme model, the light curve of SN 2019bkc is still significantly faster and bluer in its $g-r$ colour. The model spectra of the ELDD-L model at 8 d past maximum are shown alongside our models in Fig.~\ref{fig:models}. The spectrum peaks at $\sim 5500$ \AA\ and is defined by copious narrow absorption features, mostly \TiII\ and \CaII. This is not unlike the model spectra produced from the explosion models given in \citet{Waldman2011}. Comparison with our spectra at similar phases, however, shows very few similarities. Our spectra peak towards 4000 \AA, and while there are several unidentified narrow features in the blue the spectrum has a strong continuum and is rather sparse of features blueward of 6000 \AA.

\subsubsection{SN 2019bkc as a tidal disruption event by an intermediate-mass black hole}
There have been suggestions of a link between Ca-rich events and the tidal disruption of a white dwarf by an intermediate mass black holes (IMBH) \citep[e.g.][]{Luminet:1989a,rosswog2009, Sell2015}. A prediction of white dwarf-IMBH tidal disruption models is that a thermonuclear explosion can be triggered during the disruption that can produce a low-luminosity optical transient \citep{rosswog2009,Clausen:2011,Macleod2014,macleod2016_opt} and if the impact parameter is relatively low then the ejecta may be Ca-rich \citep{Kawana2018,Anninos2018}. The burning efficiency, however, may be too low to produce Ca in large quantities and depends on how much is required to explain the late-time spectra of Ca-rich transients. 

Other predictions of white dwarf and IMBH tidal disruption models are that there should be soft X-ray emission resulting from the accretion disk and potentially X-ray emission from a relativistic jet \citep{macleod2016_opt}. Emission from a relativistic jet is strongly viewing angle dependent and is not expected to be visible in all cases. The X-ray emission from the accretion disk, however, should be visible from any viewing angle and is expected to have a luminosity of $\sim 10^{41}-10^{43}$ \ergs\ that last for months to a year after the event. A handful of Ca-rich events have been searched for X-ray signatures but no high-energy counterpart has been identified to date \citep[e.g.][]{Sell2015,sell2018_16hnk}. Our Swift XRT upper limit of $<6.3\times10^{40}$ \ergs\ was obtained $\sim 42$ d after explosion, which would rule out most of the model range for X-ray emission from an accretion disk for SN 2019bkc.

These models also predict strong viewing angle dependencies in the observed optical properties due to a bulk orbital motion \citep{macleod2016_opt}. It is predicted that the absorption profiles in the photospheric phase could have an additional offset of up to 10\,000 \kms\ to bluer or redder wavelengths. This is not seen in the early spectra of SN 2019bkc that were found through spectral modelling to be consistent with a stripped envelope SN or an unusual thermonuclear event with no velocity offset needing to be applied.  The line velocity shifts were only seen in the nebular-phase spectra and this is not explained by the bulk orbital motion scenario in tidal disruption event of a white dwarf by an IMBH.

\subsubsection{SN 2019bkc as a merger between a white dwarf and neutron star or stellar-mass black hole}
\citet{Metzger2012} modelled the merger of a white dwarf with a neutron star or stellar mass black hole and looked at the resulting properties of the accretion disk. The observables include a light curve that rises to a peak on a timescale of $\sim5$ days before decaying rapidly, reaching peak bolometric luminosities of $10^{39}-10^{41.5}$ \ergs, and an ejecta that contains Ar with a total ejecta mass is between 0.3-1 \msun\ and ejecta velocities $\sim 10^4$ \kms.
The properties of the explosion model are similar to the observed SN 2019bkc, especially through the light curve, although SN 2019bkc is more luminous at $10^{42.3}$ \ergs. 
The argument against this model is that the ejecta is Ca-deficient and so couldn't explain the later appearance of [\CaII] and \CaII\ NIR. Ca is an efficient coolant however, and would give rise to these lines even if small amounts were present in the ejecta of such a merger.

\subsubsection{SN 2019bkc as an ultra-stripped SN}
Ultra-stripped SNe \citep{Tauris2013} arise from the core-collapse of massive He stars that have been stripped of virtually all the material above their Chandrasekhar-mass core through binary interaction with a companion, most likely a neutron star. The low ejecta mass of a few $0.1$ \msun\ and low \ek\ of $1\times 10^{50} - 1\times 10^{51}$ erg leads to dim, rapidly evolving light curves as the \mni\ synthesised is a few $0.01$ \msun, the diffusion time is short \citep{Tauris2015,Moriya2017}. The pre-explosion abundance structure should be rich in the elements produced in stellar nucleosynthesis; He, C, O, Mg, Ca, Si, S. At core-collapse most of this material is left unburnt and contributes to spectroscopic line formation.

\citet{De2018} interpreted iPTF14gqr as an ultra-stripped type Ic SN. The spectra of this object and SN 2019bkc have many similarities, including the nebular-phase spectra that are undoubtedly Ca-rich in appearance and unlike those more usually seen in SNe Ic where the \Oneb\ \lam\lam 6300, 6363 line dominates. In terms of the light curve, SN 2019bkc is much faster, but the overall colour evolution and temperature evolution are not too dissimilar. iPTF14gqr had a double peaked light curve, interpreted as shock-cooling from recently ejected He-rich material from the companion star. Comparatively, we inferred that the early $g$-band light curve of SN 2019bkc showed a steep rise and a rapid levelling off. There is no evidence of He in the spectra of SN 2019bkc, although this may be because spectral observations began too late.
The results of our spectroscopic modelling can neither confirm nor deny that this transient is an ultra-stripped SNe. Aside from the possible presence of Ar at early times the model still remains viable for a CC event.

However, the explosion of a massive star apparently so far from a star-forming region is much harder to explain with a CC event. Assuming that the life time of the progenitor star is $<40$ Myr \citep{Tauris2015} and that the host is the nearby elliptical galaxy NGC 3090, then this object must have travelled at least $\sim 95$ kpc in this time,  placing a lower limit on the velocity of $\sim 2300$ \kms. If the progenitor system was not ejected immediately, and has any radial velocity component, then this velocity must be significantly higher. It also stands to reason that this requires NGC 3090 to have had some star formation in the recent past in order to produce a NS/He-star binary system.

\section{Conclusions}\label{sec:conc}
In this work, we present the discovery, observations, and analysis of the extraordinary Ca-rich transient ATLAS19dqr/SN 2019bkc.
The event took place in the intracluster space of a galaxy group with $z\sim 0.02$. There is no source underlying the transient in pre-explosion DECaLS images ruling out the presence of a faint host or globular cluster at approximately $M >-10$ mag. 

A robust non-detection $0.8$ d prior to the first observation reveals a short rise, while the multi-colour light curves show a clear peak at $\sim -17$ mag approximately $5 - 6$ d after the non-detection. These components give a light curve shape that is not consistent with a blast-wave approximation to the luminosity.
The light curves then begin a remarkably rapid decline, with $\Delta m_{15} (g) \sim 5$ mag, making SN 2019bkc one of the most rapidly evolving transients known. 
A pseudo-bolometric light curve constructed from $ugriz$ photometry reached a peak luminosity of \lp\ $ = (1.9\pm{0.1})\times 10^{42}$ \ergs in $5.0\pm{0.4}$ d, giving \mni\ $=0.038\pm{0.001}$ \msun.

The spectroscopic observations proved to be equally unusual. 
In the photospheric phase we find similarities to SNe Ic but with considerable complexity in the spectra in the form of weak and narrow features blueward of 5500 \AA. 
In the late photospheric phase ($+5.1$ d) we see an unusual situation with indications of high velocity IMEs but low velocity Fe-group elements. There is evidence of an increase in line velocity and line broadening in the \SiII\ \lam 6355 line at these times.
The nebular phase spectra display strong lines which we identify to be [\CaII] \lam\lam 7291, 7324 and \CaII\ NIR, but with a measured blueshift of 10\,000 -- 12\,000 \kms. 
Assuming that the peak of the emission line defines the velocity then we see a change in [\CaII] velocity from $v=12\,000$ \kms\ on day $+12.9$ to $v=9500$ \kms\ on day $+19.6$. The velocity of the \CaII\ NIR line remains unchanged.

There is no detection of \Oneb\ \lam6300, 6363 at either the rest wavelength or $\sim 10\,000$ \kms\ in the nebular spectra. Consequently we measure only upper limits for the flux of this line and the resulting [\CaII]/\Oneb\ ratio places the object firmly in the region occupied by Ca-rich transients. We speculate that the object may be O-deficient, or the ejecta density may be too high and so it may be too early for the line to appear.

Three photospheric phase spectra were modelled using a 1D radiative transfer code and reveal a low ejecta mass of \mej\ $\sim 0.2-0.4$ \msun\ (consistent with $0.4\pm^{0.1}_{0.2}$ \msun\ calculated from the light curve), and \ek\ $\sim (2-4)\times 10^{50}$ erg. Abundance profiles for SNe Ic \citep{Sauer2006,Mazzali2017} and He-detonation models \citep{Waldman2011} were used as a starting basis for defining the input abundance of the models. We find that the early spectra are mostly IMEs with $\sim 3-4 \%$ as a combination of \Nifs, \Fefs, Ti, Cr at velocities $>14\,000$ \kms. A broad absorption feature, which is present in the earliest two spectra can be explained by a combination of \ArII\ lines along the abundances found in the \citet{Waldman2011} explosion models. We note difficulties in modelling the $+5.1$ d spectrum due to the odd velocity profile.
The tentative detection of an \ArII\ line in the maximum light spectrum marginally preferences a thermonuclear explosion (possibly a He detonation) over a explosion of CC origin. However, with a low ejecta mass and specific kinetic energy of the ejecta \eom\ $\sim1$ [$10^{51}$ erg]/\msun, it is more similar to ultra-stripped Ca-rich iPTF14gqr than models of He-detonation and double-detonation events.

SN 2019bkc was a highly unusual event with one of the fastest ever light curve decline rate for an extragalactic transient.  It has been our hope that as we find more Ca-rich transients, we will better understand their nature. In particular, thanks to the extreme velocity shift seen in the early nebular lines this object has added to the overall diversity and now the picture seems less clear. Our results show that the transient cannot be conclusively attributed to a massive star or white dwarf progenitor and more detailed modelling of the nebular-phase spectra, in particular, is required to identify the most likely explosion scenario.

\begin{acknowledgements}

We thank Stuart Sim and Steve Schulze for useful discussions, Takashi Moriya for sharing the density profile used to model the spectra of iPTF14qgr, and Stefano Valenti for providing the reference [\CaII]/\Oneb\ data.
SJP, KM, MRM, and KS are supported by H2020 ERC grant no.~758638.
MF is supported by a Royal Society - Science Foundation Ireland University Research Fellowship.
L.G. was funded by the European Union’s Horizon 2020 research and innovation programme under the Marie Sk\l{}odowska-Curie grant agreement No. 839090.
MG is supported by the Polish NCN MAESTRO grant 2014/14/A/ST9/00121.
CPG acknowledges support from EU/FP7-ERC grant no. [615929].
GL was supported by a research grant (19054) from VILLUM FONDEN.
AF acknowledges the support of an ESO Studentship. 
TWC acknowledgments the funding provided by the Alexander von Humboldt Foundation.
PGJ acknowledges funding from the European Research Council under ERC 
Consolidator Grant agreement no 647208.
The Liverpool Telescope is operated on the island of La Palma by Liverpool John Moores University in the Spanish Observatorio del Roque de los Muchachos of the Instituto de Astrofisica de Canarias with financial support from the UK Science and Technology Facilities Council. 
This work has made use of data from the Asteroid Terrestrial-impact Last Alert System (ATLAS) project. ATLAS is primarily funded to search
  for near earth asteroids through NASA grants NN12AR55G, 80NSSC18K0284,
  and 80NSSC18K1575; byproducts of the NEO search include images and
  catalogues from the survey area.  The ATLAS science products have been
  made possible through the contributions of the University of Hawaii
  Institute for Astronomy, the Queen's University Belfast, the Space
  Telescope Science Institute, and the South African Astronomical Observatory.

\end{acknowledgements}

\bibliographystyle{aa}
\bibliography{allbib}

\begin{thebibliography}{117}
\expandafter\ifx\csname natexlab\endcsname\relax\def\natexlab#1{#1}\fi

\bibitem[{{Ahn} {et~al.}(2012){Ahn}, {Alexandroff}, {Allende Prieto},
  {Anderson}, {Anderton}, {Andrews}, {Aubourg}, {Bailey}, {Balbinot}, {Barnes},
  \& et~al.}]{Ahn2012}
{Ahn}, C.~P., {Alexandroff}, R., {Allende Prieto}, C., {et~al.} 2012, \apjs,
  203, 21

\bibitem[{{Anninos} {et~al.}(2018){Anninos}, {Fragile}, {Olivier}, {Hoffman},
  {Mishra}, \& {Camarda}}]{Anninos2018}
{Anninos}, P., {Fragile}, P.~C., {Olivier}, S.~S., {et~al.} 2018, \apj, 865, 3

\bibitem[{{Arnett}(1980)}]{Arnett1980}
{Arnett}, W.~D. 1980, \apj, 237, 541

\bibitem[{{Arnett}(1982)}]{Arnett1982}
{Arnett}, W.~D. 1982, \apj, 253, 785

\bibitem[{{Ashall} {et~al.}(2014){Ashall}, {Mazzali}, {Bersier}, {Hachinger},
  {Phillips}, {Percival}, {James}, \& {Maguire}}]{Ashall2014}
{Ashall}, C., {Mazzali}, P., {Bersier}, D., {et~al.} 2014, \mnras, 445, 4427

\bibitem[{{Barnes} {et~al.}(2018){Barnes}, {Duffell}, {Liu}, {Modjaz},
  {Bianco}, {Kasen}, \& {MacFadyen}}]{Barnes2018}
{Barnes}, J., {Duffell}, P.~C., {Liu}, Y., {et~al.} 2018, \apj, 860, 38

\bibitem[{{Bellm} {et~al.}(2019){Bellm}, {Kulkarni}, {Graham}, {Dekany},
  {Smith}, {Riddle}, {Masci}, {Helou}, {Prince}, {Adams}, {Barbarino},
  {Barlow}, {Bauer}, {Beck}, {Belicki}, {Biswas}, {Blagorodnova}, {Bodewits},
  {Bolin}, {Brinnel}, {Brooke}, {Bue}, {Bulla}, {Burruss}, {Cenko}, {Chang},
  {Connolly}, {Coughlin}, {Cromer}, {Cunningham}, {De}, {Delacroix}, {Desai},
  {Duev}, {Eadie}, {Farnham}, {Feeney}, {Feindt}, {Flynn}, {Franckowiak},
  {Frederick}, {Fremling}, {Gal-Yam}, {Gezari}, {Giomi}, {Goldstein},
  {Golkhou}, {Goobar}, {Groom}, {Hacopians}, {Hale}, {Henning}, {Ho}, {Hover},
  {Howell}, {Hung}, {Huppenkothen}, {Imel}, {Ip}, {Ivezi{\'c}}, {Jackson},
  {Jones}, {Juric}, {Kasliwal}, {Kaspi}, {Kaye}, {Kelley}, {Kowalski},
  {Kramer}, {Kupfer}, {Landry}, {Laher}, {Lee}, {Lin}, {Lin}, {Lunnan},
  {Giomi}, {Mahabal}, {Mao}, {Miller}, {Monkewitz}, {Murphy}, {Ngeow},
  {Nordin}, {Nugent}, {Ofek}, {Patterson}, {Penprase}, {Porter}, {Rauch},
  {Rebbapragada}, {Reiley}, {Rigault}, {Rodriguez}, {van Roestel}, {Rusholme},
  {van Santen}, {Schulze}, {Shupe}, {Singer}, {Soumagnac}, {Stein}, {Surace},
  {Sollerman}, {Szkody}, {Taddia}, {Terek}, {Van Sistine}, {van Velzen},
  {Vestrand}, {Walters}, {Ward}, {Ye}, {Yu}, {Yan}, \& {Zolkower}}]{Bellm2019}
{Bellm}, E.~C., {Kulkarni}, S.~R., {Graham}, M.~J., {et~al.} 2019, \pasp, 131,
  018002

\bibitem[{{Buzzoni} {et~al.}(1984){Buzzoni}, {Delabre}, {Dekker}, {Dodorico},
  {Enard}, {Focardi}, {Gustafsson}, {Nees}, {Paureau}, \&
  {Reiss}}]{Buzzoni1984}
{Buzzoni}, B., {Delabre}, B., {Dekker}, H., {et~al.} 1984, The Messenger, 38, 9

\bibitem[{{Cappellaro} {et~al.}(1997){Cappellaro}, {Mazzali}, {Benetti},
  {Danziger}, {Turatto}, {della Valle}, \& {Patat}}]{Cappellaro1997}
{Cappellaro}, E., {Mazzali}, P.~A., {Benetti}, S., {et~al.} 1997, \aap, 328,
  203

\bibitem[{{Chen} {et~al.}(2020){Chen}, {Dong}, {Stritzinger}, {Holmbo},
  {Strader}, {Kochanek}, {Peng}, {Benetti}, {Bersier}, {Brownsberger},
  {Buckley}, {Gromadzki}, {Moran}, {Pastorello}, {Aydi}, {Bose}, {Connor},
  {Boutsia}, {Mille}, {Elias-Rosa}, {French}, {Holoien}, {Mattila}, {Shappee},
  {Stark}, \& {Swihart}}]{Chen2020}
{Chen}, P., {Dong}, S., {Stritzinger}, M.~D., {et~al.} 2020, \apjl, 889, L6

\bibitem[{{Chen} {et~al.}(2018){Chen}, {Inserra}, {Fraser}, {Moriya}, {Schady},
  {Schweyer}, {Filippenko}, {Perley}, {Ruiter}, {Seitenzahl}, {Sollerman},
  {Taddia}, {Anderson}, {Foley}, {Jerkstrand}, {Ngeow}, {Pan}, {Pastorello},
  {Points}, {Smartt}, {Smith}, {Taubenberger}, {Wiseman}, {Young}, {Benetti},
  {Berton}, {Bufano}, {Clark}, {Della Valle}, {Galbany}, {Gal-Yam},
  {Gromadzki}, {Guti{\'e}rrez}, {Heinze}, {Kankare}, {Kilpatrick},
  {Kuncarayakti}, {Leloudas}, {Lin}, {Maguire}, {Mazzali}, {McBrien},
  {Prentice}, {Rau}, {Rest}, {Siebert}, {Stalder}, {Tonry}, \& {Yu}}]{Chen2018}
{Chen}, T.-W., {Inserra}, C., {Fraser}, M., {et~al.} 2018, \apjl, 867, L31

\bibitem[{{Chugai}(2000)}]{Chugai2000}
{Chugai}, N.~N. 2000, Astronomy Letters, 26, 797

\bibitem[{{Clausen} \& {Eracleous}(2011)}]{Clausen:2011}
{Clausen}, D. \& {Eracleous}, M. 2011, \apj, 726, 34

\bibitem[{{Colless} {et~al.}(2001){Colless}, {Dalton}, {Maddox}, {Sutherland},
  {Norberg}, {Cole}, {Bland-Hawthorn}, {Bridges}, {Cannon}, {Collins}, {Couch},
  {Cross}, {Deeley}, {De Propris}, {Driver}, {Efstathiou}, {Ellis}, {Frenk},
  {Glazebrook}, {Jackson}, {Lahav}, {Lewis}, {Lumsden}, {Madgwick}, {Peacock},
  {Peterson}, {Price}, {Seaborne}, \& {Taylor}}]{Colless2001}
{Colless}, M., {Dalton}, G., {Maddox}, S., {et~al.} 2001, \mnras, 328, 1039

\bibitem[{{Coughlin} {et~al.}(2018){Coughlin}, {Darbha}, {Kasen}, \&
  {Quataert}}]{Coughlin2018}
{Coughlin}, E.~R., {Darbha}, S., {Kasen}, D., \& {Quataert}, E. 2018, \apjl,
  863, L24

\bibitem[{{De} {et~al.}(2018{\natexlab{a}}){De}, {Kasliwal}, {Cantwell}, {Cao},
  {Cenko}, {Gal-Yam}, {Johansson}, {Kong}, {Kulkarni}, {Lunnan}, {Masci},
  {Matuszewski}, {Mooley}, {Neill}, {Nugent}, {Ofek}, {Perrott},
  {Rebbapragada}, {Rubin}, {O' Sullivan}, \& {Yaron}}]{De2018b}
{De}, K., {Kasliwal}, M.~M., {Cantwell}, T., {et~al.} 2018{\natexlab{a}}, \apj,
  866, 72

\bibitem[{{De} {et~al.}(2018{\natexlab{b}}){De}, {Kasliwal}, {Ofek}, {Moriya},
  {Burke}, {Cao}, {Cenko}, {Doran}, {Duggan}, {Fender}, {Fransson}, {Gal-Yam},
  {Horesh}, {Kulkarni}, {Laher}, {Lunnan}, {Manulis}, {Masci}, {Mazzali},
  {Nugent}, {Perley}, {Petrushevska}, {Piro}, {Rumsey}, {Sollerman},
  {Sullivan}, \& {Taddia}}]{De2018}
{De}, K., {Kasliwal}, M.~M., {Ofek}, E.~O., {et~al.} 2018{\natexlab{b}},
  Science, 362, 201

\bibitem[{{Dessart} \& {Hillier}(2015)}]{Dessart2015}
{Dessart}, L. \& {Hillier}, D.~J. 2015, \mnras, 447, 1370

\bibitem[{{Dey} {et~al.}(2019){Dey}, {Schlegel}, {Lang}, {Blum}, {Burleigh},
  {Fan}, {Findlay}, {Finkbeiner}, {Herrera}, {Juneau}, {Landriau}, {Levi},
  {McGreer}, {Meisner}, {Myers}, {Moustakas}, {Nugent}, {Patej}, {Schlafly},
  {Walker}, {Valdes}, {Weaver}, {Y{\`e}che}, {Zou}, {Zhou}, {Abareshi},
  {Abbott}, {Abolfathi}, {Aguilera}, {Alam}, {Allen}, {Alvarez}, {Annis},
  {Ansarinejad}, {Aubert}, {Beechert}, {Bell}, {BenZvi}, {Beutler}, {Bielby},
  {Bolton}, {Brice{\~n}o}, {Buckley-Geer}, {Butler}, {Calamida}, {Carlberg},
  {Carter}, {Casas}, {Castander}, {Choi}, {Comparat}, {Cukanovaite}, {Delubac},
  {DeVries}, {Dey}, {Dhungana}, {Dickinson}, {Ding}, {Donaldson}, {Duan},
  {Duckworth}, {Eftekharzadeh}, {Eisenstein}, {Etourneau}, {Fagrelius},
  {Farihi}, {Fitzpatrick}, {Font-Ribera}, {Fulmer}, {G{\"a}nsicke},
  {Gaztanaga}, {George}, {Gerdes}, {Gontcho}, {Gorgoni}, {Green}, {Guy},
  {Harmer}, {Hernandez}, {Honscheid}, {Huang}, {James}, {Jannuzi}, {Jiang},
  {Joyce}, {Karcher}, {Karkar}, {Kehoe}, {Kneib}, {Kueter-Young}, {Lan},
  {Lauer}, {Le Guillou}, {Le Van Suu}, {Lee}, {Lesser}, {Perreault Levasseur},
  {Li}, {Mann}, {Marshall}, {Mart{\'{\i}}nez-V{\'a}zquez}, {Martini}, {du Mas
  des Bourboux}, {McManus}, {Meier}, {M{\'e}nard}, {Metcalfe},
  {Mu{\~n}oz-Guti{\'e}rrez}, {Najita}, {Napier}, {Narayan}, {Newman}, {Nie},
  {Nord}, {Norman}, {Olsen}, {Paat}, {Palanque-Delabrouille}, {Peng},
  {Poppett}, {Poremba}, {Prakash}, {Rabinowitz}, {Raichoor}, {Rezaie},
  {Robertson}, {Roe}, {Ross}, {Ross}, {Rudnick}, {Safonova}, {Saha},
  {S{\'a}nchez}, {Savary}, {Schweiker}, {Scott}, {Seo}, {Shan}, {Silva},
  {Slepian}, {Soto}, {Sprayberry}, {Staten}, {Stillman}, {Stupak}, {Summers},
  {Sien Tie}, {Tirado}, {Vargas-Maga{\~n}a}, {Vivas}, {Wechsler}, {Williams},
  {Yang}, {Yang}, {Yapici}, {Zaritsky}, {Zenteno}, {Zhang}, {Zhang}, {Zhou}, \&
  {Zhou}}]{Dey2019}
{Dey}, A., {Schlegel}, D.~J., {Lang}, D., {et~al.} 2019, \aj, 157, 168

\bibitem[{{Drout} {et~al.}(2014){Drout}, {Chornock}, {Soderberg}, {Sanders},
  {McKinnon}, {Rest}, {Foley}, {Milisavljevic}, {Margutti}, {Berger},
  {Calkins}, {Fong}, {Gezari}, {Huber}, {Kankare}, {Kirshner}, {Leibler},
  {Lunnan}, {Mattila}, {Marion}, {Narayan}, {Riess}, {Roth}, {Scolnic},
  {Smartt}, {Tonry}, {Burgett}, {Chambers}, {Hodapp}, {Jedicke}, {Kaiser},
  {Magnier}, {Metcalfe}, {Morgan}, {Price}, \& {Waters}}]{Drout2014}
{Drout}, M.~R., {Chornock}, R., {Soderberg}, A.~M., {et~al.} 2014, \apj, 794,
  23

\bibitem[{{Drout} {et~al.}(2017){Drout}, {Piro}, {Shappee}, {Kilpatrick},
  {Simon}, {Contreras}, {Coulter}, {Foley}, {Siebert}, {Morrell}, {Boutsia},
  {Di Mille}, {Holoien}, {Kasen}, {Kollmeier}, {Madore}, {Monson},
  {Murguia-Berthier}, {Pan}, {Prochaska}, {Ramirez-Ruiz}, {Rest}, {Adams},
  {Alatalo}, {Ba{\~n}ados}, {Baughman}, {Beers}, {Bernstein}, {Bitsakis},
  {Campillay}, {Hansen}, {Higgs}, {Ji}, {Maravelias}, {Marshall}, {Moni Bidin},
  {Prieto}, {Rasmussen}, {Rojas-Bravo}, {Strom}, {Ulloa},
  {Vargas-Gonz{\'a}lez}, {Wan}, \& {Whitten}}]{Drout2017}
{Drout}, M.~R., {Piro}, A.~L., {Shappee}, B.~J., {et~al.} 2017, Science, 358,
  1570

\bibitem[{{Drout} {et~al.}(2013){Drout}, {Soderberg}, {Mazzali}, {Parrent},
  {Margutti}, {Milisavljevic}, {Sanders}, {Chornock}, {Foley}, {Kirshner},
  {Filippenko}, {Li}, {Brown}, {Cenko}, {Chakraborti}, {Challis}, {Friedman},
  {Ganeshalingam}, {Hicken}, {Jensen}, {Modjaz}, {Perets}, {Silverman}, \&
  {Wong}}]{Drout2013}
{Drout}, M.~R., {Soderberg}, A.~M., {Mazzali}, P.~A., {et~al.} 2013, \apj, 774,
  58

\bibitem[{{Faber} {et~al.}(1989){Faber}, {Wegner}, {Burstein}, {Davies},
  {Dressler}, {Lynden-Bell}, \& {Terlevich}}]{Faber1989}
{Faber}, S.~M., {Wegner}, G., {Burstein}, D., {et~al.} 1989, \apjs, 69, 763

\bibitem[{{Filippenko} {et~al.}(1995){Filippenko}, {Barth}, {Matheson},
  {Armus}, {Brown}, {Espey}, {Fan}, {Goodrich}, {Ho}, {Junkkarinen}, {Koo},
  {Lehnert}, {Martel}, {Mazzarella}, {Miller}, {Smith}, {Tytler}, \&
  {Wirth}}]{Filippenko1995}
{Filippenko}, A.~V., {Barth}, A.~J., {Matheson}, T., {et~al.} 1995, \apjl, 450,
  L11

\bibitem[{{Foley}(2015)}]{Foley2015}
{Foley}, R.~J. 2015, \mnras, 452, 2463

\bibitem[{{Foley} {et~al.}(2003){Foley}, {Papenkova}, {Swift}, {Filippenko},
  {Li}, {Mazzali}, {Chornock}, {Leonard}, \& {Van Dyk}}]{Foley2003}
{Foley}, R.~J., {Papenkova}, M.~S., {Swift}, B.~J., {et~al.} 2003, \pasp, 115,
  1220

\bibitem[{{Freudling} {et~al.}(2013){Freudling}, {Romaniello}, {Bramich},
  {Ballester}, {Forchi}, {Garc{\'{\i}}a-Dabl{\'o}}, {Moehler}, \&
  {Neeser}}]{Freudling2013}
{Freudling}, W., {Romaniello}, M., {Bramich}, D.~M., {et~al.} 2013, \aap, 559,
  A96

\bibitem[{{Frohmaier} {et~al.}(2018){Frohmaier}, {Sullivan}, {Maguire}, \&
  {Nugent}}]{Frohmaier2018}
{Frohmaier}, C., {Sullivan}, M., {Maguire}, K., \& {Nugent}, P. 2018, \apj,
  858, 50

\bibitem[{{Gal-Yam}(2017)}]{Gal-Yam2017}
{Gal-Yam}, A. 2017, {Observational and Physical Classification of Supernovae},
  ed. A.~W. {Alsabti} \& P.~{Murdin}, 195

\bibitem[{{Gal-Yam} {et~al.}(2013){Gal-Yam}, {Mazzali}, {Manulis}, \&
  {Bishop}}]{GalYam2013}
{Gal-Yam}, A., {Mazzali}, P.~A., {Manulis}, I., \& {Bishop}, D. 2013, \pasp,
  125, 749

\bibitem[{{Galbany} {et~al.}(2019){Galbany}, {Ashall}, {H{\"o}flich},
  {Gonz{\'a}lez-Gait{\'a}n}, {Taubenberger}, {Stritzinger}, {Hsiao}, {Mazzali},
  {Baron}, {Blondin}, {Bose}, {Bulla}, {Burke}, {Burns}, {Cartier}, {Chen},
  {Della Valle}, {Diamond}, {Guti{\'e}rrez}, {Harmanen}, {Hiramatsu},
  {Holoien}, {Hosseinzadeh}, {Howell}, {Huang}, {Inserra}, {de Jaeger}, {Jha},
  {Kangas}, {Kromer}, {Lyman}, {Maguire}, {Marion}, {Milisavljevic},
  {Prentice}, {Razza}, {Reynolds}, {Sand}, {Shappee}, {Shekhar}, {Smartt},
  {Stassun}, {Sullivan}, {Valenti}, {Villanueva}, {Wang}, {Wheeler}, {Zhai}, \&
  {Zhang}}]{Galbany2019}
{Galbany}, L., {Ashall}, C., {H{\"o}flich}, P., {et~al.} 2019, \aap, 630, A76

\bibitem[{{Greiner} {et~al.}(2008){Greiner}, {Bornemann}, {Clemens}, {Deuter},
  {Hasinger}, {Honsberg}, {Huber}, {Huber}, {Krauss}, {Kr{\"u}hler},
  {K{\"u}pc{\"u} Yolda{\c s}}, {Mayer-Hasselwander}, {Mican}, {Primak},
  {Schrey}, {Steiner}, {Szokoly}, {Th{\"o}ne}, {Yolda{\c s}}, {Klose}, {Laux},
  \& {Winkler}}]{Greiner2008}
{Greiner}, J., {Bornemann}, W., {Clemens}, C., {et~al.} 2008, \pasp, 120, 405

\bibitem[{{Hinshaw} {et~al.}(2013){Hinshaw}, {Larson}, {Komatsu}, {Spergel},
  {Bennett}, {Dunkley}, {Nolta}, {Halpern}, {Hill}, {Odegard}, {Page}, {Smith},
  {Weiland}, {Gold}, {Jarosik}, {Kogut}, {Limon}, {Meyer}, {Tucker}, {Wollack},
  \& {Wright}}]{Hinshaw2013}
{Hinshaw}, G., {Larson}, D., {Komatsu}, E., {et~al.} 2013, \apjs, 208, 19

\bibitem[{{Inserra} {et~al.}(2015){Inserra}, {Sim}, {Wyrzykowski}, {Smartt},
  {Fraser}, {Nicholl}, {Shen}, {Jerkstrand}, {Gal-Yam}, {Howell}, {Maguire},
  {Mazzali}, {Valenti}, {Taubenberger}, {Benitez-Herrera}, {Bersier},
  {Blagorodnova}, {Campbell}, {Chen}, {Elias-Rosa}, {Hillebrandt},
  {Kostrzewa-Rutkowska}, {Koz{\l}owski}, {Kromer}, {Lyman}, {Polshaw},
  {R{\"o}pke}, {Ruiter}, {Smith}, {Spiro}, {Sullivan}, {Yaron}, {Young}, \&
  {Yuan}}]{Inserra2015}
{Inserra}, C., {Sim}, S.~A., {Wyrzykowski}, L., {et~al.} 2015, \apjl, 799, L2

\bibitem[{{Iwamoto} {et~al.}(1994){Iwamoto}, {Nomoto}, {H{\"o}flich},
  {Yamaoka}, {Kumagai}, \& {Shigeyama}}]{Iwamoto1994}
{Iwamoto}, K., {Nomoto}, K., {H{\"o}flich}, P., {et~al.} 1994, \apjl, 437, L115

\bibitem[{{Jerkstrand} {et~al.}(2015){Jerkstrand}, {Ergon}, {Smartt},
  {Fransson}, {Sollerman}, {Taubenberger}, {Bersten}, \&
  {Spyromilio}}]{Jerkstrand2015}
{Jerkstrand}, A., {Ergon}, M., {Smartt}, S.~J., {et~al.} 2015, \aap, 573, A12

\bibitem[{{Jerkstrand} {et~al.}(2014){Jerkstrand}, {Smartt}, {Fraser},
  {Fransson}, {Sollerman}, {Taddia}, \& {Kotak}}]{Jerkstrand2014}
{Jerkstrand}, A., {Smartt}, S.~J., {Fraser}, M., {et~al.} 2014, \mnras, 439,
  3694

\bibitem[{{Jha} {et~al.}(2006){Jha}, {Branch}, {Chornock}, {Foley}, {Li},
  {Swift}, {Casebeer}, \& {Filippenko}}]{Jha2006}
{Jha}, S., {Branch}, D., {Chornock}, R., {et~al.} 2006, \aj, 132, 189

\bibitem[{{Kasliwal} {et~al.}(2012){Kasliwal}, {Kulkarni}, {Gal-Yam}, {Nugent},
  {Sullivan}, {Bildsten}, {Yaron}, {Perets}, {Arcavi}, {Ben-Ami}, {Bhalerao},
  {Bloom}, {Cenko}, {Filippenko}, {Frail}, {Ganeshalingam}, {Horesh}, {Howell},
  {Law}, {Leonard}, {Li}, {Ofek}, {Polishook}, {Poznanski}, {Quimby},
  {Silverman}, {Sternberg}, \& {Xu}}]{Kasliwal2012}
{Kasliwal}, M.~M., {Kulkarni}, S.~R., {Gal-Yam}, A., {et~al.} 2012, \apj, 755,
  161

\bibitem[{{Kasliwal} {et~al.}(2010){Kasliwal}, {Kulkarni}, {Gal-Yam}, {Yaron},
  {Quimby}, {Ofek}, {Nugent}, {Poznanski}, {Jacobsen}, {Sternberg}, {Arcavi},
  {Howell}, {Sullivan}, {Rich}, {Burke}, {Brimacombe}, {Milisavljevic},
  {Fesen}, {Bildsten}, {Shen}, {Cenko}, {Bloom}, {Hsiao}, {Law}, {Gehrels},
  {Immler}, {Dekany}, {Rahmer}, {Hale}, {Smith}, {Zolkower}, {Velur},
  {Walters}, {Henning}, {Bui}, \& {McKenna}}]{Kasliwal2010}
{Kasliwal}, M.~M., {Kulkarni}, S.~R., {Gal-Yam}, A., {et~al.} 2010, \apjl, 723,
  L98

\bibitem[{{Kawana} {et~al.}(2018){Kawana}, {Tanikawa}, \&
  {Yoshida}}]{Kawana2018}
{Kawana}, K., {Tanikawa}, A., \& {Yoshida}, N. 2018, \mnras, 477, 3449

\bibitem[{{Khatami} \& {Kasen}(2019)}]{Khatami2019}
{Khatami}, D.~K. \& {Kasen}, D.~N. 2019, \apj, 878, 56

\bibitem[{{Kraft} {et~al.}(1991){Kraft}, {Burrows}, \& {Nousek}}]{Kraft1991}
{Kraft}, R.~P., {Burrows}, D.~N., \& {Nousek}, J.~A. 1991, \apj, 374, 344

\bibitem[{{Kr{\"u}hler} {et~al.}(2008){Kr{\"u}hler}, {K{\"u}pc{\"u} Yolda{\c
  s}}, {Greiner}, {Clemens}, {McBreen}, {Primak}, {Savaglio}, {Yolda{\c s}},
  {Szokoly}, \& {Klose}}]{Kruhler2008}
{Kr{\"u}hler}, T., {K{\"u}pc{\"u} Yolda{\c s}}, A., {Greiner}, J., {et~al.}
  2008, \apj, 685, 376

\bibitem[{{Lucy}(1991)}]{Lucy1991}
{Lucy}, L.~B. 1991, \apj, 383, 308

\bibitem[{{Lucy}(1999)}]{Lucy1999}
{Lucy}, L.~B. 1999, \aap, 345, 211

\bibitem[{Luminet \& Pichon(1989)}]{Luminet:1989a}
Luminet, J. \& Pichon, B. 1989, \aap, 209, 103

\bibitem[{{Lunnan} {et~al.}(2017){Lunnan}, {Kasliwal}, {Cao}, {Hangard},
  {Yaron}, {Parrent}, {McCully}, {Gal-Yam}, {Mulchaey}, {Ben-Ami},
  {Filippenko}, {Fremling}, {Fruchter}, {Howell}, {Koda}, {Kupfer}, {Kulkarni},
  {Laher}, {Masci}, {Nugent}, {Ofek}, {Yagi}, \& {Yan}}]{Lunnan2017}
{Lunnan}, R., {Kasliwal}, M.~M., {Cao}, Y., {et~al.} 2017, \apj, 836, 60

\bibitem[{{Lyman} {et~al.}(2013){Lyman}, {James}, {Perets}, {Anderson},
  {Gal-Yam}, {Mazzali}, \& {Percival}}]{Lyman2013}
{Lyman}, J.~D., {James}, P.~A., {Perets}, H.~B., {et~al.} 2013, \mnras, 434,
  527

\bibitem[{{Lyman} {et~al.}(2014){Lyman}, {Levan}, {Church}, {Davies}, \&
  {Tanvir}}]{Lyman2014b}
{Lyman}, J.~D., {Levan}, A.~J., {Church}, R.~P., {Davies}, M.~B., \& {Tanvir},
  N.~R. 2014, \mnras, 444, 2157

\bibitem[{{Lyman} {et~al.}(2016){Lyman}, {Levan}, {James}, {Angus}, {Church},
  {Davies}, \& {Tanvir}}]{Lyman2016b}
{Lyman}, J.~D., {Levan}, A.~J., {James}, P.~A., {et~al.} 2016, \mnras, 458,
  1768

\bibitem[{{MacLeod} {et~al.}(2014){MacLeod}, {Goldstein}, {Ramirez-Ruiz},
  {Guillochon}, \& {Samsing}}]{Macleod2014}
{MacLeod}, M., {Goldstein}, J., {Ramirez-Ruiz}, E., {Guillochon}, J., \&
  {Samsing}, J. 2014, \apj, 794, 9

\bibitem[{{MacLeod} {et~al.}(2016){MacLeod}, {Guillochon}, {Ramirez-Ruiz},
  {Kasen}, \& {Rosswog}}]{macleod2016_opt}
{MacLeod}, M., {Guillochon}, J., {Ramirez-Ruiz}, E., {Kasen}, D., \& {Rosswog},
  S. 2016, \apj, 819, 3

\bibitem[{{Maeda} {et~al.}(2010){Maeda}, {Taubenberger}, {Sollerman},
  {Mazzali}, {Leloudas}, {Nomoto}, \& {Motohara}}]{Maeda2010}
{Maeda}, K., {Taubenberger}, S., {Sollerman}, J., {et~al.} 2010, \apj, 708,
  1703

\bibitem[{{Maguire} {et~al.}(2018){Maguire}, {Sim}, {Shingles}, {Spyromilio},
  {Jerkstrand}, {Sullivan}, {Chen}, {Cartier}, {Dimitriadis}, {Frohmaier},
  {Galbany}, {Guti{\'e}rrez}, {Hosseinzadeh}, {Howell}, {Inserra}, {Rudy}, \&
  {Sollerman}}]{Maguire2018}
{Maguire}, K., {Sim}, S.~A., {Shingles}, L., {et~al.} 2018, \mnras, 477, 3567

\bibitem[{{Margalit} \& {Metzger}(2016)}]{Margalit2016}
{Margalit}, B. \& {Metzger}, B.~D. 2016, \mnras, 461, 1154

\bibitem[{{Mazzali} {et~al.}(2018){Mazzali}, {Ashall}, {Pian}, {Stritzinger},
  {Gall}, {Phillips}, {H{\"o}flich}, \& {Hsiao}}]{Mazzali2018}
{Mazzali}, P.~A., {Ashall}, C., {Pian}, E., {et~al.} 2018, \mnras, 476, 2905

\bibitem[{{Mazzali} {et~al.}(2000){Mazzali}, {Iwamoto}, \&
  {Nomoto}}]{Mazzali2000}
{Mazzali}, P.~A., {Iwamoto}, K., \& {Nomoto}, K. 2000, \apj, 545, 407

\bibitem[{{Mazzali} \& {Lucy}(1993)}]{Mazzali1993}
{Mazzali}, P.~A. \& {Lucy}, L.~B. 1993, \aap, 279, 447

\bibitem[{{Mazzali} {et~al.}(2011){Mazzali}, {Maurer}, {Stritzinger},
  {Taubenberger}, {Benetti}, \& {Hachinger}}]{Mazzali2011}
{Mazzali}, P.~A., {Maurer}, I., {Stritzinger}, M., {et~al.} 2011, \mnras, 416,
  881

\bibitem[{{Mazzali} {et~al.}(2010){Mazzali}, {Maurer}, {Valenti}, {Kotak}, \&
  {Hunter}}]{Mazzali2010}
{Mazzali}, P.~A., {Maurer}, I., {Valenti}, S., {Kotak}, R., \& {Hunter}, D.
  2010, \mnras, 408, 87

\bibitem[{{Mazzali} {et~al.}(2017){Mazzali}, {Sauer}, {Pian}, {Deng},
  {Prentice}, {Ben Ami}, {Taubenberger}, \& {Nomoto}}]{Mazzali2017}
{Mazzali}, P.~A., {Sauer}, D.~N., {Pian}, E., {et~al.} 2017, \mnras, 469, 2498

\bibitem[{{Mazzali} {et~al.}(2015){Mazzali}, {Sullivan}, {Filippenko},
  {Garnavich}, {Clubb}, {Maguire}, {Pan}, {Shappee}, {Silverman}, {Benetti},
  {Hachinger}, {Nomoto}, \& {Pian}}]{Mazzali2015}
{Mazzali}, P.~A., {Sullivan}, M., {Filippenko}, A.~V., {et~al.} 2015, \mnras,
  450, 2631

\bibitem[{{Mazzali} {et~al.}(2014){Mazzali}, {Sullivan}, {Hachinger}, {Ellis},
  {Nugent}, {Howell}, {Gal-Yam}, {Maguire}, {Cooke}, {Thomas}, {Nomoto}, \&
  {Walker}}]{Mazzali2014}
{Mazzali}, P.~A., {Sullivan}, M., {Hachinger}, S., {et~al.} 2014, \mnras, 439,
  1959

\bibitem[{{McCully} {et~al.}(2014){McCully}, {Jha}, {Foley}, {Chornock},
  {Holtzman}, {Balam}, {Branch}, {Filippenko}, {Frieman}, {Fynbo}, {Galbany},
  {Ganeshalingam}, {Garnavich}, {Graham}, {Hsiao}, {Leloudas}, {Leonard}, {Li},
  {Riess}, {Sako}, {Schneider}, {Silverman}, {Sollerman}, {Steele}, {Thomas},
  {Wheeler}, \& {Zheng}}]{McCully2014}
{McCully}, C., {Jha}, S.~W., {Foley}, R.~J., {et~al.} 2014, \apj, 786, 134

\bibitem[{{Metzger}(2012)}]{Metzger2012}
{Metzger}, B.~D. 2012, \mnras, 419, 827

\bibitem[{{Milisavljevic} {et~al.}(2017){Milisavljevic}, {Patnaude}, {Raymond},
  {Drout}, {Margutti}, {Kamble}, {Chornock}, {Guillochon}, {Sanders},
  {Parrent}, {Lovisari}, {Chilingarian}, {Challis}, {Kirshner}, {Penny},
  {Itagaki}, {Eldridge}, \& {Moriya}}]{Mili2017}
{Milisavljevic}, D., {Patnaude}, D.~J., {Raymond}, J.~C., {et~al.} 2017, \apj,
  846, 50

\bibitem[{{Modigliani} {et~al.}(2010){Modigliani}, {Goldoni}, {Royer},
  {Haigron}, {Guglielmi}, {Fran{\c c}ois}, {Horrobin}, {Bristow}, {Vernet},
  {Moehler}, {Kerber}, {Ballester}, {Mason}, \& {Christensen}}]{Modigliani2010}
{Modigliani}, A., {Goldoni}, P., {Royer}, F., {et~al.} 2010, in \procspie, Vol.
  7737, Observatory Operations: Strategies, Processes, and Systems III, 773728

\bibitem[{{Modjaz} {et~al.}(2014){Modjaz}, {Blondin}, {Kirshner}, {Matheson},
  {Berlind}, {Bianco}, {Calkins}, {Challis}, {Garnavich}, {Hicken}, {Jha},
  {Liu}, \& {Marion}}]{Modjaz2014}
{Modjaz}, M., {Blondin}, S., {Kirshner}, R.~P., {et~al.} 2014, \aj, 147, 99

\bibitem[{{Morgan} {et~al.}(1975){Morgan}, {Kayser}, \& {White}}]{Morgan1975}
{Morgan}, W.~W., {Kayser}, S., \& {White}, R.~A. 1975, \apj, 199, 545

\bibitem[{{Moriya} {et~al.}(2017){Moriya}, {Mazzali}, {Tominaga}, {Hachinger},
  {Blinnikov}, {Tauris}, {Takahashi}, {Tanaka}, {Langer}, \&
  {Podsiadlowski}}]{Moriya2017}
{Moriya}, T.~J., {Mazzali}, P.~A., {Tominaga}, N., {et~al.} 2017, \mnras, 466,
  2085

\bibitem[{{Mulchaey} {et~al.}(2014){Mulchaey}, {Kasliwal}, \&
  {Kollmeier}}]{Mulchaey2014}
{Mulchaey}, J.~S., {Kasliwal}, M.~M., \& {Kollmeier}, J.~A. 2014, \apjl, 780,
  L34

\bibitem[{Nagy(2018)}]{Nagy2018}
Nagy, A.~P. 2018, ApJ, 862, 143

\bibitem[{{Nicholl} {et~al.}(2019){Nicholl}, {Berger}, {Blanchard}, {Gomez}, \&
  {Chornock}}]{Nicholl2019}
{Nicholl}, M., {Berger}, E., {Blanchard}, P.~K., {Gomez}, S., \& {Chornock}, R.
  2019, \apj, 871, 102

\bibitem[{{Nomoto} {et~al.}(1994){Nomoto}, {Yamaoka}, {Pols}, {van den Heuvel},
  {Iwamoto}, {Kumagai}, \& {Shigeyama}}]{Nomoto1994}
{Nomoto}, K., {Yamaoka}, H., {Pols}, O.~R., {et~al.} 1994, \nat, 371, 227

\bibitem[{{Oke}(1990)}]{Oke1990}
{Oke}, J.~B. 1990, \aj, 99, 1621

\bibitem[{{Patat} {et~al.}(2001){Patat}, {Cappellaro}, {Danziger}, {Mazzali},
  {Sollerman}, {Augusteijn}, {Brewer}, {Doublier}, {Gonzalez}, {Hainaut},
  {Lidman}, {Leibundgut}, {Nomoto}, {Nakamura}, {Spyromilio}, {Rizzi},
  {Turatto}, {Walsh}, {Galama}, {van Paradijs}, {Kouveliotou}, {Vreeswijk},
  {Frontera}, {Masetti}, {Palazzi}, \& {Pian}}]{Patat2001}
{Patat}, F., {Cappellaro}, E., {Danziger}, J., {et~al.} 2001, \apj, 555, 900

\bibitem[{{Perets} {et~al.}(2010){Perets}, {Gal-Yam}, {Mazzali}, {Arnett},
  {Kagan}, {Filippenko}, {Li}, {Arcavi}, {Cenko}, {Fox}, {Leonard}, {Moon},
  {Sand}, {Soderberg}, {Anderson}, {James}, {Foley}, {Ganeshalingam}, {Ofek},
  {Bildsten}, {Nelemans}, {Shen}, {Weinberg}, {Metzger}, {Piro}, {Quataert},
  {Kiewe}, \& {Poznanski}}]{Perets2010}
{Perets}, H.~B., {Gal-Yam}, A., {Mazzali}, P.~A., {et~al.} 2010, \nat, 465, 322

\bibitem[{{Piascik} {et~al.}(2014){Piascik}, {Steele}, {Bates}, {Mottram},
  {Smith}, {Barnsley}, \& {Bolton}}]{Piascik2014}
{Piascik}, A.~S., {Steele}, I.~A., {Bates}, S.~D., {et~al.} 2014, in \procspie,
  Vol. 9147, Ground-based and Airborne Instrumentation for Astronomy V, 91478H

\bibitem[{{Poznanski} {et~al.}(2010){Poznanski}, {Chornock}, {Nugent}, {Bloom},
  {Filippenko}, {Ganeshalingam}, {Leonard}, {Li}, \& {Thomas}}]{Poznanski2010}
{Poznanski}, D., {Chornock}, R., {Nugent}, P.~E., {et~al.} 2010, Science, 327,
  58

\bibitem[{{Prentice} {et~al.}(2019){Prentice}, {Ashall}, {James}, {Short},
  {Mazzali}, {Bersier}, {Crowther}, {Barbarino}, {Chen}, {Copperwheat},
  {Darnley}, {Denneau}, {Elias-Rosa}, {Fraser}, {Galbany}, {Gal-Yam},
  {Harmanen}, {Howell}, {Hosseinzadeh}, {Inserra}, {Kankare}, {Karamehmetoglu},
  {Lamb}, {Limongi}, {Maguire}, {McCully}, {Olivares E}, {Piascik}, {Pignata},
  {Reichart}, {Rest}, {Reynolds}, {Rodr{\'{\i}}guez}, {Saario}, {Schulze},
  {Smartt}, {Smith}, {Sollerman}, {Stalder}, {Sullivan}, {Taddia}, {Valenti},
  {Vergani}, {Williams}, \& {Young}}]{Prentice2019}
{Prentice}, S.~J., {Ashall}, C., {James}, P.~A., {et~al.} 2019, \mnras, 485,
  1559

\bibitem[{{Prentice} {et~al.}(2018{\natexlab{a}}){Prentice}, {Ashall},
  {Mazzali}, {Zhang}, {James}, {Wang}, {Vink{\'o}}, {Percival}, {Short},
  {Piascik}, {Huang}, {Mo}, {Rui}, {Wang}, {Xiang}, {Xin}, {Yi}, {Yu}, {Zhai},
  {Zhang}, {Hosseinzadeh}, {Howell}, {McCully}, {Valenti}, {Cseh}, {Hanyecz},
  {Kriskovics}, {P{\'a}l}, {S{\'a}rneczky}, {S{\'o}dor}, {Szak{\'a}ts},
  {Sz{\'e}kely}, {Varga-Vereb{\'e}lyi}, {Vida}, {Bradac}, {Reichart}, {Sand},
  \& {Tartaglia}}]{Prentice2018a}
{Prentice}, S.~J., {Ashall}, C., {Mazzali}, P.~A., {et~al.} 2018{\natexlab{a}},
  \mnras, 478, 4162

\bibitem[{{Prentice} {et~al.}(2018{\natexlab{b}}){Prentice}, {Maguire},
  {Smartt}, {Magee}, {Schady}, {Sim}, {Chen}, {Clark}, {Colin}, {Fulton},
  {McBrien}, {O{'}Neill}, {Smith}, {Ashall}, {Chambers}, {Denneau},
  {Flewelling}, {Heinze}, {Holoien}, {Huber}, {Kochanek}, {Mazzali}, {Prieto},
  {Rest}, {Shappee}, {Stalder}, {Stanek}, {Stritzinger}, {Thompson}, \&
  {Tonry}}]{Prentice2018b}
{Prentice}, S.~J., {Maguire}, K., {Smartt}, S.~J., {et~al.} 2018{\natexlab{b}},
  \apjl, 865, L3

\bibitem[{{Prentice} \& {Mazzali}(2017)}]{Prentice2017}
{Prentice}, S.~J. \& {Mazzali}, P.~A. 2017, \mnras, 469, 2672

\bibitem[{{Prentice} {et~al.}(2016){Prentice}, {Mazzali}, {Pian}, {Gal-Yam},
  {Kulkarni}, {Rubin}, {Corsi}, {Fremling}, {Sollerman}, {Yaron}, {Arcavi},
  {Zheng}, {Kasliwal}, {Filippenko}, {Cenko}, {Cao}, \&
  {Nugent}}]{Prentice2016}
{Prentice}, S.~J., {Mazzali}, P.~A., {Pian}, E., {et~al.} 2016, \mnras, 458,
  2973

\bibitem[{Pursiainen {et~al.}(2018)Pursiainen, Childress, Smith, Prajs,
  Sullivan, Davis, Foley, Asorey, Calcino, Carollo, \& et~al.}]{Pursiainen2018}
Pursiainen, M., Childress, M., Smith, M., {et~al.} 2018, MNRAS, 481, 894

\bibitem[{{Richmond} {et~al.}(1994){Richmond}, {Treffers}, {Filippenko},
  {Paik}, {Leibundgut}, {Schulman}, \& {Cox}}]{Richmond1994}
{Richmond}, M.~W., {Treffers}, R.~R., {Filippenko}, A.~V., {et~al.} 1994, \aj,
  107, 1022

\bibitem[{{Rosswog} {et~al.}(2009){Rosswog}, {Kasen}, {Guillochon}, \&
  {Ramirez-Ruiz}}]{rosswog2009}
{Rosswog}, S., {Kasen}, D., {Guillochon}, J., \& {Ramirez-Ruiz}, E. 2009,
  \apjl, 705, L128

\bibitem[{{Rosswog} {et~al.}(2008){Rosswog}, {Ramirez-Ruiz}, \&
  {Hix}}]{Rosswog2008}
{Rosswog}, S., {Ramirez-Ruiz}, E., \& {Hix}, W.~R. 2008, \apj, 679, 1385

\bibitem[{{Sauer} {et~al.}(2006){Sauer}, {Mazzali}, {Deng}, {Valenti},
  {Nomoto}, \& {Filippenko}}]{Sauer2006}
{Sauer}, D.~N., {Mazzali}, P.~A., {Deng}, J., {et~al.} 2006, \mnras, 369, 1939

\bibitem[{{Schlafly} \& {Finkbeiner}(2011)}]{Schlafly2011}
{Schlafly}, E.~F. \& {Finkbeiner}, D.~P. 2011, \apj, 737, 103

\bibitem[{{Sell} {et~al.}(2018){Sell}, {Arur}, {Maccarone}, {Kotak}, {Knigge},
  {Sand}, \& {Valenti}}]{sell2018_16hnk}
{Sell}, P.~H., {Arur}, K., {Maccarone}, T.~J., {et~al.} 2018, \mnras, 475, L111

\bibitem[{{Sell} {et~al.}(2015){Sell}, {Maccarone}, {Kotak}, {Knigge}, \&
  {Sand}}]{Sell2015}
{Sell}, P.~H., {Maccarone}, T.~J., {Kotak}, R., {Knigge}, C., \& {Sand}, D.~J.
  2015, \mnras, 450, 4198

\bibitem[{{Shen} {et~al.}(2010){Shen}, {Kasen}, {Weinberg}, {Bildsten}, \&
  {Scannapieco}}]{Shen2010}
{Shen}, K.~J., {Kasen}, D., {Weinberg}, N.~N., {Bildsten}, L., \&
  {Scannapieco}, E. 2010, \apj, 715, 767

\bibitem[{{Sim} {et~al.}(2012){Sim}, {Fink}, {Kromer}, {R{\"o}pke}, {Ruiter},
  \& {Hillebrandt}}]{Sim2012}
{Sim}, S.~A., {Fink}, M., {Kromer}, M., {et~al.} 2012, \mnras, 420, 3003

\bibitem[{{Sim} {et~al.}(2007){Sim}, {Sauer}, {R{\"o}pke}, \&
  {Hillebrandt}}]{Sim2007}
{Sim}, S.~A., {Sauer}, D.~N., {R{\"o}pke}, F.~K., \& {Hillebrandt}, W. 2007,
  \mnras, 378, 2

\bibitem[{{Smartt} {et~al.}(2017){Smartt}, {Chen}, {Jerkstrand}, {Coughlin},
  {Kankare}, {Sim}, {Fraser}, {Inserra}, {Maguire}, {Chambers}, {Huber},
  {Kr{\"u}hler}, {Leloudas}, {Magee}, {Shingles}, {Smith}, {Young}, {Tonry},
  {Kotak}, {Gal-Yam}, {Lyman}, {Homan}, {Agliozzo}, {Anderson}, {Angus},
  {Ashall}, {Barbarino}, {Bauer}, {Berton}, {Botticella}, {Bulla}, {Bulger},
  {Cannizzaro}, {Cano}, {Cartier}, {Cikota}, {Clark}, {De Cia}, {Della Valle},
  {Denneau}, {Dennefeld}, {Dessart}, {Dimitriadis}, {Elias-Rosa}, {Firth},
  {Flewelling}, {Fl{\"o}rs}, {Franckowiak}, {Frohmaier}, {Galbany},
  {Gonz{\'a}lez-Gait{\'a}n}, {Greiner}, {Gromadzki}, {Guelbenzu},
  {Guti{\'e}rrez}, {Hamanowicz}, {Hanlon}, {Harmanen}, {Heintz}, {Heinze},
  {Hernandez}, {Hodgkin}, {Hook}, {Izzo}, {James}, {Jonker}, {Kerzendorf},
  {Klose}, {Kostrzewa-Rutkowska}, {Kowalski}, {Kromer}, {Kuncarayakti},
  {Lawrence}, {Lowe}, {Magnier}, {Manulis}, {Martin-Carrillo}, {Mattila},
  {McBrien}, {M{\"u}ller}, {Nordin}, {O'Neill}, {Onori}, {Palmerio},
  {Pastorello}, {Patat}, {Pignata}, {Podsiadlowski}, {Pumo}, {Prentice}, {Rau},
  {Razza}, {Rest}, {Reynolds}, {Roy}, {Ruiter}, {Rybicki}, {Salmon}, {Schady},
  {Schultz}, {Schweyer}, {Seitenzahl}, {Smith}, {Sollerman}, {Stalder},
  {Stubbs}, {Sullivan}, {Szegedi}, {Taddia}, {Taubenberger}, {Terreran}, {van
  Soelen}, {Vos}, {Wainscoat}, {Walton}, {Waters}, {Weiland}, {Willman},
  {Wiseman}, {Wright}, {Wyrzykowski}, \& {Yaron}}]{Smartt2017}
{Smartt}, S.~J., {Chen}, T.-W., {Jerkstrand}, A., {et~al.} 2017, \nat, 551, 75

\bibitem[{{Smartt} {et~al.}(2015){Smartt}, {Valenti}, {Fraser}, {Inserra},
  {Young}, {Sullivan}, {Pastorello}, {Benetti}, {Gal-Yam}, {Knapic},
  {Molinaro}, {Smareglia}, {Smith}, {Taubenberger}, {Yaron}, {Anderson},
  {Ashall}, {Balland}, {Baltay}, {Barbarino}, {Bauer}, {Baumont}, {Bersier},
  {Blagorodnova}, {Bongard}, {Botticella}, {Bufano}, {Bulla}, {Cappellaro},
  {Campbell}, {Cellier-Holzem}, {Chen}, {Childress}, {Clocchiatti},
  {Contreras}, {Dall'Ora}, {Danziger}, {de Jaeger}, {De Cia}, {Della Valle},
  {Dennefeld}, {Elias-Rosa}, {Elman}, {Feindt}, {Fleury}, {Gall},
  {Gonzalez-Gaitan}, {Galbany}, {Morales Garoffolo}, {Greggio}, {Guillou},
  {Hachinger}, {Hadjiyska}, {Hage}, {Hillebrandt}, {Hodgkin}, {Hsiao}, {James},
  {Jerkstrand}, {Kangas}, {Kankare}, {Kotak}, {Kromer}, {Kuncarayakti},
  {Leloudas}, {Lundqvist}, {Lyman}, {Hook}, {Maguire}, {Manulis}, {Margheim},
  {Mattila}, {Maund}, {Mazzali}, {McCrum}, {McKinnon}, {Moreno-Raya},
  {Nicholl}, {Nugent}, {Pain}, {Pignata}, {Phillips}, {Polshaw}, {Pumo},
  {Rabinowitz}, {Reilly}, {Romero-Ca{\~n}izales}, {Scalzo}, {Schmidt},
  {Schulze}, {Sim}, {Sollerman}, {Taddia}, {Tartaglia}, {Terreran},
  {Tomasella}, {Turatto}, {Walker}, {Walton}, {Wyrzykowski}, {Yuan}, \&
  {Zampieri}}]{Smartt2015}
{Smartt}, S.~J., {Valenti}, S., {Fraser}, M., {et~al.} 2015, \aap, 579, A40

\bibitem[{{Smith} {et~al.}(2019){Smith}, {Williams}, {Young}, {Ibsen},
  {Smartt}, {Lawrence}, {Morris}, {Voutsinas}, \& {Nicholl}}]{Smith2019}
{Smith}, K.~W., {Williams}, R.~D., {Young}, D.~R., {et~al.} 2019, Research
  Notes of the American Astronomical Society, 3, 26

\bibitem[{{Steele} {et~al.}(2004){Steele}, {Smith}, {Rees}, {Baker}, {Bates},
  {Bode}, {Bowman}, {Carter}, {Etherton}, {Ford}, {Fraser}, {Gomboc}, {Lett},
  {Mansfield}, {Marchant}, {Medrano-Cerda}, {Mottram}, {Raback}, {Scott},
  {Tomlinson}, \& {Zamanov}}]{Steele2004}
{Steele}, I.~A., {Smith}, R.~J., {Rees}, P.~C., {et~al.} 2004, in \procspie,
  Vol. 5489, Ground-based Telescopes, ed. J.~M. {Oschmann}, Jr., 679--692

\bibitem[{{Stritzinger} \& {Leibundgut}(2005)}]{Stritzinger2005}
{Stritzinger}, M. \& {Leibundgut}, B. 2005, \aap, 431, 423

\bibitem[{{Suh} {et~al.}(2011){Suh}, {Yoon}, {Jeong}, \& {Yi}}]{Suh2011}
{Suh}, H., {Yoon}, S.-c., {Jeong}, H., \& {Yi}, S.~K. 2011, \apj, 730, 110

\bibitem[{{Summa} {et~al.}(2018){Summa}, {Janka}, {Melson}, \&
  {Marek}}]{Summa2018}
{Summa}, A., {Janka}, H.-T., {Melson}, T., \& {Marek}, A. 2018, \apj, 852, 28

\bibitem[{{Taddia} {et~al.}(2018){Taddia}, {Stritzinger}, {Bersten}, {Baron},
  {Burns}, {Contreras}, {Holmbo}, {Hsiao}, {Morrell}, {Phillips}, {Sollerman},
  \& {Suntzeff}}]{Taddia2018}
{Taddia}, F., {Stritzinger}, M.~D., {Bersten}, M., {et~al.} 2018, \aap, 609,
  A136

\bibitem[{{Taubenberger} {et~al.}(2013){Taubenberger}, {Kromer}, {Pakmor},
  {Pignata}, {Maeda}, {Hachinger}, {Leibundgut}, \& {Hillebrand
  t}}]{Taubenberger2013}
{Taubenberger}, S., {Kromer}, M., {Pakmor}, R., {et~al.} 2013, \apjl, 775, L43

\bibitem[{{Taubenberger} {et~al.}(2009){Taubenberger}, {Valenti}, {Benetti},
  {Cappellaro}, {Della Valle}, {Elias-Rosa}, {Hachinger}, {Hillebrandt},
  {Maeda}, {Mazzali}, {Pastorello}, {Patat}, {Sim}, \&
  {Turatto}}]{Taubenberger2009}
{Taubenberger}, S., {Valenti}, S., {Benetti}, S., {et~al.} 2009, \mnras, 397,
  677

\bibitem[{{Tauris} {et~al.}(2013){Tauris}, {Langer}, {Moriya}, {Podsiadlowski},
  {Yoon}, \& {Blinnikov}}]{Tauris2013}
{Tauris}, T.~M., {Langer}, N., {Moriya}, T.~J., {et~al.} 2013, \apjl, 778, L23

\bibitem[{{Tauris} {et~al.}(2015){Tauris}, {Langer}, \&
  {Podsiadlowski}}]{Tauris2015}
{Tauris}, T.~M., {Langer}, N., \& {Podsiadlowski}, P. 2015, \mnras, 451, 2123

\bibitem[{{Tonry} {et~al.}(2019){Tonry}, {Denneau}, {Heinze}, {Weiland},
  {Flewelling}, {Stalder}, {Rest}, {Stubbs}, {Smith}, {Smartt}, {Young},
  {Maguire}, {Prentice}, {McBrien}, {O'Neill}, {Clark}, {Magee}, {Fulton},
  {Mccormack}, \& {Wright}}]{Tonry19bkcTNS}
{Tonry}, J., {Denneau}, L., {Heinze}, A., {et~al.} 2019, Transient Name Server
  Discovery Report, 310

\bibitem[{{Tonry} {et~al.}(2018){Tonry}, {Denneau}, {Heinze}, {Stalder},
  {Smith}, {Smartt}, {Stubbs}, {Weiland}, \& {Rest}}]{Tonry2018}
{Tonry}, J.~L., {Denneau}, L., {Heinze}, A.~N., {et~al.} 2018, \pasp, 130,
  064505

\bibitem[{{Valenti} {et~al.}(2008){Valenti}, {Benetti}, {Cappellaro}, {Patat},
  {Mazzali}, {Turatto}, {Hurley}, {Maeda}, \& {Gal-Yam}}]{Valenti2008}
{Valenti}, S., {Benetti}, S., {Cappellaro}, E., {et~al.} 2008, \mnras, 383,
  1485

\bibitem[{{Valenti} {et~al.}(2014){Valenti}, {Yuan}, {Taubenberger}, {Maguire},
  {Pastorello}, {Benetti}, {Smartt}, {Cappellaro}, {Howell}, {Bildsten},
  {Moore}, {Stritzinger}, {Anderson}, {Benitez-Herrera}, {Bufano},
  {Gonzalez-Gaitan}, {McCrum}, {Pignata}, {Fraser}, {Gal-Yam}, {Le Guillou},
  {Inserra}, {Reichart}, {Scalzo}, {Sullivan}, {Yaron}, \&
  {Young}}]{Valenti2014}
{Valenti}, S., {Yuan}, F., {Taubenberger}, S., {et~al.} 2014, \mnras, 437, 1519

\bibitem[{{Vernet} {et~al.}(2011){Vernet}, {Dekker}, {D'Odorico}, {Kaper},
  {Kjaergaard}, {Hammer}, {Randich}, {Zerbi}, {Groot}, {Hjorth}, {Guinouard},
  {Navarro}, {Adolfse}, {Albers}, {Amans}, {Andersen}, {Andersen}, {Binetruy},
  {Bristow}, {Castillo}, {Chemla}, {Christensen}, {Conconi}, {Conzelmann},
  {Dam}, {de Caprio}, {de Ugarte Postigo}, {Delabre}, {di Marcantonio},
  {Downing}, {Elswijk}, {Finger}, {Fischer}, {Flores}, {Fran{\c c}ois},
  {Goldoni}, {Guglielmi}, {Haigron}, {Hanenburg}, {Hendriks}, {Horrobin},
  {Horville}, {Jessen}, {Kerber}, {Kern}, {Kiekebusch}, {Kleszcz}, {Klougart},
  {Kragt}, {Larsen}, {Lizon}, {Lucuix}, {Mainieri}, {Manuputy}, {Martayan},
  {Mason}, {Mazzoleni}, {Michaelsen}, {Modigliani}, {Moehler}, {M{\o}ller},
  {Norup S{\o}rensen}, {N{\o}rregaard}, {P{\'e}roux}, {Patat}, {Pena}, {Pragt},
  {Reinero}, {Rigal}, {Riva}, {Roelfsema}, {Royer}, {Sacco}, {Santin},
  {Schoenmaker}, {Spano}, {Sweers}, {Ter Horst}, {Tintori}, {Tromp}, {van
  Dael}, {van der Vliet}, {Venema}, {Vidali}, {Vinther}, {Vola}, {Winters},
  {Wistisen}, {Wulterkens}, \& {Zacchei}}]{vernet2011}
{Vernet}, J., {Dekker}, H., {D'Odorico}, S., {et~al.} 2011, \aap, 536, A105

\bibitem[{{Waldman} {et~al.}(2011){Waldman}, {Sauer}, {Livne}, {Perets},
  {Glasner}, {Mazzali}, {Truran}, \& {Gal-Yam}}]{Waldman2011}
{Waldman}, R., {Sauer}, D., {Livne}, E., {et~al.} 2011, \apj, 738, 21

\bibitem[{{Wongwathanarat} {et~al.}(2013){Wongwathanarat}, {Janka}, \&
  {M{\"u}ller}}]{Wongwathanrat2013}
{Wongwathanarat}, A., {Janka}, H.-T., \& {M{\"u}ller}, E. 2013, \aap, 552, A126

\bibitem[{{Woosley} \& {Kasen}(2011)}]{Woosley2011}
{Woosley}, S.~E. \& {Kasen}, D. 2011, \apj, 734, 38

\bibitem[{{Yuan} {et~al.}(2013){Yuan}, {Kobayashi}, {Schmidt}, {Podsiadlowski},
  {Sim}, \& {Scalzo}}]{Yuan2013}
{Yuan}, F., {Kobayashi}, C., {Schmidt}, B.~P., {et~al.} 2013, \mnras, 432, 1680

\end{thebibliography}


\end{document}